\documentclass[longauth]{aa}  

\usepackage{txfonts}
\usepackage[T1]{fontenc}
\pdfoutput=1 
\usepackage{ae,aecompl}
\usepackage{amssymb,amsmath,amsfonts}
\usepackage{graphicx}
\graphicspath{{./figures/}}
\usepackage{hyperref}
\hypersetup{colorlinks=true,linkcolor=blue,citecolor=blue,filecolor=blue,urlcolor=blue}

\usepackage{siunitx}    

\usepackage{makecell}

\newcommand\Tstrut{\rule{0pt}{2.6ex}}         
\newcommand\Bstrut{\rule[-0.9ex]{0pt}{0pt}}   

\usepackage{textcomp}
\usepackage{threeparttable}

\usepackage{subcaption}

\usepackage{multirow}

\begin{document}
   \title{The LOFAR Two Metre Sky Survey: Deep Fields Data Release 1}

   \subtitle{III. Host-galaxy identifications and value added catalogues}
   
\authorrunning{Kondapally et~al.}
\titlerunning{LoTSS Deep DR1: Host-Galaxy Identifications}

\author{R.~Kondapally\inst{\ref{i:ed}}\fnmsep\thanks{E-mail: rohitk@roe.ac.uk}\and P.~N.~Best\inst{\ref{i:ed}}\and
M.~J.~Hardcastle\inst{\ref{i:herts}}\and D.~Nisbet\inst{\ref{i:ed}}\and M.~Bonato\inst{\ref{i:inaf},\ref{i:alma_bologna},\ref{i:padova}}\and J.~Sabater\inst{\ref{i:ed}}\and K.~J.~Duncan\inst{\ref{i:leiden},\ref{i:ed}}\and I.~McCheyne\inst{\ref{i:sussex}}\and R.~K.~Cochrane\inst{\ref{i:harvard}}\and R.~A.~A.~Bowler\inst{\ref{i:oxford}}\and W.~L.~Williams\inst{\ref{i:leiden}}\and T.~W.~Shimwell\inst{\ref{i:astron},\ref{i:leiden}}\and C.~Tasse\inst{\ref{i:gepi},\ref{i:rhodes}}\and 
J.~H.~Croston\inst{\ref{i:ou}} \and A.~Goyal\inst{\ref{i:aoj}}\and M.~Jamrozy\inst{\ref{i:aoj}}\and M.~J.~Jarvis\inst{\ref{i:oxford},\ref{i:wcape}}\and V.~H.~Mahatma\inst{\ref{i:herts}}\and H.~J.~A.~R\"{o}ttgering\inst{\ref{i:leiden}}\and D.~J.~B.~Smith\inst{\ref{i:herts}}\and A.~Wo\l{}owska\inst{\ref{i:ncu}}\and 
M.~Bondi\inst{\ref{i:inaf}}\and 
M.~Brienza\inst{\ref{i:bologna},\ref{i:inaf}}\and 
M.~J.~I.~Brown\inst{\ref{i:monash}}\and 
M.~Br\"{u}ggen\inst{\ref{i:hamburg}}\and 
K.~Chambers\inst{\ref{i:hawaii}}\and 
M.~A.~Garrett\inst{\ref{i:manchester},\ref{i:leiden}}\and 
G.~G\"urkan\inst{\ref{i:csiro}}\and 
M.~Huber\inst{\ref{i:hawaii}}\and 
M.~Kunert-Bajraszewska\inst{\ref{i:ncu}}\and 
E.~Magnier\inst{\ref{i:hawaii}}\and 
B.~Mingo\inst{\ref{i:ou}}\and 
R.~Mostert\inst{\ref{i:leiden},\ref{i:astron}}\and 
B.~Nikiel-Wroczy\'nski\inst{\ref{i:aoj}}\and 
S.~P.~O'Sullivan\inst{\ref{i:dublin}}\and 
R.~Paladino\inst{\ref{i:inaf}}\and 
T.~Ploeckinger\inst{\ref{i:leiden}}\and 
I.~Prandoni\inst{\ref{i:inaf}}\and 
M.~J.~Rosenthal\inst{\ref{i:leiden}}\and 
D.~J.~Schwarz\inst{\ref{i:bielefeld}}\and 
A.~Shulevski\inst{\ref{i:amsterdam}}\and 
J.~D.~Wagenveld\inst{\ref{i:maxplanck}}\and 
L.~Wang\inst{\ref{i:groningen},\ref{i:kapteyn}}}

\institute{SUPA, Institute for Astronomy, Royal Observatory, Blackford Hill, Edinburgh, EH9 3HJ, UK\label{i:ed} \and 
Centre for Astrophysics Research, School of Physics, Astronomy and Mathematics, University of Hertfordshire, College Lane, Hatfield AL10 9AB, UK\label{i:herts} \and
INAF -- Istituto di Radioastronomia, via Gobetti 101, 40129 Bologna, Italy\label{i:inaf} \and
Italian ALMA Regional Centre, Via Gobetti 101, I-40129, Bologna, Italy\label{i:alma_bologna} \and
INAF-Osservatorio Astronomico di Padova, Vicolo dell'Osservatorio 5, I-35122, Padova, Italy\label{i:padova} \and
Leiden Observatory, Leiden University, PO Box 9513, NL-2300 RA Leiden, The Netherlands\label{i:leiden} \and
Astronomy Centre, Department of Physics \& Astronomy, University of Sussex, Brighton, BN1 9QH, England, UK\label{i:sussex} \and
Harvard-Smithsonian Center for Astrophysics, 60 Garden St, Cambridge, MA 02138, USA\label{i:harvard} \and
Astrophysics, University of Oxford, Keble Road, Oxford, OX1 3RH\label{i:oxford} \and
ASTRON, Netherlands Institute for Radio Astronomy, Oude Hoogeveensedijk 4, 7991 PD, Dwingeloo, The Netherlands\label{i:astron} \and 
GEPI, Observatoire de Paris, CNRS, Universite Paris Diderot, 5 place Jules Janssen, 92190 Meudon, France\label{i:gepi} \and
Department of Physics \& Electronics, Rhodes University, PO Box 94, Grahamstown, 6140, South Africa\label{i:rhodes} \and
School of Physical Sciences, The Open University, Walton Hall, Milton Keynes, MK7 6AA, UK\label{i:ou} \and
Astronomical Observatory, Jagiellonian University, ul. Orla 171, 30–244 Krak\'ow, Poland\label{i:aoj} \and
Department of Physics \& Astronomy, University of the Western Cape, Private Bag X17, Bellville, Cape Town, 7535, South Africa\label{i:wcape} \and
Institute of Astronomy, Faculty of Physics, Astronomy and Informatics, NCU, Grudziadzka 5, 87-100 Toru\'n, Poland\label{i:ncu} \and
Dipartimento di Fisica e Astronomia, Università di Bologna, via P. Gobetti 93/2, 40129, Bologna, Italy\label{i:bologna} \and
School of Physics and Astronomy, Monash University, Clayton, Victoria 3800, Australia\label{i:monash} \and
Hamburger Sternwarte, University of Hamburg, Gojenbergsweg 112, 21029 Hamburg, Germany\label{i:hamburg} \and
Institute for Astronomy, University of Hawaii at Manoa, Honolulu, HI 96822, USA\label{i:hawaii} \and
Jodrell Bank Centre for Astrophysics, University of Manchester, Alan Turing Building, Oxford Road, M13 9PL, UK\label{i:manchester} \and
CSIRO Astronomy and Space Science, PO Box 1130, Bentley WA 6102, Australia\label{i:csiro} \and
School of Physical Sciences and Centre for Astrophysics \& Relativity, Dublin City University, Glasnevin, D09 W6Y4, Ireland\label{i:dublin} \and
Fakult\"{a}t f\"{u}r Physik, Universit\"{a}t Bielefeld, Postfach 100131, 33501 Bielefeld, Germany\label{i:bielefeld} \and
Anton Pannekoek Institute for Astronomy, University of Amsterdam, Postbus 94249, 1090 GE Amsterdam, The Netherlands\label{i:amsterdam} \and
Max-Planck Institut f\"{u}r Radioastronomie, Auf dem H\"{u}gel 69, 53121 Bonn, Germany\label{i:maxplanck} \and
SRON Netherlands Institute for Space Research, Landleven 12, 9747 AD, Groningen, The Netherlands\label{i:groningen} \and
Kapteyn Astronomical Institute, University of Groningen, Postbus 800, 9700 AV Groningen, The Netherlands\label{i:kapteyn}
}

   \date{Received July 01, 2020; accepted October 26, 2020}

\abstract{We present the source associations, cross-identifications, and multi-wavelength properties of the faint radio source population detected in the deep tier of the LOFAR Two Metre Sky Survey (LoTSS): the LoTSS Deep Fields. The first LoTSS Deep Fields data release consists of deep radio imaging at 150~MHz of the ELAIS-N1, Lockman Hole, and Bo\"{o}tes fields, down to RMS sensitives of around 20, 22, and 32$~\mu$Jy\,beam$^{-1}$, respectively. These fields are some of the best studied extra-galactic fields in the northern sky, with existing deep, wide-area panchromatic photometry from X-ray to infrared wavelengths, covering a total of $\approx$~26~\mbox{deg$^{2}$}. We first generated improved multi-wavelength catalogues in ELAIS-N1 and Lockman Hole; combined with the existing catalogue for Bo\"{o}tes, we present forced, matched aperture photometry for over 7.2 million sources across the three fields. We identified multi-wavelength counterparts to the radio detected sources, using a combination of the Likelihood Ratio method and visual classification, which greatly enhances the scientific potential of radio surveys and allows for the characterisation of the photometric redshifts and the physical properties of the host galaxies. The final radio-optical cross-matched catalogue consists of 81\,951 radio-detected sources, with counterparts identified and multi-wavelength properties presented for 79\,820 ($>$97\%) sources. We also examine the properties of the host galaxies, and through stacking analysis find that the radio population with no identified counterpart is likely dominated by AGN at $z\sim3-4$. This dataset contains one of the largest samples of radio-selected star-forming galaxies and active galactic nuclei (AGN) at these depths, making it ideal for studying the history of star-formation, and the evolution of galaxies and AGN across cosmic time.
}


   \maketitle
%

\defcitealias{tasse2020_inpress}{Paper-I}
\defcitealias{sabater2020_inpress}{Paper-II}
\defcitealias{duncan2020_inpress}{Paper-IV}
\defcitealias{best2020_inpress}{Paper-V}
\defcitealias{2019A&A...622A...2W}{W19}
\defcitealias{tasse2020_inpress,sabater2020_inpress}{Papers I \& II}

\section{Introduction}\label{sec:intro}
Radio wavelengths offer a unique window to study both the build-up of stars and the formation and growth of supermassive black holes (SMBHs) across cosmic time. In the nearby Universe, large-area radio surveys such as the National Radio Astronomy Observatory (NRAO) Very Large Array (VLA) Sky Survey (NVSS; \citealt{condon1998nvss}) and the Faint Images of the Radio Sky at Twenty centimetres (FIRST; \citealt{becker1995first}) have been instrumental in allowing the selection of large, robust statistical samples of both radio-active galactic nuclei (AGN) and star-forming galaxies. The combination of these radio surveys with complementary multi-wavelength and spectroscopic surveys, such as the Sloan Digital Sky Survey (SDSS; \citealt{york2000sdss}), the Two-Micron All Sky Survey (2MASS; \citealt{skrutskie2006AJ_2mass}), the Two-degree Field Galaxy Redshift Survey (2dFGRS; \citealt{colless2001_2dfgrs}), and successors, has dramatically improved our understanding of the formation and evolution of galaxies, enabling studies of AGN physics, the properties of the host galaxies (e.g. stellar mass, black hole mass, age, morphology, environment) of radio AGN and their role in regulating star-formation and the growth of galaxies (e.g. \citealt{sadler2002_nvss_2dfgrs,2005MNRAS.362...25B,best2005_sdss_xmatch,mauch2007_radio_lf,2007MNRAS.379..894B,donso2009_radio_agn_evol,2012MNRAS.421.1569B}; see review by \citealt{2014ARA&A..52..589H}). The local radio luminosity function (LF) has also been used to estimate the star formation rate density (SFRD; e.g. \citealt{yun2001_radio_iras_sfr,condon2002_radio_sf,sadler2002_nvss_2dfgrs, mauch2007_radio_lf}).

Extending these analyses to higher redshifts to study the history of both star-formation and AGN activity to beyond the cosmic noon remain key objectives in galaxy formation and evolution studies. However, such studies are typically limited to small-area fields with deep multi-wavelength and spectroscopic datasets, such as VLA-GOODS-N \citep{morrison2010goodsn_vla}, VVDS-VLA \citep{bondi2003vvds}, XMM-LSS \citep{tasse2006xmmlss}, and VLA-COSMOS \citep{schinnerer2007vla_cosmos_14G,smolcic2017_vla_cosmos_3G}. These deep surveys have helped to trace the history of star-formation, in a manner unaffected by dust absorption, thus constraining the dust-unbiased SFRD (e.g. \citealt{novak2017sfr_history}). They have also enabled the first studies of the evolution of the low luminosity AGN (e.g. \citealt{2014MNRAS.445..955B,2016MNRAS.460....2P,smolcic2017agn_evol_vla,butler2019xxls_agn}) as well as allowing the detection and characterisation of dust-obscured AGN \citep[e.g.][]{webster1995red_quasars,gregg2002redqso}. However, even fields as large as COSMOS ($\sim$2 \mbox{deg$^{2}$}) are subject to limited source statistics and cosmic variance effects; surveys covering large areas across many sight-lines are required to minimise these effects and to detect statistical samples of rare objects.

In the near future, the advent of the next generation of radio telescopes, such as the Square Kilometre Array (SKA; \citealt{dewdney2009ska}) and its pathfinders, in conjunction with other multi-wavelength facilities, such as Euclid \citep{amendola2018_euclid} and the Vera C. Rubin Legacy Survey of Space and Time (LSST; \citealt{ivezic2019_lsst}), will provide a revolutionary increase in survey speed, sensitivity, and source counts. The combination of these datasets will transform our understanding of the faint radio source population over the next decades, detecting orders of magnitude of more sources over large sky areas, down to sensitivities below what is even possible in the current small-area deep fields. The Low Frequency Array (LOFAR; \citealt{2013A&A...556A...2V}) Two Metre Sky Survey (LoTSS; \citealt{2017A&A...598A.104S,2019A&A...622A...1S}) Deep Fields project aims to bridge this gap between the current deep narrow-area and future ultra-deep, wide-area radio surveys.

LoTSS is currently mapping all of the northern sky to a high sensitivity and resolution ($\mathrm{S_{150MHz}}\sim$0.1mJy \mbox{beam$^{-1}$} and FWHM $\sim$6$\arcsec$) at the relatively unexplored 120-168~MHz frequencies. In parallel with this, LOFAR is also undertaking deep observations of best studied multi-wavelength, degree scale fields in the northern sky, as part of the deep tier of LoTSS: the LoTSS Deep Fields (\citealt{tasse2020_inpress} and \citealt{sabater2020_inpress}; hereafter \citetalias{tasse2020_inpress} and \citetalias{sabater2020_inpress}). The first three LoTSS Deep Fields are the European Large-Area ISO Survey-North 1 (ELAIS-N1; \citealt{oliver2000elais}), Lockman Hole, and Bo\"{o}tes \citep{1999ASPC..191..111J}; these were chosen to have extensive multi-wavelength coverage from past and ongoing deep, wide-area surveys sampling the X-ray (e.g. \citealt{Brandt2001chandra_dfn_lh,hasinger2001_xmmnewton_lh,manners2003_chandra_en1,murray2005_xbootes}), ultra-violet (UV; e.g. \citealt{2005ApJ...619L...1M,2007ApJS..173..682M})
to optical (e.g. \citealt{1999ASPC..191..111J,2007ApJS..169...21C,2009ApJ...698.1934M,2009ApJ...698.1943W,2016arXiv161205560C,2017AAS...22923706H,2018PASJ...70S...8A}) and to infrared (IR; e.g. \citealt{2003PASP..115..897L,2007MNRAS.379.1599L,2009ApJ...701..428A,whitaker2011newfirm,2012PASP..124..714M,oliver2012hermes}) wavelengths; this is ideal for a wide range of our scientific objectives. These fields also benefit from additional radio observations at higher frequencies from the Giant Metrewave Radio Telescope (GMRT; e.g. \citealt{garn2008_gmrt_lh,garn2008_en1_gmrt_610,sirothia2009_en1_gmrt_325,intema2011_gmrt_bootes,ocran2019gmrten1_ids,iswharachandra2020gmrt_en1_wide}) and the VLA (e.g. \citealt{ciliegi1999_vla_en1_small,ibar2009_vla_lh}). The current LoTSS Deep Fields dataset, covering $\sim$~26 \mbox{deg$^{2}$} (including multi-wavelength coverage) and reaching an unprecedented depth of $\rm{S_{150MHz}} \sim$20 \mbox{$\mu$Jy\,beam$^{-1}$}, is comparable in depth to the deepest existing radio continuum surveys (e.g. VLA-COSMOS) but with more than an order of magnitude larger sky-area coverage. With this combination of deep, high-quality radio and multi-wavelength data over tens of square degrees, and along multiple sight-lines, the LoTSS Deep Fields are now able to probe a cosmological volume large enough to sample all galaxy environments to beyond $z\sim1$, minimise the effects of cosmic variance (to an estimated level of) $\sim$ 4\% for $0.5 < z < 1.0$; \citealt{driver2010cosvariance}), and build statistical radio-selected samples of AGN and star-forming galaxies, even when simultaneously split by various physical parameters.

Identifying multi-wavelength counterparts of radio sources is vital in maximising the scientific potential of radio surveys. This allows for the classification of radio sources, characterisation of their hosts, and spectral energy distribution (SED) fitting to determine photometric redshifts and many redshift-dependent physical parameters such as luminosities, stellar masses, and star-formation rates. Extensive cross-matching efforts are therefore common for deep radio surveys, for example, the LoTSS Data Release 1 \citep{2019A&A...622A...2W,2019A&A...622A...3D}, VLA-COSMOS 3GHz Large Project \citep{smolcic2017_vla_3ghz_counterparts}, XXL-S Survey \citep{ciliegi2018xxls_optids}.

The identification of radio source counterparts and subsequent SED fitting and photometric redshift estimates rely upon having a complete, homogeneous sample of objects measured across all optical to IR wavelengths. To achieve this, we build a forced, matched aperture, multi-wavelength catalogue in each field spanning the UV to mid-infrared wavelengths using the latest deep datasets. This higher quality multi-wavelength catalogue is then used for cross-identification of radio sources in this paper, for photometric redshift estimates (see \citealt{duncan2020_inpress}; hereafter \citetalias{duncan2020_inpress}) and, for detailed SED fitting to allow source classification and characterisation (see \citealt{best2020_inpress}; hereafter \citetalias{best2020_inpress}).

The identification of genuine counterparts to radio sources as opposed to random background objects is a challenging task. Emission from radio sources can be extended and the typically lower resolution of the radio data can lead to poor positional accuracy (and large, asymmetric positional uncertainties). This is compounded by the high source density of deep optical and infrared (IR) surveys, meaning that the genuine counterpart could lie anywhere within a large region around the radio source, with multiple potential counterparts within this region. For this reason, a simple nearest neighbour (NN) search is not always reliable, producing significant numbers of false identifications. Moreover, radio surveys detect many classes of sources (e.g. star-forming galaxies, radio quiet quasars, radio-loud AGN, etc.) with a wide variety of morphologies which complicates this effort. For example, source extraction algorithms may split extended radio sources into multiple components, and sources nearby in sky-projections may be blended together. Automatic association of the components and the identification of the genuine counterpart for such complex sources is difficult.

In this paper, we utilise the properties of a radio source and its neighbours to develop a decision tree to identify radio sources that are correctly associated, with secure radio positions and are hence suitable for an automated, statistical approach of cross-identification. For these sources, we use the Likelihood Ratio (LR) method \citep{1977A&AS...28..211D,1992MNRAS.259..413S}, which is a commonly used statistical technique to identify real counterparts of sources detected at different wavelengths \citep[e.g.][]{2011MNRAS.416..857S,2012MNRAS.423..132M,2012MNRAS.423.2407F}. In particular, we use the colour-based adaptation of the LR method, developed by \cite{nisbet2018role} and used in the LoTSS-DR1 \citep{2019A&A...622A...2W}. This method incorporates positional uncertainties of the radio sources along with the magnitude and colour information of potential counterparts to generate a highly reliable and complete sample of cross-identifications. For sources where the decision tree indicates that the LR method is not suitable, we make use of a visual classification scheme to identify counterparts and perform accurate source association.

For this first LoTSS Deep Fields data release, in this paper, we present and release the value added radio-optical cross-matched catalogues along with the full forced, matched aperture multi-wavelength catalogues for the three fields. The paper is structured as follows. In Sect.~\ref{sec:data}, we first summarise the radio data that is presented in more detail in \citetalias{tasse2020_inpress} and \citetalias{sabater2020_inpress}. Then, the multi-wavelength data used for catalogue generation and radio-optical cross-matching is described. Section~\ref{sec:cata_creation} describes the process of generating pixel-matched images, and the creation of forced, matched-aperture multi-wavelength catalogues. Section~\ref{sec:radio_cross-match} describes both the statistical LR and the visual classification methods employed to find multi-wavelength counterparts to radio detected sources. Section~\ref{sec:final_radio_cata} details the properties and contents of the final cross-matched value-added catalogue released. Section~\ref{sec:science} presents the properties of the host-galaxies of radio sources in these deep fields. Section~\ref{sec:conclusions} presents our conclusions and discusses future prospects.

Throughout the paper and in the catalogues released, magnitudes are in the AB system \citep{1983ApJ...266..713O}, unless otherwise stated. Where appropriate, we use a cosmology with $\Omega_{\rm{M}}=0.3,~\Omega_{\rm{\Lambda}}=0.7$ and $H_{0} = 70~\mathrm{km~s^{-1}~Mpc^{-1}}$.


\section{Description of the data}\label{sec:data}
\subsection{Radio data}\label{sec:radio_data}
The details of the LOFAR observations used, along with the calibration and source extraction methods employed, are described in detail in \citetalias{tasse2020_inpress} and \citetalias{sabater2020_inpress}. Here, we summarise these steps and list the key properties of the radio data released (see Table~\ref{tab:init_radio}).

The LOFAR observations for the LoTSS Deep Fields were taken with the High Band Antenna (HBA) array, with frequencies between 114.9--177.4~MHz. The ELAIS-N1 data were obtained from LOFAR observation cycles 0, 2, and 4, consisting of 22 visits of $\sim$~8 hour integrations (total $\sim$~164 hours). The Lockman Hole data were obtained from cycles 3 and 10, with 12 visits of $\sim$~8 hour integrations (total $\sim$~112 hours). The Bo\"{o}tes dataset was obtained from cycles 3 and 8 with total integration time of $\sim$~100 hours. The total exposure times, pointing centres and root mean square (RMS) sensitivities of calibrated data are listed in Table~\ref{tab:init_radio}.

The calibration of interferometric data at these low frequencies is a challenging task, in particular due to direction dependent effects caused by the ionosphere and the station beam \citep{intema2009ionosphere}. These direction dependent effects (DDEs) are corrected using a facet based calibration, where the entire field-of-view is divided into small facets and the solutions computed for each facet individually (see \citealt{2019A&A...622A...1S} and \citetalias{tasse2020_inpress} for details). The overall calibration pipeline involves solving first for direction-independent \citep{2016ApJS..223....2V,williams2016facet_bootes,deGasperin2019dieffects} and then for direction-dependent effects, as described for the LoTSS DR1 \citep{2017A&A...598A.104S,2019A&A...622A...1S}, but with an updated version of the pipeline applied to the LoTSS Deep Fields \citepalias{tasse2020_inpress} that is more robust against un-modelled flux absorption and artefacts around bright sources. Finally, the imaging was carried out using \textsc{DDFacet} \citep{tasse2018ddfacet} to generate a high resolution (6$\arcsec$) Stokes I image for all fields, reaching unprecedented RMS depths of $\rm{S_{150MHz}} \sim$ 20, 22, and 32 $\mu$Jy beam$^{-1}$ at the field centres in ELAIS-N1, Lockman Hole, and Bo\"{o}tes, respectively (see Table~\ref{tab:init_radio}). The current imaging data released includes data from the Dutch baselines only; international station data are available and will be included in future data releases.

\begin{table*}
    \centering
    \caption{Summary of the radio data properties in the current data release of the LoTSS Deep Fields. The radio data have an angular resolution of 6$\arcsec$ and cover around 68 \mbox{deg$^{2}$} out to the primary beam 30\% power point.\label{tab:init_radio}}
    \begin{tabular}{lccc}
    \hline\hline
    {} & ELAIS-N1 & Lockman Hole & Bo\"{o}tes \Tstrut \Bstrut\\
    \hline
    Centre RA, DEC [deg] & 242.75, 55.00 & 161.75, 58.083 & 218.0, 34.50 \Tstrut \\
    Central frequency [MHz] & 146 & 144 & 144 \\
    Central RMS \mbox{[$\mathrm{\mu}$Jy/beam]} & 20 & 22 & 32\\
    Integration Time [hrs] & 164 & 112 & 80 \\
    N\textsuperscript{\underline{o}} \textsc{PyBDSF} radio sources & 84\,862 & 50\,112 & 36\,767 \\
    Reference & \citetalias{sabater2020_inpress} & \citetalias{tasse2020_inpress} & \citetalias{tasse2020_inpress} \Bstrut\\
    \hline
    \end{tabular}
\end{table*}

Source extraction is performed on the Stokes I radio image in each field using Python Blob Detector and Source Finder (\textsc{PyBDSF}; \citealt{2015ascl.soft02007M}). We refer the reader to \cite{2015ascl.soft02007M} for a detailed description of the software, and to \citetalias{tasse2020_inpress} and \citetalias{sabater2020_inpress} for details of the detection parameters used to generate the \textsc{PyBDSF} radio catalogues. In summary, sources are extracted by first identifying islands of emission (using island and peak detection thresholds of 3 and 5$\sigma$, respectively). The islands are then decomposed into Gaussians, which are then grouped together to form a source. An island of emission may contain single or multiple Gaussians and sources may be formed of either only one Gaussian or by grouping multiple Gaussians. For unintentional historic reasons, source extraction in Lockman Hole and Bo\"{o}tes were performed with slightly different parameters than in ELAIS-N1, leading to a higher fraction of \textsc{PyBDSF} sources being split into multiple Gaussians; however, after these are correctly grouped using our visual classification schemes (see Sect.~\ref{sec:general_vis}) this should have little or no effect on the final cross-matched catalogue. We summarise some key properties of the radio data and the \textsc{PyBDSF} catalogues for each field in Table~\ref{tab:init_radio}.

\begin{figure}
    \centering
    \begin{subfigure}{\columnwidth}
    \centering
    \includegraphics[width=0.9\columnwidth,height=0.242\textheight,keepaspectratio]{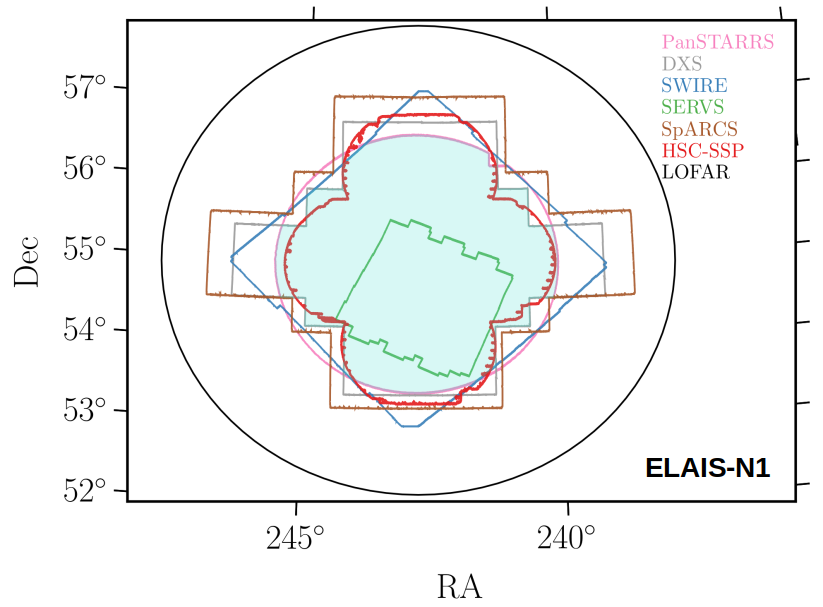}\vspace{-2pt}
    \end{subfigure}
    \begin{subfigure}{\columnwidth}
    \centering
    \includegraphics[width=0.9\columnwidth,height=0.242\textheight,keepaspectratio]{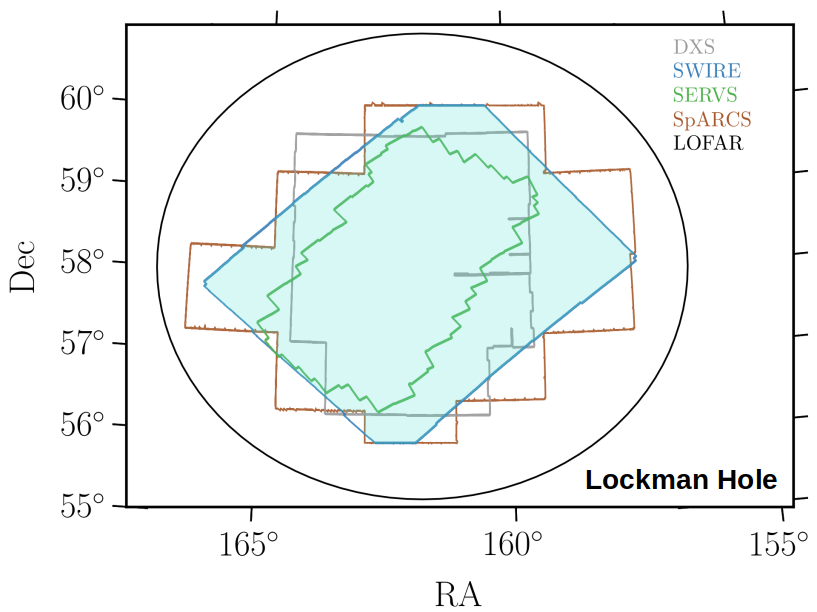}\vspace{-2pt}
    \end{subfigure}
    \begin{subfigure}{\columnwidth}
    \centering
    \includegraphics[width=0.9\columnwidth,height=0.242\textheight,keepaspectratio]{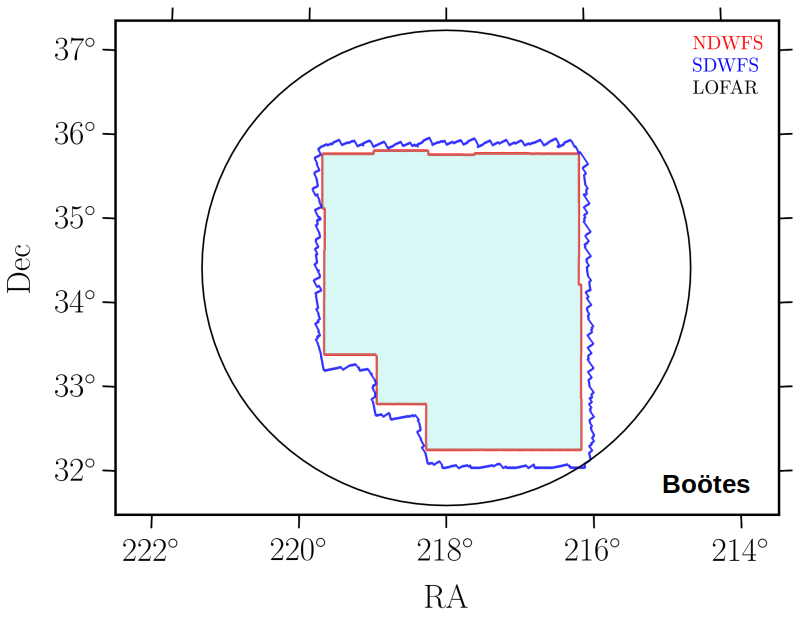}\vspace{-10pt}
    \end{subfigure}
    \caption{\label{fig:en1_lh_bootes_footprint} Footprint (north up, east left) for ELAIS-N1 (\textit{top}) Lockman Hole (\textit{middle}) and Bo\"{o}tes (\textit{bottom}) showing the coverage of multi-wavelength data from various surveys in optical and IR bands described in Sect.~\ref{sec:data_en1}~-~\ref{sec:data_bootes}. The LOFAR radio coverage is also shown in black. The shaded light blue region shows the selected area of overlap that is used for the radio-optical cross-match in this paper for ELAIS-N1 ($\sim$7.15~\mbox{deg$^{2}$}), Lockman Hole ($\sim$10.73~\mbox{deg$^{2}$}) and Bo\"{o}tes ($\sim$9.5\mbox{deg$^{2}$}), as described in Sects.~\ref{sec:en1_overlap_area},~\ref{sec:lh_overlap area} and~\ref{sec:bootes_overlap_area}, respectively, with slightly reduced area after bright-star masking.}
\end{figure}

\subsection{Multi-wavelength data in ELAIS-N1}\label{sec:data_en1}
The ELAIS-N1 field has deep multi-wavelength (0.15\,$\mathrm{\mu}$m - 500\,$\mathrm{\mu}$m) observations taken as part of many different surveys, covering up to 10~\mbox{deg$^{2}$}. The ELAIS-N1 footprint illustrated in Fig.~\ref{fig:en1_lh_bootes_footprint} (\textit{top}) shows the coverage of some of the key optical-IR surveys used, as well as the region imaged by LOFAR (plot limited to the 30\% power of the primary beam). In total, we generate photometry from 20 UV to mid-IR filters, with additional far-IR data from Spitzer and Herschel. The typical depths and areas covered by the multi-wavelength imaging datasets are listed in Table~\ref{tab:en1_lh_description}.

\subsubsection{UV to mid-infrared data in ELAIS-N1}\label{sec:en1_optical_data}
Optical data for ELAIS-N1 comes from Panoramic Survey Telescope and Rapid Response System (Pan-STARRS-1; \citealt{2010SPIE.7733E..0EK}). Pan-STARRS 1 (PS1) is installed on the peak of Haleakala on the island of Maui in the Hawaiian island chain. The PS1 system uses a 1.8m diameter telescope together with a 1.4 gigapixel CCD camera with a 7 \mbox{deg$^{2}$} field-of-view. A full description of the PS1 system is provided by~\cite{2010SPIE.7733E..0EK} and the PS1 optical design is described in~\cite{2004AN....325..636H}. The PS1 photometry is in the AB system~\citep{1983ApJ...266..713O} and the photometric system is described in detail by~\cite{2012ApJ...750...99T}. The PS1 data in ELAIS-N1 consists of broadband optical (g, r, i, z and y) imaging from the Medium Deep Survey (MDS), one of the PS1 surveys \citep{2016arXiv161205560C}. As part of the MDS, ELAIS-N1 (and the other fields) was visited on an almost nightly basis to obtain deep, high cadence images, with each epoch consisting of eight dithered exposures. This PS1 dataset provides the deepest wide-area imaging at redder optical wavelengths across ELAIS-N1.

Additional optical data were taken from the Hyper-Suprime-Cam Subaru Strategic Program (HSC-SSP) survey. ELAIS-N1 is one of the `deep' fields of the HSC-SSP survey, covering a total of $\sim$ 7.7 \mbox{deg$^{2}$} in optical filters g, r, i, z, y, and the narrow-band NB921, taken over four HSC pointings. The images were acquired from the first HSC-SSP data release \citep{2018PASJ...70S...8A}.\footnote{We note that DR2 of HSC-SSP was released in May 2019 \citep{2019arXiv190512221A}. At this time, our optical catalogues had been finalised and the visual cross-identification process was in progress. Processing the new HSC-SSP DR2 data to modify the optical catalogues would have been unfeasible, leading to delays in the visual identification process. However, we plan on including new HSC-SSP data releases for future deep fields data releases.} The HSC data have higher angular resolution than the PS1 data, and are of comparable depths at bluer wavelengths. The use of both HSC and PS1 data allows the advantages of each survey to be present in the catalogues, and in addition, provides complementary photometric data points for SED fitting.

The broadband u-band data were obtained from the Spitzer Adaptation of the Red-sequence Cluster Survey (SpARCS; \citealt{2009ApJ...698.1943W,2009ApJ...698.1934M}). SpARCS is a follow-up of the Spitzer Wide-area Infra-Red Extragalactic (SWIRE) survey fields 
taken using the MegaCAM instrument on the Canada-France-Hawaii Telescope (CFHT). In ELAIS-N1, the data were taken over 12 CFHT pointings (1 \mbox{deg$^{2}$} each) covering $\sim$ 12~\mbox{deg$^{2}$} in total.

The UV data were obtained from the Release 6 and 7 of the Deep Imaging Survey (DIS) taken with the Galaxy Evolution Explorer (GALEX) space telescope \citep{2005ApJ...619L...1M,2007ApJS..173..682M}. GALEX observations were taken in the near-UV (NUV) and far-UV (FUV) spanning 1350\AA~-~2800\AA~ and have a field-of-view $\approx$ 1.5 \mbox{deg$^{2}$} per pointing, covering around 13.5 \mbox{deg$^{2}$} in total.

The near-infrared (NIR) J and K band data come from the UK Infrared Deep Sky Survey (UKIDSS) Deep Extragalactic Survey (DXS) DR10 \citep{2007MNRAS.379.1599L}. Observations were taken using the WFCAM instrument \citep{casali2007wfcam} on the UK Infrared Telescope (UKIRT) in Hawaii as part of the 7 year DXS survey plan and cover $\sim$ 8.9 \mbox{deg$^{2}$} of the ELAIS-N1 field. The photometric system is described in \citet{hewett2006ukidss_photsys}.

The mid-infrared (MIR) 3.6\,$\mathrm{\mu}$m, 4.5\,$\mathrm{\mu}$m, 5.8\,$~\mathrm{\mu}$m and 8.0\,$~\mathrm{\mu}$m data were acquired from the IRAC instrument \citep{fazio2004irac} on board the Spitzer Space Telescope \citep{werner2004spitzer}. We use two Spitzer surveys that cover the ELAIS-N1 field: the SWIRE \citep{2003PASP..115..897L} survey and the Spitzer Extragalactic Representative Volume Survey (SERVS; \citealt{2012PASP..124..714M}). The SWIRE data were taken in January 2004 and cover an area of $\sim$ 10 \mbox{deg$^{2}$} in all four IRAC channels. The SERVS project imaged a small part of the ELAIS-N1 field, covering around 2.4 \mbox{deg$^{2}$} in only two channels (3.6\,$\mu$m and 4.5\,$\mu$m) during Spitzer's warm mission but reaching $\sim$ 1 mag deeper than SWIRE.

\subsubsection{Additional far-infrared data in ELAIS-N1}\label{sec:en1_additional_data}
Longer wavelength data at 24\,$\mu$m comes from the Multiband Imaging Photometer for Spitzer (MIPS; \citealt{rieke2004mips}) instrument on-board Spitzer. Data were also taken from Herschel Multi-tiered Extragalactic Survey (HerMES; \citealt{oliver2012hermes}) by the Herschel Space Observatory \citep{pilbratt2010herschel}, using the Spectral and Photometric Imaging Receiver (SPIRE; \citealt{griffin2010spire}) instrument at 250\,$\mu$m, 360\,$\mu$m and 520\,$\mu$m, and Photodetector Array Camera and Spectrometer (PACS; \citealt{poglitsch2010pacs}) at 100\,$\mu$m and 160\,$\mu$m. The three fields are all part of Level 5 or 6 deep tiers of HerMES, comprising one of the deepest, large-area Herschel surveys available. The 70\,$\mu$m data from MIPS or PACS are not included in our catalogues (and nor within HELP) due to their poorer sensitivity.

In part due to their low angular resolution, these FIR data are not used to generate the forced, matched aperture catalogues. Instead, FIR fluxes are added from existing catalogues from the SPIRE and PACS maps, generated by the Herschel Extragalactic Legacy Project (HELP; Oliver et~al. 2020, in prep.). FIR fluxes from HELP were incorporated by performing a cross-match between our multi-wavelength catalogue and HELP catalogues using a 1.5$\arcsec$ cross-match. If no match was found within the HELP catalogues, FIR fluxes were extracted using the XID+ software \citep{hurley2017xid}, incorporating the radio (or optical) positions into the list of priors. The details of the process of generating and adding FIR fluxes is described by \cite{mccheyne2020_inpress}.

\subsubsection{Selected survey area in ELAIS-N1}\label{sec:en1_overlap_area}
The radio data cover a significantly larger area than the accompanying multi-wavelength data. We therefore define the area used for cross-matching in this paper for the ELAIS-N1 field as the overlapping area between PanSTARRS, UKIDSS, and SWIRE, covering $\sim$ 7.15 \mbox{deg$^{2}$}. This overlap area is indicated by the blue shaded region in the ELAIS-N1 footprint shown in Fig.~\ref{fig:en1_lh_bootes_footprint} (\textit{top}). At the largest extent of this selected area from the radio field centre, the radio primary beam correction factor is $\sim$0.65, resulting in a noise level approximately 50\% higher than in the centre. There is thus a moderate variation in the depth of the radio data across the survey region.

\subsection{Multi-wavelength data in Lockman Hole}\label{sec:data_lh}
Lockman Hole also possesses deep multi-wavelength (0.15\,$\mathrm{\mu}$m--500\,$\mathrm{\mu}$m) data and is the field with the largest area of multi-wavelength coverage, as shown by the footprint in Fig.~\ref{fig:en1_lh_bootes_footprint} (\textit{middle}). The typical depths and areas covered by the multi-wavelength and radio imaging datasets are listed in Table~\ref{tab:en1_lh_description}.

\subsubsection{UV to mid-infrared data in Lockman Hole}\label{sec:lh_optical_data}
The optical data in Lockman Hole come from two surveys taken by the CFHT-MegaCam instrument: SpARCS and the Red Cluster Sequence Lensing Survey (RCSLenS; \citealt{2016MNRAS.463..635H}). The SpARCS data in Lockman Hole consist of broadband u, g, r, z filter images taken using 14 pointings of the CFHT, covering around 13.3 \mbox{deg$^{2}$} of the field. The RCSLenS data consist of g, r, i, z observations covering around 16 \mbox{deg$^{2}$}. The coverage from RCSLenS however, is not contiguous, with gaps between different pointings.  

Similar to ELAIS-N1, the NUV and FUV imaging data come from the GALEX DIS Release 6 and 7, and, the NIR data is obtained from the J and K bands of the UKIDSS-DXS DR10, covering a maximum area of around 8 \mbox{deg$^{2}$}. Observations of the Lockman Hole field were also taken in IRAC channels as part of SWIRE and SERVS, reaching similar depths as in ELAIS-N1 but over much larger areas. The SWIRE data in all four IRAC channels cover around 11 \mbox{deg$^{2}$} whereas the deeper SERVS data in the two IRAC channels (3.6\,$\mathrm{\mu}$m and 4.5\,$\mathrm{\mu}$m) cover around 5.6 \mbox{deg$^{2}$}.

\subsubsection{Additional far-infrared data in Lockman Hole}\label{sec:lh_additional_data}
Lockman Hole is also covered by both Spitzer MIPS and HerMES observations. These FIR fluxes were added using catalogues generated by HELP and by running XID+, following the same method as for ELAIS-N1 (see \citealt{mccheyne2020_inpress}).

\subsubsection{Selected survey area for Lockman Hole}\label{sec:lh_overlap area}
In this paper, for radio-optical cross-matching, we use the overlapping area between the SpARCS r-band and the SWIRE survey which covers $\approx$~10.73 \mbox{deg$^{2}$}. As such, Lockman Hole is the largest deep field released with respect to the accompanying multi-wavelength data. This overlap area in Lockman Hole is also illustrated by the blue shaded region in the footprint in Fig.~\ref{fig:en1_lh_bootes_footprint} (\textit{middle}). At the largest extent of this selected area from the radio field centre, the radio primary beam correction factor is $\sim$0.42.

\subsection{Multi-wavelength data in Bo\"{o}tes}\label{sec:data_bootes}
In Bo\"{o}tes, we make use of existing PSF matched I-band and 4.5\,$\mu$m band catalogues \citep{2007ApJ...654..858B,2008ApJ...682..937B} built using imaging data from the NOAO Deep Wide Field Survey (NDWFS; \citealt{1999ASPC..191..111J}) and follow-up imaging campaigns in other filters. This catalogue contains 15 multi-wavelength bands (0.14\,$\mathrm{\mu}$m--24\,$\mathrm{\mu}$m) from different surveys. Fig.~\ref{fig:en1_lh_bootes_footprint} (\textit{bottom}) shows the footprint of the key surveys covering the Bo\"{o}tes field. Typical 3$\sigma$ depths estimated using variance from random apertures for each filter are listed in Table~\ref{tab:bootes_extcorr}.

In summary, deep optical photometry in the B\textsubscript{W}, R, and I filters comes from NDWFS \citep{1999ASPC..191..111J}. Photometry in the NUV and FUV comes from GALEX surveys. Additional z-band data covering the full NDWFS field comes from the zBo\"{o}tes survey \citep{2007ApJS..169...21C} taken with the Bok 90Prime imager, and additional data from the Subaru z-band (PI: Yen-Ting, Lin). Additional optical imaging in the U$_{\rm{spec}}$ and the Y bands comes from the Large Binocular Telescope \citep{bian2013lbt_bootes}. NIR data in J, H, and K\textsubscript{s} comes from \citet{gonzalez2010newfirmbootes}. In the MIR, Spitzer surveyed $\sim$10 \mbox{deg$^{2}$} of the NDWFS field at 3.6, 4.5, 5.8 and 8.0\,$\mathrm{\mu}$m across 5 epochs. Primarily, the data consist of 4 epochs from the Spitzer Deep Wide Field Survey (SDWFS; \citealt{2009ApJ...701..428A}), a subset of which is the IRAC Shallow Survey \citep{eisenhardt2004iracshallow} and, the fifth epoch from the Decadal IRAC Bo\"{o}tes Survey (M.L.N. Ashby PI, PID 10088).

The full details of the data used and the catalogue generation process are provided in \cite{2007ApJ...654..858B,2008ApJ...682..937B}. In summary, images in all filters were first moved on to a common pixel scale and then sources detected using \textsc{SExtractor} \citep{1996A&AS..117..393B}. Forced photometry was then performed on optical-NIR filters smoothed to a common PSF. The common PSF was chosen to be a Moffat profile with $\beta=2.5$ and a FWHM of 1.35$\arcsec$ (B\textsubscript{W}, R, I, Y, H and K), a FWHM of 1.6$\arcsec$ (u, z, J) and a FWHM of 0.68$\arcsec$ for the Subaru z-band. Aperture corrections based on the chosen Moffat profile were then applied to account for the different FWHM choices in PSF smoothing.

In Bo\"{o}tes, the FIR data from HerMES and MIPS were obtained by a similar method to ELAIS-N1 and Lockman Hole (see \citealt{mccheyne2020_inpress}), and form a new addition to the existing catalogues of \cite{2007ApJ...654..858B,2008ApJ...682..937B}.

\subsubsection{Selected survey area for Bo\"{o}tes}\label{sec:bootes_overlap_area}
In this paper, subsequent analysis is performed for the overlap of the NDWFS and SDWFS datasets, covering $\sim$ 9.5 \mbox{deg$^{2}$}, as shown in Fig.~\ref{fig:en1_lh_bootes_footprint} (\textit{bottom}). This area was chosen as the largest area with coverage in most of the optical-IR bands. At the largest extent of the selected area from the radio field centre, the primary beam correction factor is $\sim$0.39.

\begin{table*}
	\centering
 	\caption{\label{tab:en1_lh_description} Key properties of the multi-wavelength data in ELAIS-N1 and Lockman Hole. For each filter, we include the Vega-AB conversion factor (if any) used for generating pixel-matched mosaics (see Sect.~\ref{sec:resamp}), the average PSF FWHM, and the approximate area covered by each survey. The 3$\sigma$ depths (in AB system) in each filter estimated from the variance of empty, source free 3$\arcsec$ apertures, and the filter dependent Galactic extinction values, $\mathrm{A_{band}}/E(B-V)$ are listed.} 
	\begin{tabular}{lcccccr} 
		\hline\hline
		{Field/Survey} & {Band} & {Vega-AB} & {PSF} & {$3 \sigma$ depth} & {$\mathrm{A_{band}}/E(B-V)$} & {Area} \Tstrut \\
		{} & {} & {[mag]} & {[arcsec]} & {[mag]} & {} & {[\mbox{deg$^{2}$}]} \Bstrut\\
		\hline
		\bf{ELAIS-N1} & {} & {} & {} & {} & {} & {} \Tstrut\\
		\it{SpARCS} & {} & {} & {} & {} & {} & {11.81}\\
		{} & {u} & {-} & {0.9} & {25.4} & {4.595} & {}\\
		\it{PanSTARRS} & {} & {} & {} & {} & {} & {8.05}\\
		{} & g & - & {1.2} & {25.5} & {3.612} & {}\\
		{} & r & - & {1.1} & {25.2} & {2.569} & {}\\
		{} & i & - & {1.0} & {25.0} & {1.897} & {}\\
		{} & z & - & {0.9} & {24.6} & {1.495} & {}\\
		{} & y & - & {1.0} & {23.4} & {1.248} & {}\\
		\it{HSC} & {} & {} & {} & {} & {} & {7.70}\\
		{} & G & - & {0.5} & {25.6} & {3.659} & {}\\
		{} & R & - & {0.7} & {25.0} & {2.574} & {}\\
		{} & I & - & {0.5} & {24.6} & {1.840} & {}\\
		{} & Z & - & {0.7} & {24.2} & {1.428} & {}\\
		{} & Y & - & {0.6} & {23.4} & {1.213} & {}\\
		{} & NB921 & - & {0.6} & {24.3} & {1.345} & {}\\
		\it{UKIDSS-DXS} & {} & {} & {} & {} & {} & {8.87}\\
		{} & J & 0.938 & {0.8} & {23.2} & {0.797} & {}\\
		{} & K & 1.900 & {0.9} & {22.7} & {0.340} & {}\\
		\it{SWIRE} & {} & {} & {} & {} & {} & {9.32}\\
		{} & 3.6\,$\mathrm{\mu}$m & 2.788 & {1.66} & {23.4} & {0.184} & {}\\
		{} & 4.5\,$\mathrm{\mu}$m & 3.255 & {1.72} & {22.9} & {0.139} & {}\\
		{} & 5.8\,$\mathrm{\mu}$m & 3.743 & {1.88} & {21.2} & {0.106} & {}\\
		{} & 8\,$\mathrm{\mu}$m & 4.372 & {1.98} & {21.3} & {0.075} & {}\\
		\it{SERVS} & {} & {} & {} & {} & {} & {2.39}\\
		{} & 3.6\,$\mathrm{\mu}$m & 2.788 & {1.66} & {24.1} & {0.184} & {}\\
		{} & 4.5\,$\mathrm{\mu}$m & 3.255 & {1.72} & {24.1} & {0.139} & {}\\
		\hline
		\bf{Lockman Hole} & {} & {} & {} & {} & {} & {} \Tstrut\\
		\it{SpARCS} & {} & {} & {} & {} & {} & {13.32}\\
		{} & {u} & {-} & {1.06} & {25.5} & {4.595} & {}\\
		{} & {g} & {-} & {1.13} & {25.8} & {3.619} & {}\\
		{} & {r} & {-} & {0.76} & {25.1} & {2.540} & {}\\
		{} & {z} & {-} & {0.69} & {23.5} & {1.444} & {}\\
		\it{RCSLenS} & {} & {} & {} & {} & {} & {16.63}\\
		{} & {g} & {-} & {0.78} & {25.1} & {3.619} & {}\\
		{} & {r} & {-} & {0.68} & {24.8} & {2.540} & {}\\
		{} & {i} & {-} & {0.60} & {23.8} & {1.898} & {}\\
		{} & {z} & {-} & {0.65} & {22.4} & {1.444} & {}\\
		\it{UKIDSS} & {} & {} & {} & {} & {} & {8.16}\\
		{} & J & 0.938 & {0.76} & {23.4} & {0.797} & {}\\
		{} & K & 1.900 & {0.88} & {22.8} & {0.340} & {}\\
		\it{SWIRE} & {} & {} & {} & {} & {} & {10.95}\\
		{} & 3.6\,$\mathrm{\mu}$m & 2.788 & {1.66} & {23.4} & {0.184} & {}\\
		{} & 4.5\,$\mathrm{\mu}$m & 3.255 & {1.72} & {22.9} & {0.139} & {}\\
		{} & 5.8\,$\mathrm{\mu}$m & 3.743 & {1.88} & {21.2} & {0.106} & {}\\
		{} & 8\,$\mathrm{\mu}$m & 4.372 & {1.98} & {21.2} & {0.075} & {}\\
		\it{SERVS} & {} & {} & {} & {} & {} & {5.58}\\
		{} & 3.6\,$\mathrm{\mu}$m & 2.788 & {1.66} & {24.1} & {0.184} & {}\\
		{} & 4.5\,$\mathrm{\mu}$m & 3.255 & {1.72} & {24.0} & {0.139} & {}\\
		\hline
	\end{tabular} 
\end{table*}

\section{Creation of multi-wavelength catalogues}\label{sec:cata_creation}
For both ELAIS-N1 and Lockman Hole, individual catalogues already exist in each filter generated by each survey. However, catalogue combination issues, such as when sources are blended in lower resolution catalogues, or only detected in a subset of filters, present significant challenges. Furthermore, the usefulness of existing catalogues for photometric redshifts is limited due to the varying catalogue creation methods. For example, magnitudes were typically measured within different apertures and with different methods of correcting to total magnitudes, leading to colours that are not sufficiently robust. In addition, for the sources that were detected in only a subset of filters, the lack of information or application of a generic limiting magnitude in other filters, would lead to a loss of information on galaxy colours compared to a forced photometry measurement. This can have a significant impact on the accuracy of SED fitting and therefore the photometric redshifts. To alleviate these issues, we have created pixel-matched images and built matched aperture, multi-wavelength catalogues with forced photometry spanning the UV to mid-infrared wavelengths in ELAIS-N1 and Lockman Hole. This provides high quality catalogues for radio cross-matching and photometric redshift estimates. This section describes the creation of the pixel-matched images and the generation of the new multi-wavelength catalogues in both ELAIS-N1 and Lockman Hole.

The Bo\"{o}tes field already possesses PSF-matched forced photometry catalogues created using an I-band and a 4.5\,$\mathrm{\mu}$m band detected catalogue \citep{2007ApJ...654..858B,2008ApJ...682..937B}. To generate a similar multi-wavelength catalogue in Bo\"{o}tes as the other two fields for radio-optical cross-matching, we apply only the final steps of our catalogue generation process, namely, the masking around stars (see Sect.~\ref{sec:star_masking}), the merging of the I-band and 4.5\,$\mathrm{\mu}$m detected catalogues (see Sect.~\ref{sec:cata_merging}), and the Galactic extinction corrections.

\subsection{Creation of the pixel-matched images}\label{sec:resamp}
The images from different instruments had different pixel scales and therefore all of the images needed to be re-gridded (resampled) onto the same pixel scale to perform matched aperture photometry across all filters. Observations in most filters consisted of many overlapping exposures of the total area. We obtained reduced images from survey archives for all filters and used \textsc{SWarp} \citep{2002ASPC..281..228B} to both resample the individual images in each filter to a common pixel scale of 0.2$\arcsec$ per pixel and then to combine (co-add) these resampled images to make a single large mosaic in each filter. We make no attempt to perform point-spread function (PSF) homogenisation of these observations; instead, we account for the varying PSF in each filter by performing aperture corrections (see Sect.~\ref{sec:aper_corr}).

Changes to the astrometric projection or the photometric calibration were performed during the resampling process by \textsc{SWarp}. During this step, the contribution to the flux from the background/sky is subtracted before the resampling and co-addition process to avoid artefacts resulting from image combination. The flux scale of the images was also adjusted using each input frame's zero-point magnitude, exposure time and any Vega-AB conversion factors (see Table~\ref{tab:en1_lh_description}) to shift the zero-point magnitude of all the images to 30 mag (in the AB system). The resampled images in each filter were then co-added in a `weighted' manner to take into account the relative exposure time/noise per pixel in multiple input frames and in overlapping frames. Table~\ref{tab:en1_lh_description} also lists the typical PSF full-width half-maximum (FWHM) for each filter in ELAIS-N1 and Lockman Hole. We compared photometry in fixed apertures for given sources in both the resampled frames and the final mosaics to ensure that the photometry is consistent with the original images.

\subsection{Source detection}\label{sec:source_detection}
Source detection is performed using \textsc{SExtractor} \citep{1996A&AS..117..393B}. We ran \textsc{SExtractor} in `dual-mode', using a deep image for detecting sources and then performing photometry using these detections on all of the filters. To produce as complete a catalogue as possible, we built our deep detection image by using \textsc{SWarp} to create deep $\chi^{2}$ images \citep{1999AJ....117...68S} by combining observations from multiple filters. Specifically, due to the significantly worse angular resolution of the Spitzer data, we built two $\chi^{2}$ images, one using optical and NIR filters and a separate $\chi^{2}$ image using only the Spitzer-IRAC data. In ELAIS-N1, the optical $\chi^{2}$ image was created using SpARCS-u, PS1-griz and UKIDSS-DXS-JK filters (the PS1 y-band is not included due to its shallower depth and lower sensitivity relative to the adjacent filters). In Lockman Hole, we used SpARCS-ugrz, RCSLenS-i and UKIDSS-DXS-JK filters. The Spitzer $\chi^{2}$ images in both ELAIS-N1 and Lockman Hole are built from the IRAC 3.6\,$\mathrm{\mu}$m and 4.5\,$\mathrm{\mu}$m bands from both SWIRE and SERVS. The longer wavelength Spitzer data are not included in the $\chi^{2}$ detection images due to a further decrease in angular resolution.

The key detection parameters in \textsc{SExtractor} are the ones concerning deblending, the detection threshold and minimum detection area. These key parameters are listed in Table~\ref{tab:sextractor_params} for the optical-NIR and Spitzer $\mathrm{\chi^{2}}$ images. We fine-tuned these parameters for each $\chi^{2}$ image by adjusting their values and inspecting the resulting catalogue overlaid on the $\chi^{2}$ images. 

Although more sophisticated tools exist for performing multi-band photometry that allow model-fitting of detections (e.g. The Tractor; \citealt{lang2016tractor,nyland2017tractor} and T-PHOT; \citealt{merlin2015tphot,merlin2016tphotv2}), \textsc{SExtractor} is a flexible and easily scalable tool that allows both source detection and forced, matched aperture photometry to be performed in a practical and robust manner over $\sim$30 bands across $>$ 18 deg$^{2}$ for the two fields. Moreover, the use of \textsc{SExtractor} for ELAIS-N1 and Lockman Hole also provides consistency with the method that was adopted to generate the existing Bo\"{o}tes catalogues.

\begin{table}
    \centering
    \caption{Key \textsc{SExtractor} detection and deblending parameters used for the optical-NIR and Spitzer $\chi^{2}$ detection images in ELAIS-N1 and Lockman Hole.}
    \label{tab:sextractor_params}
    \begin{tabular}{lll}
    \hline\hline
    \multicolumn{1}{l}{Parameter} & \multicolumn{2}{c}{Value} \Tstrut\\
    {} & Optical-NIR $\chi^{2}$ & Spitzer $\chi^{2}$ \Bstrut\\
    \hline
    \textsc{detect\_minarea} & 8 & 10 \Tstrut \\
    \textsc{detect\_thresh} & 1 & 2.5\\
    \textsc{deblend\_nthresh} & 64 & 64\\
    \textsc{deblend\_mincont} & 0.0001 & 0.001 \Bstrut\\
    \hline
    \end{tabular}
\end{table}

\begin{table}
    \caption{Area flagged by the star masks. Two star mask images are generated using the Spitzer- and optical-detected catalogue (see Sect~\ref{sec:star_masking}).}
    \label{tab:star_mask_flags}
    \centering
    \begin{tabular}{lll}
    \hline\hline
    Field & Spitzer-mask area & Optical-mask area \Tstrut \\
    {} & [\mbox{deg$^{2}$}] & [\mbox{deg$^{2}$}] \Bstrut \\
    \hline
    ELAIS-N1 & 0.40 & 0.61 \Tstrut \\
    Lockman Hole & 0.31 & 0.85 \\
    Bo\"{o}tes & 0.87 & 1.18\\
    \hline
    \end{tabular}
\end{table}

\subsection{Photometric measurements}\label{sec:photometry}
Running \textsc{SExtractor} in dual mode, we measure fluxes in all of the filters using both the optical-NIR and Spitzer $\chi^{2}$ images; this includes Spitzer fluxes from sources detected on the optical-NIR $\chi^{2}$ images and vice versa. We extract fluxes from a wide variety of aperture sizes in each filter, specifically 1$\arcsec$~--~7$\arcsec$ diameter (in 1$\arcsec$ steps) and also 10$\arcsec$ diameter apertures in each filter.

\subsubsection{Aperture and Galactic extinction corrections}\label{sec:aper_corr}
Fluxes from fixed apertures were corrected to total fluxes using aperture corrections based on the curve of growth estimated from our full range of aperture measurements (assuming all of the flux from a source is contained within the 10$\arcsec$ aperture). We compute median correction factors for each aperture size using relatively isolated ($>$5$\arcsec$ from nearest neighbour) sources of moderate magnitude (e.g. i-band 19 $<$ i $<$ 20.5), chosen to have high sky density but also sufficient signal-to-noise ratio (S/N) even in the larger apertures. These sources are typical of moderately distant galaxies, with this selection driven by the primary scientific aims of the LOFAR surveys. It is important to note that the resulting correction factors are found to be not sensitive to the exact choice of magnitude used in selecting sources used for calibrating the aperture corrections. The full list of aperture corrections are provided in Appendix~\ref{sec:appendix1:apcorr} (Table~\ref{tab:apcorr}). In addition, we also provide in Table~\ref{tab:apcorr}, a list of aperture corrections calibrated based on stars with 18 $<$ Gmag $<$ 20 in GAIA Data Release 2 \citep[GAIA DR2;][]{2016A&A...595A...1G,2018A&A...616A...1G,2018A&A...616A...3R,2018A&A...616A...4E}. In deriving the aperture corrections, we assume that the PSF variations between different images of a given filter are insignificant compared to the PSF variation across different filters (see discussion in Sect.~\ref{sec:cata_validation}).

Galactic extinction corrections are computed at the position of each object using the map of \cite{1998Schlegeldustmap} \footnote{Performed using the \textsc{dustmaps} package \citep{green2018dustmaps} for \textsc{Python}.}. We provide a column of $E(B-V)$ reddening values computed from \cite{1998Schlegeldustmap} for each source, which is then multiplied by the filter dependent factor (listed in Table~\ref{tab:en1_lh_description} and \ref{tab:bootes_extcorr}) derived from the filter transmission curve and the Milky Way extinction curve \citep{fitzpatrick1999mwext}. The raw photometry in any aperture can be corrected for both aperture and extinction using the method described in Appendix~\ref{sec:appendix1:apcorr}.

\subsubsection{Computation of photometric errors}\label{sec:ferr_adjust}
We find that the flux uncertainties reported by \textsc{SExtractor} typically underestimate the total uncertainties. This is a well-known issue and occurs as \textsc{SExtractor} only takes into account photon and detector noise, and does not account for background subtraction errors or correlated noise arising from image combination. We estimate the additional flux error term using the same method used by \cite{2012A&A...545A..23B} and \cite{2016ApJS..224...24L}. Firstly, fluxes were measured in random isolated apertures (with the same sized apertures as our flux measurements) on a background-subtracted image. Then, to remove the contribution from sources to the flux in the random apertures, an iterative sigma clipping of the measured flux distribution is performed. Finally, the standard deviation of the clipped distribution is taken to be the additional contribution to the flux uncertainties from correlated noise and background subtraction errors, and is then added in quadrature with the uncertainties reported by \textsc{SExtractor} on a source-by-source basis to compute the total photometric errors. The magnitude errors were then updated accordingly. The 3$\sigma$ magnitude depths estimated from the variance of empty, source free 3$\arcsec$ apertures in each filter are listed in Table~\ref{tab:en1_lh_description}.

\subsection{Catalogue cleaning and merging}\label{sec:cat_clean}
In this section, we describe the key steps used to clean the ELAIS-N1 and Lockman Hole catalogues of spurious sources and low-significance detections. We then discuss masking around bright stars and merging of the optical-NIR and Spitzer detected catalogues in all three fields.

\begin{table*}
\centering
\caption{Properties of the initial \textsc{PyBDSF} catalogues and the final multi-wavelength catalogues in ELAIS-N1, Lockman Hole, and Bo\"{o}tes. We also list here the overlapping multi-wavelength coverage area, the number of radio and multi-wavelength sources within this region, and the overlap bit flag, \textsc{flag\_overlap} for each field, which can be used to select both radio and multi-wavelength sources within our chosen area.}
\label{tab:prop_final-opt_pybdsf}
\begin{tabular}{lccc}
\hline\hline
{} & {ELAIS-N1} & Lockman Hole & Bo\"{o}tes \\
\hline
N\textsuperscript{\underline{o}} \textsc{PyBDSF} radio sources & 84\,862 & 50\,112 & 36\,767 \\
N\textsuperscript{\underline{o}} multi-wavelength sources & 2\,106\,293 & 3\,041\,956 & 2\,214\,358 \\
\textsc{flag\_overlap}\tablefootmark{a} & 7 & 3 & 1 \\
Overlap Area [\mbox{deg$^{2}$}]\tablefootmark{b} & 6.74 & 10.28 & 8.63\\
N\textsuperscript{\underline{o}} \textsc{PyBDSF} radio sources overlap\tablefootmark{c} & 31\,059 & 29\,784 & 18\,766 \\
N\textsuperscript{\underline{o}} optical sources overlap\tablefootmark{c} & 1\,470\,968 & 1\,906\,317 & 1\,911\,929 \\
Multi-wavelength catalogue sky density [arcsec$^{-2}$] & 0.0168 & 0.0143 & 0.0171 \\
\textsc{PyBDSF} radio catalogue sky density [arcsec$^{-2}$] & $\mathrm{3.6\times 10^{-4}}$ & $\mathrm{2.2\times 10^{-4}}$ & $\mathrm{1.7\times 10^{-4}}$ \\
\hline
\end{tabular}\vspace{-0.1cm}
\tablefoot{
\tablefoottext{a}{Overlap bit flag (\textsc{flag\_overlap}) provided in the full multi-wavelength catalogues and the radio cross-match catalogues indicating the coverage of each source. The overlap flag value in this table should be used to select sources in the overlapping multi-wavelength area defined in Sect.~\ref{sec:data}.}\\
\tablefoottext{b}{The overlap area listed covers the overlapping multi-wavelength coverage (based on \textsc{flag\_overlap}) and excludes the region masked based on the Spitzer star mask. Radio-optical cross-matching is only performed for sources in this overlap area.}\\
\tablefoottext{c}{Number of radio (in initial \textsc{PyBDSF} list) and optical sources in the overlap area above can be selected using the flag combination of \textsc{flag\_clean} $\neq$ 3 and the respective \textsc{flag\_overlap} listed above.}
}
\end{table*} 

\subsubsection{Cross-talk removal}\label{sec:cross_talk_removal}
Cross-talks are non-astronomical artefacts that appear on the UKIDSS (J or K) images at fixed offsets from bright stars due to readout patterns; these may appear in the $\chi^{2}$ detection image. Cross-talks may have extreme colours due to their non-astrophysical nature and, therefore, we use the flux measurements (or lack thereof) in the optical and NIR filters to identify and remove cross-talks from the catalogue. Specifically, we searched for catalogued detections within 2$\arcsec$ of the expected cross-talk positions to identify (and remove) detections that have either extreme optical-NIR colours (ie. (i~-~K) $>$ 4) or, have low significance (S/N $<$ 3) measurements in multiple optical bands and a NIR magnitude that is more than 6~mag fainter than the `host' star. These criteria were confirmed by visual inspection of detections that were removed and retained (i.e. sources present at the expected cross-talk positions but not satisfying other criteria above). Furthermore, the radial distribution of detected objects around bright stars showed narrow peaks at the radii expected for cross-talk artefacts: after application of these cross-talk removal techniques, these peaks were eliminated (without over-removal).

\subsubsection{Cleaning low significance detections}\label{sec:low_snr_clean}
In the final cleaning step, we removed any sources which have a S/N less than 3 in all apertures of all filters. Such low significance detections may have a S/N $<$ 3 in each of the single band images but could end up in the catalogue due to the use of $\chi^{2}$ detection images which combines the signal from multiple bands. Although probably genuine, such sources are of limited scientific value as none of their flux measurements are sufficiently reliable. This step removes $\sim$15\% and 27\% of the sources from the ELAIS-N1 and Lockman Hole catalogues, respectively. The higher fraction of low-significance sources removed in Lockman Hole are largely located near the edge of the field where the $\chi^{2}$ image contains few filters with variable relative depth: ELAIS-N1 possesses both deeper optical data, and also coverage from most filters across a higher fraction of the total area of the field.

\subsubsection{Masking sources near bright stars}\label{sec:star_masking}
Next, we created a mask image by masking regions around bright stars and flagging sources within these regions in our catalogue, in each of the ELAIS-N1, Lockman Hole, and Bo\"{o}tes fields. The rationale behind this is twofold. Firstly, in regions around stars, \textsc{SExtractor} may detect additional spurious sources or miss other sources nearby or behind the star in sky projection. Secondly, the photometry of objects near bright stars will not be reliable. Masking such regions therefore allows scientific analysis to be restricted to areas where there is reliable coverage. This is crucial for some science cases, for example, clustering analysis. 

To select the stars around which regions must be masked, we cross-matched our catalogue to stars with Gmag $<$ 16.5~mag in GAIA DR2. Then, we split the stars into narrow magnitude bins and select the radius to mask around stars in each bin by using a plot of the sky density of the sources as a function of the radius from the star. An appropriate radius was chosen where neither the `holes' in the detections nor a `ring' of additional spurious sources near the star were affecting the detections (e.g. \citealt{coupon2018hsc_starmask}). We validated the choice of the magnitude dependent radii using careful visual inspection, with the values listed in Appendix~\ref{sec:appendix2:starmask}.

Detections around stars are affected less by this issue in the Spitzer $\chi^{2}$ image (and catalogue), allowing us to mask a smaller area. Therefore, in practice, we create two such masks, one for the optical-NIR $\chi^{2}$ image (a conservative mask) and, one for the Spitzer $\chi^{2}$ image (an optimistic mask), 
with both masks being applied to both the optical-NIR and the Spitzer detected catalogues. Using detections from the Spitzer-detected catalogue masks a smaller area around stars, recovering some genuine sources detected in the Spitzer image that are not affected by source extraction biases. However, photometry of these sources in the optical-NIR images may be less reliable due to stellar emission, and moreover, any optical-only detected sources may be missing from this extra recovered area. The area masked in each field using both the optical-NIR (conservative) and the Spitzer (optimistic) mask is given in Table~\ref{tab:star_mask_flags}. For convenience, we include a flag (\textsc{flag\_clean}) column in both the multi-wavelength catalogues and the radio cross-match catalogues which indicates if a source is within the two masked areas. For readers requiring a clean homogeneous catalogue, we recommend using \textsc{flag\_clean} = 1 to select sources that are not in either the optical or Spitzer star mask region. Instead, if the largest sample of sources is required, with photometry not critical, we recommend using \textsc{flag\_clean} $\mathrm{\neq}$ 3 to exclude only sources in the smaller Spitzer star mask. We note that this should be used in conjunction with \textsc{flag\_overlap} to select sources with reliable photometry in the majority of the bands (see Sect.~\ref{sec:final_optcat_properties} and Table~\ref{tab:prop_final-opt_pybdsf}).

\subsubsection{Merging optical and Spitzer catalogues}\label{sec:cata_merging}
After applying our cleaning steps, the optical-NIR detected catalogue was merged with the Spitzer detected catalogue in each of the three fields. Many of the Spitzer-detected sources, especially those with blue colours, will already be present in the optical-NIR catalogue. We therefore merge the two catalogues by appending `Spitzer-only' sources to the optical-NIR catalogue. We define a source as `Spitzer-only' if its nearest neighbour in the optical-NIR catalogue is more than 1.5$\arcsec$ away. This search radius was chosen based on both visual inspection of the `Spitzer-only' sources and by inspecting the radius above which the number of genuine matches decreases rapidly and the number of random matches starts to increase. In ELAIS-N1, we find that $\sim 15$\% of Spitzer detected sources are `Spitzer-only' sources, which make up 4.6\% of the total number of sources in the final multi-wavelength merged catalogue.

\begin{figure}
    \centering
    \includegraphics[width=\columnwidth,trim=0 0.2cm 0 0.3cm]{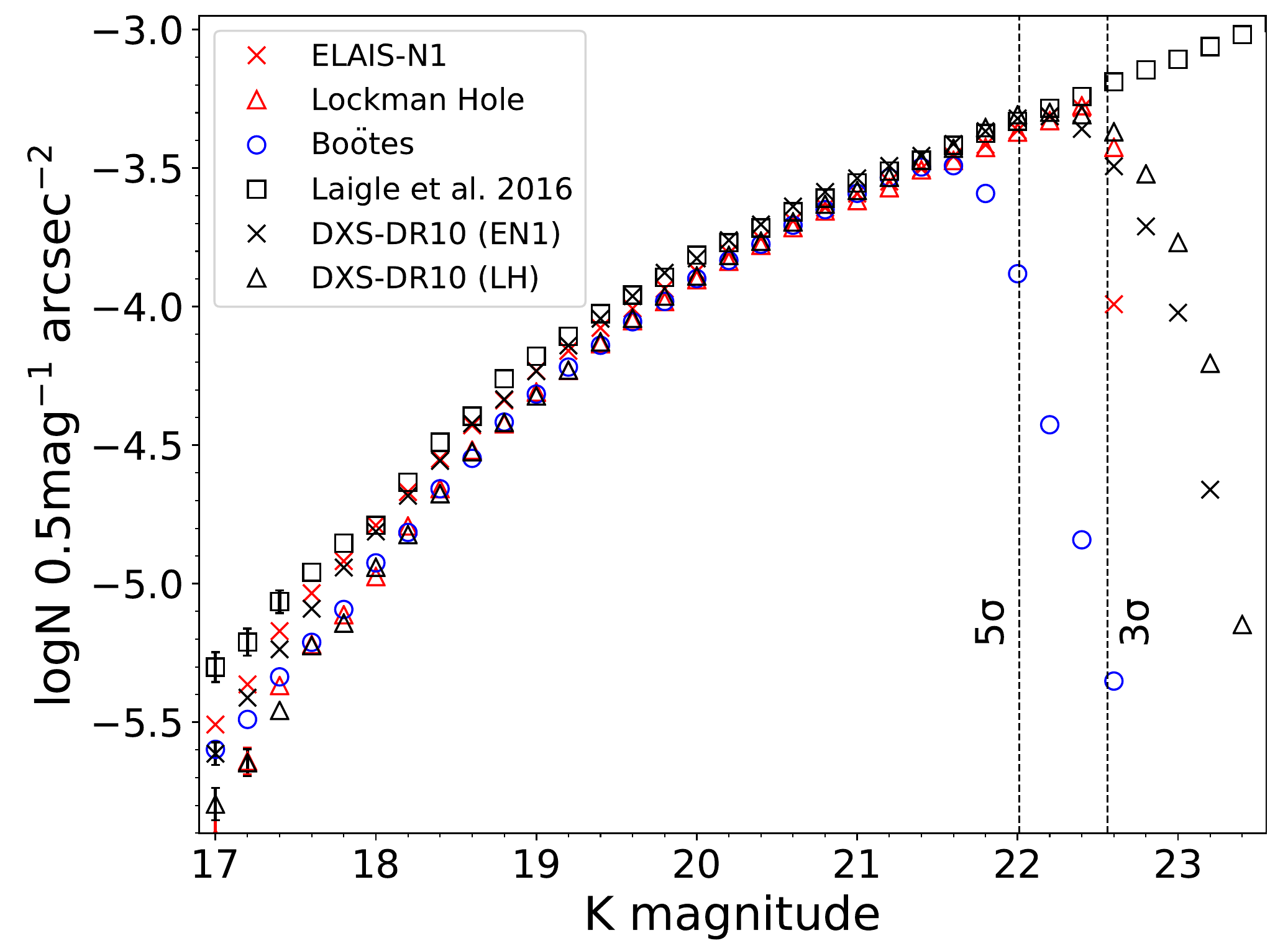}
    \caption{K-band (3$\sigma$) selected source counts (per square arcsecond per 0.5 magnitude) in ELAIS-N1 (red crosses) and Lockman Hole (red triangles) from the $\chi^{2}$ catalogue. In Bo\"{o}tes, we show source counts from the merged catalogue (blue circles). We also show the galaxy counts from the COSMOS deep area taken from \cite{2016ApJS..224...24L} for comparison (black squares). Additionally, the galaxy counts from the UKIDSS-DXS DR10 catalogues \citep{2007MNRAS.379.1599L} in both ELAIS-N1 (black crosses) and Lockman Hole (black triangles) are shown. In all cases, we have attempted to remove the contribution from foreground stars via a cross-match to GAIA DR2 catalogues in each field. Vertical lines show the 3- and 5-$\sigma$ magnitude depths in ELAIS-N1 (dashed lines) estimated from random, source free 3$\arcsec$ diameter apertures. Poissonian error bars are shown only where they are larger than the symbol size, but there may be other cosmic-variance related errors.} 
    \label{fig:nm_comp}
\end{figure}

\subsection{Catalogue validation}\label{sec:cata_validation}
To validate the catalogues generated, we compare our astrometry and photometry to publicly available catalogues in ELAIS-N1 and Lockman Hole.

\begin{figure*}
    \centering
    \includegraphics[width=\textwidth]{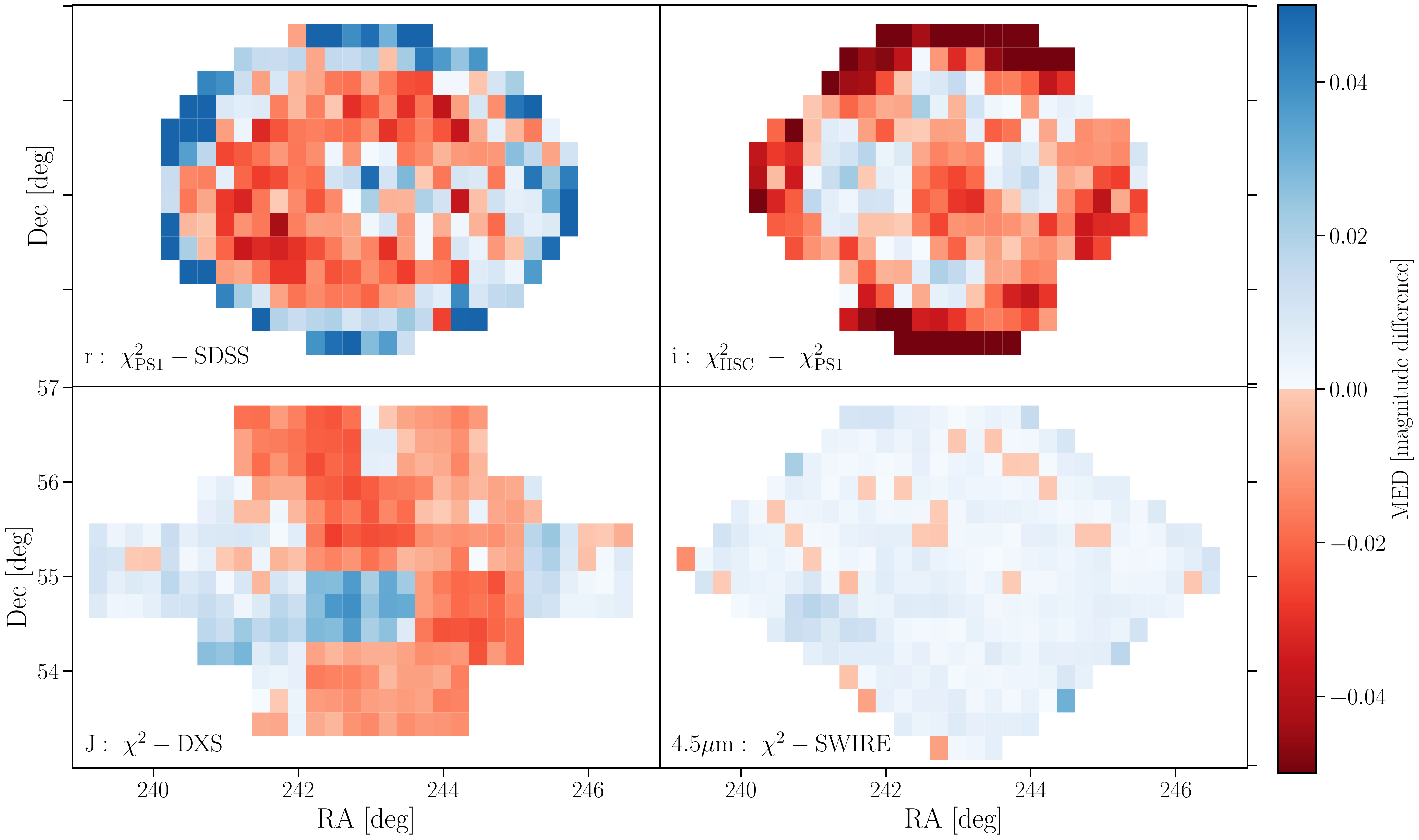}
    \caption{Comparison of photometry in the r, i, J and 4.5\,$\mu$m bands in ELAIS-N1. The colour-map shows the median magnitude difference computed over cells of 0.06 \mbox{deg$^{2}$} between our $\mathrm{\chi^{2}}$ and publicly available catalogues from SDSS DR12, DXS DR10 and the SWIRE survey for the r, J and 4.5\,$\mu$m bands, respectively. For the i-band, we compare the photometry between PS1 and HSC within our $\mathrm{\chi^{2}}$ catalogue. We use aperture corrected magnitudes based on the 3 arcsec aperture for optical-NIR bands and 4 arcsec for the 4.5\,$\mathrm{\mu}$m band. There is excellent agreement between our $\mathrm{\chi^{2}}$ and publicly released catalogues, with differences in optical bands likely driven by zero-point calibration of individual PS1 chips.} 
    \label{fig:mdiff_cat_val}
\end{figure*}

To estimate the astrometric accuracy of our mosaics, we compared the median scatter in the RA and Declination between catalogues derived from individual mosaics. We find median astrometric offsets between 0.07$\arcsec$~--~0.13$\arcsec$, all of which occur at scales smaller than the pixel size of 0.2$\arcsec$.

In Fig.~\ref{fig:nm_comp}, we plot the K-band selected source counts (per square arcsecond per 0.5 magnitude) from the ELAIS-N1 and Lockman Hole $\chi^{2}$ catalogues, along with the Ks-band selected source counts from the merged catalogue in Bo\"{o}tes. Number counts from COSMOS deep area of \cite{2016ApJS..224...24L} in the Ks-band are also shown. The number counts from the UKIDSS DR10 catalogues in both ELAIS-N1 and Lockman Hole are also shown for comparison with each field. Vertical dashed lines indicate the 3- and 5-$\sigma$ limiting magnitudes in ELAIS-N1. In all cases, contribution from foreground stars are removed by performing a cross-match to GAIA DR2 stars with Gmag $<$ 19 mag. This plot shows that there is excellent agreement between our $\mathrm{\chi^{2}}$ and the UKIDSS catalogue within each field. We also note that the difference between the ELAIS-N1 and the Lockman Hole number counts seen in our $\mathrm{\chi^{2}}$ catalogues, especially at bright magnitudes, is also seen in the UKIDSS DR10 catalogues, suggesting that this is likely due to large scale structure between the two fields. This difference is also seen with the \cite{2016ApJS..224...24L} data, which agrees well with the ELAIS-N1 data (both our $\mathrm{\chi^{2}}$ and UKIDSS DR10 catalogues) but not with the other fields at K $< 20$ mag, which is likely due to large scale structure. The plot also shows that our catalogues, especially in ELAIS-N1, reach a slightly higher completeness than the UKIDSS DR10 catalogue at S/N of 3 -- 5 due to the use of $\chi^{2}$ detection images. 

For the optical filters, we have compared our aperture corrected magnitudes with model magnitudes from SDSS DR13 \citep{Albareti2017sdssdr13} where the coverage overlaps and find very good agreement to a few percent level, well below the typical photometric uncertainties. We show a typical example for the PS1 r-band in ELAIS-N1 Fig.~\ref{fig:mdiff_cat_val} (\textit{top left}) which illustrates the median magnitude difference in cells of 0.06 \mbox{deg$^{2}$}. This is calculated by comparing our photometry for relatively bright (r $<$ 21 mag) sources with SDSS model magnitudes, accounting for the small differences in the PS1 and SDSS filters using colour terms estimated from \cite{finkbeiner2016ps1_sdss_col}. There is good agreement with SDSS for most of the PS1 footprint, however, the PS1 r-band magnitudes are too faint by 5-10\% near the edge of the PS1 footprint. We find that this trend, which is observed across all PS1 filters (albeit sometimes with smaller offset, or larger scatter), is likely driven by the zero-point calibration of the individual chips in the PS1, which gets fainter by up to $\lesssim$~10\% by $\sim$~1.5 \mbox{deg} from the field centre. 
In Fig.~\ref{fig:mdiff_cat_val} (top right) we also show a comparison in the i-band between HSC-i and PS1-i (HSC$\mathrm{_{i}}$ - PS$\mathrm{_{i}}$), both taken from our $\chi^{2}$ catalogue, which shows good agreement across the field. The median magnitude difference gets more negative (i.e. PS1 is too faint compared to HSC) near the edges of the field by $\sim$~8\%, which is also consistent with the trend in zero-point variation discussed above. This suggests that the PanSTARRS photometry near the edge of the field can typically become more uncertain by $\lesssim$~10\%. However, it is important to note that this effect is comparable to the additional 10\% flux error typically added to the photometric uncertainties before SED fitting and moreover, as this effect occurs near the edges of the PanSTARRS footprint, some of these regions will be outside our recommended multi-wavelength area, where photometry is the most reliable. 

For the NIR J and K bands, we compare our aperture corrected fluxes to the UKIDSS-DXS DR10 catalogues in both of these fields and to that of 2MASS, finding excellent agreement to within 2-3\%. As a typical example, the comparison between the ELAIS-N1 J-band and UKIDSS DR10 across the full field is shown in Fig.~\ref{fig:mdiff_cat_val} (\textit{bottom left}). There are some small systematic offsets with position across the field, driven by the varying PSF across the field between different exposures. Therefore, our assumption of a constant PSF per filter is not entirely accurate for this band; the resulting photometry is, however, affected at the $\lesssim$~5\% level, which is much smaller than the typical additional photometric uncertainties used for photometric redshifts and SED fitting.

In the Spitzer-IRAC bands, we compare photometry to the public SWIRE and SERVS catalogues, finding a remarkably good agreement to within a $\pm$ 1\% level (e.g. Fig.~\ref{fig:mdiff_cat_val}, \textit{bottom right}).

\subsection{Final multi-wavelength catalogues}\label{sec:final_optcat_properties}
The resulting multi-wavelength catalogue in ELAIS-N1 contains more than 2.1 million sources with over 1.5 million sources in the overlapping region of Pan-STARRS, UKIDSS-DXS and Spitzer-SWIRE surveys that are used for the cross-match with the radio catalogue. Similarly, the multi-wavelength catalogue in Lockman Hole consists of over 3 million sources with over 1.9 million sources in the overlapping region of SpARCS r-band and Spitzer-SWIRE coverage. Finally, the merged Bo\"{o}tes catalogue consists of over 2.2 million sources, with around 1.9 million sources in the coverage of the original NDWFS area. Some of the key properties of the multi-wavelength and initial \textsc{PyBDSF} radio catalogues are listed in Table~\ref{tab:prop_final-opt_pybdsf}.

For each field, we release the multi-wavelength catalogue over the full field coverage. For convenience, we include a \textsc{flag\_overlap} bit value for each source in both the multi-wavelength catalogues and the radio cross-matched catalogues released, which indicates which survey footprint a source falls within. In Table~\ref{tab:prop_final-opt_pybdsf}, we list the recommended \textsc{flag\_overlap} value to use for each field, to select sources that are within our selected multi-wavelength overlap area.

For ELAIS-N1 and Lockman Hole, we release the raw (uncorrected for any aperture effects or Galactic extinction) aperture fluxes and magnitudes in each filter and in addition, provide, for each filter, a flux and magnitude corrected for aperture (in our recommended aperture) and Galactic extinction. We choose the 3$\arcsec$ aperture fluxes for all optical-NIR bands and the 4$\arcsec$ aperture for all Spitzer IRAC bands as our recommended apertures. While the 3$\arcsec$ aperture may have a lower S/N than the 2$\arcsec$ aperture for compact objects, the fluxes will be less sensitive to PSF variations or astrometric uncertainties, resulting in more robust colours. The 4$\arcsec$ aperture corresponds to roughly twice the PSF FWHM of the IRAC bands, and was found by \cite{2003PASP..115..897L} to reduce scatter in colour magnitude diagrams for stars. These aperture sizes are therefore used in our radio-optical cross-matching and for the photometric redshift estimates (described in \citetalias{duncan2020_inpress}) and for the SED fitting (described in \citetalias{best2020_inpress}).

The existing Bo\"{o}tes catalogues have already been aperture corrected. We therefore apply Galactic extinction corrections to the 3 (for optical-NIR bands) and 4 (for IRAC bands) arcsec aperture fluxes and magnitudes, in the same way as for ELAIS-N1 and Lockman Hole, and only provide these recommended fluxes and magnitudes in the catalogues released for this field. The $E(B-V)$ values used for each source are also provided in an additional column; the filter dependent extinction factors are listed in Table~\ref{tab:bootes_extcorr}. We refer readers who require photometry in other apertures to \cite{2007ApJ...654..858B, 2008ApJ...682..937B}.

It is worth re-iterating the key differences between the construction of the existing Bo\"{o}tes catalogues and the new catalogues generated for ELAIS-N1 and Lockman Hole. First, unlike in Bo\"{o}tes, where sources are detected in the I- and 4.5\,$\mu$m bands, source detection in the other two fields is performed using $\chi^{2}$ images which incorporates information from a wider range of wavelengths; as such, the resultant multi-wavelength catalogue would be expected to be more complete. Second, in generating the matched-aperture photometry in ELAIS-N1 and Lockman Hole, we do not smooth the PSFs unlike in Bo\"{o}tes; the variation of the PSFs is instead accounted for by computing different aperture corrections for each filter. In Bo\"{o}tes, aperture corrections are computed based on the Moffat profile PSF smoothing. Nevertheless, despite these differences, in both cases, the catalogues are built using both optical and IR data, extracted using \textsc{SExtractor} in dual-mode, and magnitudes are aperture corrected; thus, the catalogues are expected to be broadly comparable.

We provide here an itemised description of the key properties of the multi-wavelength catalogues released. Some of the properties (e.g. raw aperture fluxes) are only released for ELAIS-N1 and Lockman Hole.
\begin{itemize}
    \item Unique source identifier for the catalogue (``ID'')
    \item Multi-wavelength source position (``ALPHA\_J2000'', ``DELTA\_J2000'')
    \item Aperture and extinction corrected flux (and flux errors) from our recommended aperture size <band>\_flux\_corr and <band>\_fluxerr\_corr in $\mathrm{\mu}$Jy
    \item Aperture and extinction corrected magnitude (and magnitude errors) from our recommended aperture size in the AB system (<band>\_mag\_corr and <band>\_magerr\_corr)
    \item Raw aperture flux (and flux errors) in 8 aperture sizes in $\mathrm{\mu}$Jy (FLUX\_APER\_<band>\_ap and FLUXERR\_APER\_<band>\_ap; excluding Bo\"{o}tes)
    \item Raw aperture magnitude (and magnitude errors) in 8 aperture sizes in the AB system (MAG\_APER\_<band>\_ap and MAGERR\_APER\_<band>\_ap; excluding Bo\"{o}tes)
    \item Overlap bit flag indicating the coverage of source across overlapping multi-wavelength surveys (``\textsc{flag\_overlap}''). See Table~\ref{tab:prop_final-opt_pybdsf} for the recommended flag values.
    \item Bright star masking flag indicating masked and un-masked regions in the Spitzer- and optical-based bright star mask (``\textsc{flag\_clean}'')
    \item Position based $\mathrm{E(B-V)}$ reddening values from \cite{1998Schlegeldustmap} dust map (``EBV'').
    \item Manual masking and duplicate source flag from \cite{2007ApJ...654..858B,2008ApJ...682..937B} I-band catalog (``\textsc{flag\_deep}''; for Bo\"{o}tes only).
\end{itemize}

We also refer the reader to the accompanying documentation for full description of all of the columns provided in the multi-wavelength catalogues. Additional value-added columns regarding photometric redshifts, rest-frame colours, absolute magnitudes and stellar masses are described in \citetalias{duncan2020_inpress}, while the far-infrared fluxes are described by \cite{mccheyne2020_inpress}. A full description of all columns provided in the multi-wavelength catalogues (both those described here, and in the value-added catalogue) can be found in the documentation accompanying the data release.

\section{Radio-optical cross-matching}\label{sec:radio_cross-match}
The identification of the multi-wavelength counterparts to the radio-detected sources is crucial in maximising the scientific output from radio surveys. In addition, while \textsc{PyBDSF} is a very useful tool for source detection and measurement, the association of islands of radio emission into distinct radio sources is not expected to be perfect for all sources in all fields. Such incorrect associations in the \textsc{PyBDSF} catalogue can occur in a few ways, as noted by \citeauthor{2019A&A...622A...2W} (\citeyear{2019A&A...622A...2W}; hereafter W19). Firstly, radio emission from physically distinct nearby sources could be associated as one \textsc{PyBDSF} source (blended sources). Such blends are much more common in these deep LOFAR data than the LoTSS-DR1. Secondly, sources with multiple components could be incorrectly grouped into separate \textsc{PyBDSF} sources due to a lack of contiguous emission between the components. For example, this can occur for sources with double radio lobes, or with large extended or diffuse radio emission. Therefore in this paper, we also aim to form correct associations of the sources and components generated by \textsc{PyBDSF}. 

In this section, we describe the methods we use to form the correct associations of the radio detected sources and, to cross-match (identify) the multi-wavelength counterparts of the radio sources. The multi-wavelength identifications were achieved by using a combination of the statistical LR method and a visual classification scheme for sources where the statistical method is not suitable, whereas source association was performed using visual classification only. Both the radio-optical cross-match and the source associations for the LOFAR Deep Fields DR1 adapt the techniques developed and presented for LoTSS-DR1 by \citetalias{2019A&A...622A...2W}. We refer the reader to that paper for details of the process; here, we summarise these methods and in particular, describe our specific adaptation and implementation of these methods to the LoTSS Deep Fields.

Firstly, to determine the sources that can be cross-matched using the statistical method and those that need to be classified visually, we develop a decision tree (workflow) in Sect.~\ref{sec:workflow}. In Sect.~\ref{sec:general_lr}, we describe the application of the statistical LR method, which allows the identification of counterparts for sources with well-defined radio positions. Section~\ref{sec:general_vis} then details the visual classification schemes performed using a combination of LOFAR Galaxy Zoo (LGZ), where source association and counterpart identification is performed using a group consensus, and, a separate workflow for specialised cases which are classified by a single expert. We note that our radio cross-matching techniques and the cross-matched catalogues released are performed for sources within the overlapping multi-wavelength coverage defined in Sect.~\ref{sec:data}, less the region of the Spitzer-based bright star mask for each field. These areas are quoted in Table~\ref{tab:prop_final-opt_pybdsf}.

\subsection{Decision tree}\label{sec:workflow}
\begin{figure*}
    \centering
    \includegraphics[clip, trim=0.9cm 4.45cm 0.7cm 0.75cm, width=\textwidth]{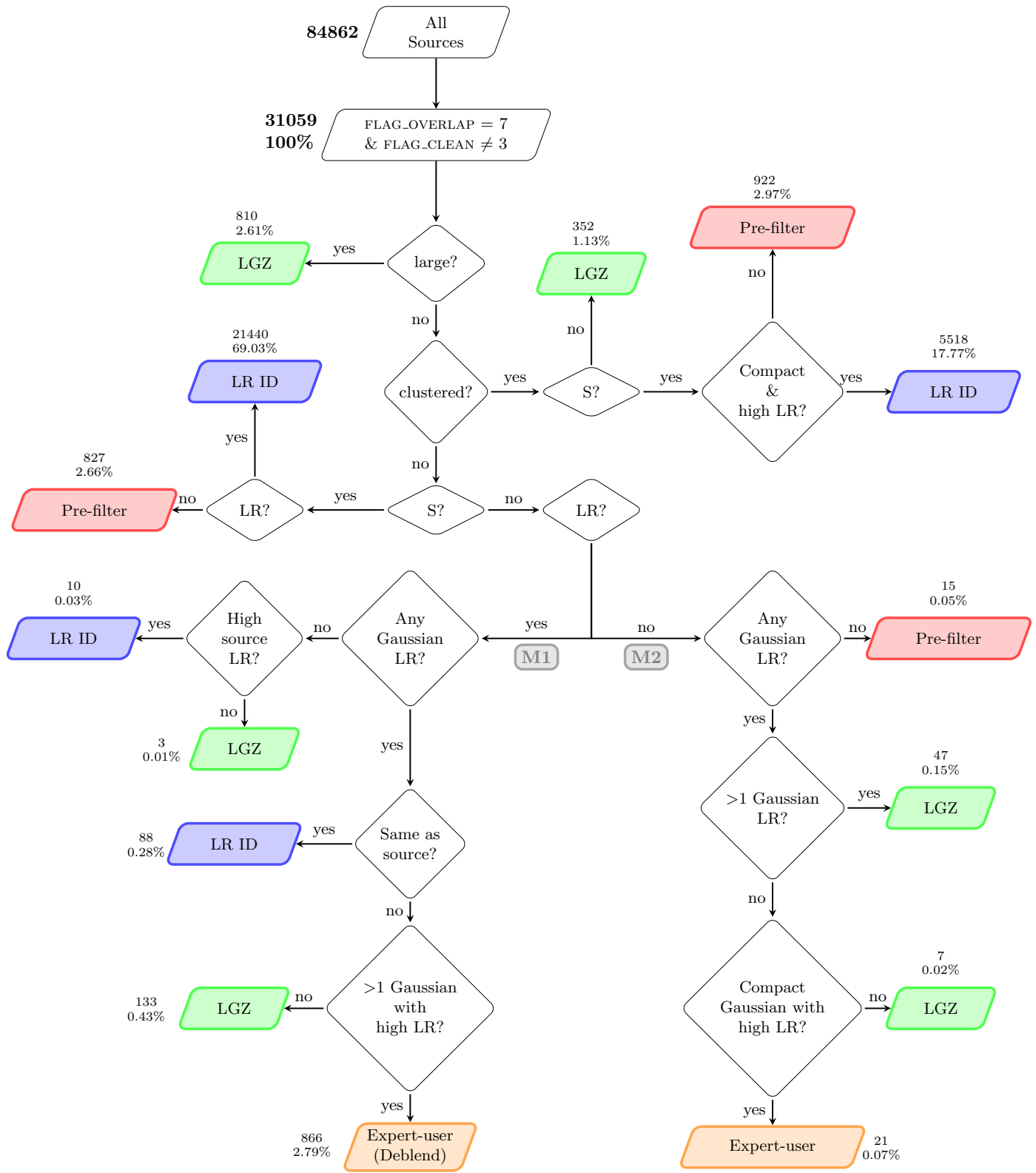}
    \caption{Flowchart developed for the deep fields to select the most appropriate method (end-point) for identification (LR or visual) based on properties of the radio source and LR-identification (if any). The `LR-ID' end-point indicates that the LR cross-match is accepted (see Sect.~\ref{sec:general_lr}). Sources with the end-point of `LGZ', `pre-filter' and `expert-user workflow' are sent to their respective visual classification and identification workflows (see Sect.~\ref{sec:general_vis}). The numbers and percentages of sources at each end-point relate to the total number of sources in the \textsc{PyBDSF} catalogue within our defined multi-wavelength area (\textsc{flag\_overlap} = 7; see Table~\ref{tab:prop_final-opt_pybdsf}) and not in the Spitzer-masked region (\textsc{flag\_clean} $\neq$ 3; see Table~\ref{tab:star_mask_flags}), corresponding to 31059 \textsc{PyBDSF} sources. Table~\ref{tab:workflow_definitions} lists the definitions used for each decision block. The decision tree is described in detail in Sect.~\ref{sec:workflow}. In ELAIS-N1, 27056 (87.1\%) of sources were selected as suitable for analysis by the statistical LR method, with the rest (4003 sources = 12.9\%) selected as requiring some form of visual classification.}
    \label{fig:en1_flowchart}
\end{figure*}

\begin{table*}
\centering
	\caption{Key criteria and definitions used in the decision blocks of the decision tree in Fig.~\ref{fig:en1_flowchart}. LR\textsubscript{th} is the LR threshold corresponding to the intersection of the completeness and reliability. This is a scalar value that varies for each field.}
	\label{tab:workflow_definitions}
\begin{tabular}{ll}
\hline\hline
Parameter & Definition \\
\hline
Large & \textsc{PyBDSF} major axis $>$ 15$\arcsec$ \\
Clustered & Distance to fourth nearest neighbour $<$ 45$\arcsec$ \\
S & `Simple' source: single Gaussian \textsc{PyBDSF} source (and only source in the island) \\
LR & LR $>$ LR\textsubscript{th} \\
High source LR & LR\textsubscript{source} $>$ 10 $\times$ LR\textsubscript{th} \\
Compact \& high LR & \textsc{PyBDSF} source major axis $<$ 10$\arcsec$ and LR\textsubscript{source} $>$ 10 $\times$ LR\textsubscript{th}\\
Same as source & The ID(s) for the Gaussian component(s) is identical to ID for the source \\
Any Gaussian LR & At least one Gaussian component with LR\textsubscript{gauss} $>$ LR\textsubscript{th}\\
> 1 Gaussian LR & More than 1 Gaussian with LR\textsubscript{gauss} $>$ LR\textsubscript{th}\\
> 1 Gaussian with high LR & More than 1 Gaussian with LR\textsubscript{gauss} $>$ 10 $\times$ LR\textsubscript{th}\\
Compact Gaussian with high LR & Gaussian major axis $<$ 10$\arcsec$ and LR\textsubscript{gauss} $>$ 10 $\times$ LR\textsubscript{th} \\
\hline
\end{tabular}
\end{table*}

The decisions for how a source would be identified and/or classified are shown pictorially in Fig.~\ref{fig:en1_flowchart}, with numbers and fractions on the plot tracking the 31059 \textsc{PyBDSF} sources in ELAIS-N1 (see also Table~\ref{tab:prop_final-opt_pybdsf}). The decisions used (many of which are the same as those of W19) are listed in Table~\ref{tab:workflow_definitions}, and were based on both the radio source properties (e.g. size, source density, etc.) and the LR cross-matches (if any) of both the \textsc{PyBDSF} source and the Gaussian component catalogues in each field. Compared to the decision tree in LoTSS-DR1 \citepalias{2019A&A...622A...2W}, some of the decision blocks could be simplified by sending sources directly for visual classification (without compromising feasibility) as the LR cross-match rate in the deep fields is significantly higher than in LoTSS-DR1, and the number of sources reaching these visual classification end-points is also much smaller (with very few extremely large/extended sources in these smaller areas). Furthermore, given the very high LR identification rates of up to 97\% (see Sect.~\ref{sec:lr_results}), it is feasible to send any source for which a counterpart cannot be determined using the LR method to visual inspection for confirmation that there is no possible counterpart.

We now describe in detail the key decision blocks and end-points of the decision tree. To select `simple' and `complex' sources in the decision tree, we use the \textsc{S\_Code} parameter from the \textsc{PyBDSF} catalogues. We define a `simple' source (`S' in Table~\ref{tab:workflow_definitions}) to only include sources that were fitted by a single Gaussian and are also the only source in the island (\textsc{S\_Code} $=$ S). Sources that were instead fitted with either multiple Gaussians (\textsc{S\_Code} $=$ M) or were fitted with a single Gaussian but were in the same island as other sources (\textsc{S\_Code} = C) are defined as being `complex'. Throughout this paper, we define a source as having a `LR identification' (or LR-ID), if the LR value of the cross-match is above the LR threshold chosen (see Sect.~\ref{sec:general_lr} and Appendix~\ref{sec:comp_rel}).

In the decision tree, we first consider the size of the radio source. Radio sources with large sizes are typically complex or have poor positional accuracy; statistical methods of cross-identification for these sources are not accurate. Moreover, large \textsc{PyBDSF} sources may be part of even larger physical sources that are not correctly associated; these sources would need to be associated visually before the correct multi-wavelength ID can be selected. We therefore directly sent all large (major axis size $>$ 15$\arcsec$) sources (810 sources = 2.6\% in ELAIS-N1) in the \textsc{PyBDSF} catalogue to LGZ (see Sect.~\ref{sec:general_lgz} for details of LGZ).

Next, for sources that are not large, we then test if they are in a region of high source density (referred to by \citetalias{2019A&A...622A...2W} as `clustered'); sources in high source density regions are more likely to be a part of some larger or complex source (although in some cases, could be just a chance occurrence due to sky projection). We define a source as `clustered' if the separation to the fourth nearest neighbour (NN) is $<$ 45$\arcsec$, the same criteria as used in LoTSS-DR1. All `clustered' sources that are `complex' were sent to LGZ since the complex nature of the sources would probably make them unsuitable for LR. Where instead these sources were `simple', we checked if the source was compact and if the LR identification found was highly secure (see  Table~\ref{tab:workflow_definitions}); if so, we accepted the secure LR identification found. Otherwise, the source was sent to the pre-filter workflow where one expert would quickly inspect the source and decide whether the LR cross-match found (or lack of a LR-ID) is correct and should be accepted as the identification (or lack of), or if the source is complex and requires additional association and identification via visual methods (see Sect.~\ref{sec:general_vis}). Section~\ref{sec:prefilt} provides a description of the pre-filter workflow.

The largest branch of the decision tree was formed of the remaining small, non-clustered sources (23457, 75.5\%). The LR analysis is most suitable for non-clustered, `simple' sources with compact radio emission so, if sources at this stage had a LR value above the LR threshold (LR\textsubscript{th}), we accept the multi-wavelength ID as found by the LR analysis (i.e. LR-ID). In ELAIS-N1, 21440 (69.0\%) sources were identified at this end-point. If instead a match is not found by the LR analysis (i.e. the LR is lower than the threshold), the source was sent to the pre-filter workflow to either confirm that there is no acceptable LR match, or to send for visual classification in the case that the multi-wavelength ID is missed by the LR analysis. In ELAIS-N1, 827 (2.66\%) sources were sent to the pre-filter workflow from this branch.

Next, the small, non-clustered sources that are `complex' instead, were treated in two separate branches based on whether the source LR value is above (`M1' branch) or below (`M2' branch) the threshold. For these sources, we considered both the LR identification of the source (LR\textsubscript{source}) and the LR identification of the Gaussian components of the source (LR\textsubscript{gauss}). 

If the `complex' source had a LR above the threshold, we decided the end-point of the source by considering the LR value and LR-ID found by the source and by the individual Gaussian components that form the source (see `M1' branch of Fig.~\ref{fig:en1_flowchart}). We do not simply accept the source LR identification for such sources as this branch may include sources that have complex emission fitted by multiple Gaussians, or cases where \textsc{PyBDSF} has incorrectly grouped Gaussians associated with multiple physical sources into a single \textsc{PyBDSF} catalogue source (i.e. blends). If the source LR-ID and all of its constituent Gaussian LR-IDs are the same, or if the \textsc{PyBDSF} source has a highly secure LR-ID and with no individual Gaussians having a LR-ID, we accepted the source LR-ID. If multiple Gaussians have secure LR-IDs, these are likely to be blended sources. We therefore sent these to the `expert user workflow' (see Sect.~\ref{sec:deblend} for details) to perform de-blending. Sources at all other end-points were sent to LGZ in the `M1' branch, as detailed in Fig.~\ref{fig:en1_flowchart}.

The non-clustered, `complex' sources that don't have a source LR match above the threshold were considered in the `M2' branch (see Fig.~\ref{fig:en1_flowchart}). In this branch, sources were sent to the pre-filter workflow if none of the Gaussians have a LR match, with the main aim of confirming the lack of a multi-wavelength counterpart. If only one of the constituent Gaussians had a LR match (which is also highly secure) and a compact size, we sent the source to the `expert user workflow' to confirm the LR match or to change this (and the \textsc{PyBDSF} Gaussian grouping; if necessary). Sources at all other end-points of the `M2' branch (comprising < 0.5\% of the total \textsc{PyBDSF} catalogued sources) were then sent for visual classification via LGZ.

Of the 31059 \textsc{PyBDSF} sources in ELAIS-N1, 27056 (87.11\%) sources were selected as suitable for the statistical LR analysis. 1352 (4.35\%) sources were sent directly to LGZ and 887 (2.86\%) sources were sent to the `expert-user workflow', with the majority of these selected as being potential blends. Finally, 1764 (5.68\%) sources were sent to the pre-filter workflow; these were appropriately flagged and then sent to the expert-user, LGZ, and LR workflows (if required). We note that the number of sources that actually underwent de-blending was different to the number of potential blends listed above, as some sources that were initially selected as blends turned out not to be genuine blends, while additional sources were input from both LGZ and pre-filter that were flagged as blends. In the rest of this section, we describe in detail how we classify and identify the host galaxies of sources that are in each of the four distinct end-points of the decision tree.

\subsection{The Likelihood Ratio method}\label{sec:general_lr}
The statistical Likelihood Ratio (LR) method \citep{1977A&AS...28..211D,1992MNRAS.259..413S} is commonly used to identify counterparts to radio and milli-metre sources (e.g. \citealt{2011MNRAS.416..857S,2012MNRAS.423.2407F,2012MNRAS.423..132M}). Defined simply, the LR is the ratio of the probability that a galaxy with a given set of properties is a genuine counterpart as opposed to the probability that it is an unrelated background object. In this paper, we use the magnitude $m$, and the colour $c$ information to compute the LR of a source. \cite{nisbet2018role} and \citetalias{2019A&A...622A...2W} show that incorporating colour into the analysis greatly benefits the LR analysis, finding that redder galaxies are more likely to host a radio source. The LR is given by
\begin{equation}\label{eq:lr}
\centering
LR = \frac{q(m,c) f(r)}{n(m,c)},
\end{equation}
where $q(m,c)$ gives the a priori probability that a source with magnitude $m$ and colour $c$ is a counterpart to the radio (LOFAR) source. $n(m,c)$ represents the sky density of all galaxies of magnitude $m$ and colour $c$. $f(r)$ is the probability distribution of the offset between the radio source and the possible counterpart, while accounting for the positional uncertainties of both of the sources. A full description of the theoretical background and method of the LR technique is given in \citetalias{2019A&A...622A...2W} and is not reproduced here. Instead, we focus mainly on the specific application of the LR technique to the LOFAR Deep Fields dataset.

\subsubsection{Calculating $n(m)$ and $n(m,c)$}\label{sec:nm_c}
The $n(m)$ corresponds to the number of objects in the multi-wavelength catalogue at a given magnitude per unit area of the sky. This is computed simply by counting the number of sources within a large representative area (typically $>$ 3.5 \mbox{deg$^{2}$} in our case) in each of the three fields. We adopt a Gaussian kernel density estimator (KDE) of width 0.5mag to smooth the $n(m)$ distribution and provide a more robust estimate when interpolated at a given magnitude. The $n(m,c)$ is then simply given by computing the $n(m)$ separately for different colour bins (see Sect.~\ref{sec:qm_c}).

\subsubsection{Calculating $f(r)$}\label{sec:fr}
The $f(r)$ term accounts for the positional difference between the radio source and a potential multi-wavelength counterpart. The form of the distribution is given as a 2D Gaussian with
\begin{equation}\label{eq:fr}
f(r) = \frac{1}{2 \pi \sigma_{maj} \sigma_{min}} \exp \left( \frac{-r^{2}}{2 \sigma_{dir}^{2}} \right)
\end{equation}
where, $\sigma_{maj}$ and $\sigma_{min}$ are the combined positional uncertainties along the major and minor axes, respectively, and $\sigma_{dir}$ is the combined positional uncertainty, projected along the direction between the radio source and the potential counterpart. The $\sigma_{maj}$ and $\sigma_{min}$ terms are a combination of the uncertainties in both the radio and the potential multi-wavelength counterpart positions, and the uncertainties in the relative astrometry of the two catalogues, calculated using the method of \cite{1997PASP..109..166C}. For the potential multi-wavelength counterparts, as the positional uncertainties from a $\chi^{2}$ detection image are unreliable, we adopt a circular positional uncertainty of $\sigma_{\mathrm{opt}} = $ 0.35$\arcsec$. Similar to \citetalias{2019A&A...622A...2W}, an additional astrometric uncertainty between the radio and multi-wavelength catalogues of $\sigma_{\mathrm{ast}} = $ 0.6$\arcsec$ was adopted. These terms were then added in quadrature for radio source and potential counterparts to derive $\sigma_{\mathrm{maj}}$ and $\sigma_{\mathrm{min}}$.

\begin{table}
    \centering
    \caption{$Q_{0}$ values in the optical (i-band for ELAIS-N1 and Bo\"{o}tes and r-band for Lockman Hole; see text) and 4.5\,$\mathrm{\mu}$m bands for the magnitude only LR run in each field.}
    \label{tab:Q0_run1}
    \begin{tabular}{lll}
        \hline\hline
        Field & $Q_{0, \mathrm{opt}}$ & $Q_{0,4.5}$ \\
        \hline
        ELAIS-N1 & 0.85 & 0.95 \\
        Bo\"{o}tes & 0.75 & 0.84 \\
        Lockman Hole & 0.78 & 0.95 \\
        \hline
    \end{tabular}
\end{table}

\subsubsection{Calculating $q(m)$ and $q(m,c)$}\label{sec:qm_c}
$q(m)$ (and $q(m,c)$) is the a priori probability distribution that a radio source has a genuine counterpart with magnitude $m$ (and colour $c$). The integral of $q(m)$ to the survey detection limit gives $Q_{0}$, the fraction of radio sources that have a genuine counterpart up to the magnitude limit of the survey. 

The LR analysis is not suitable for large or complex radio sources, and to reduce the bias introduced by such sources on the LR analysis, we initially performed the LR analysis only for radio sources with a major axis size smaller than 10$\arcsec$. In each field, this subset of radio sources was used initially to calibrate the $q(m,c)$ distributions (using the two stage method, as described below in this section). These calibrated $q(m,c)$ distributions were then used to compute the LRs for all radio sources within the multi-wavelength coverage area listed in Table~\ref{tab:prop_final-opt_pybdsf}. The decision tree described in Sect.~\ref{sec:workflow} was then used to re-select radio sources that were more suitable for the LR analysis. For this purpose, we choose to calibrate on all `simple' sources that reach the LR-ID or the pre-filter end-points of the decision tree. The $q(m,c)$ distributions were re-calibrated on this sample and then used to re-compute the LRs for all the radio sources in the field to derive the final counterparts. We found that further iterations of the decision tree made insignificant changes ($\lesssim$ 1\%) to the number of sources selected for visual analysis or LR, suggesting that the calibration was being performed on sources most suitable for the LR analysis.

Various methods have been developed to estimate $q(m)$ and $Q_{0}$ using the data itself (e.g. \citealt{2011MNRAS.416..857S,2012MNRAS.423..132M,2012MNRAS.423.2407F}), in a manner that is unbiased by the clustering of galaxies. However, as explained by \citetalias{2019A&A...622A...2W}, these methods cannot be used to estimate $q(m,c)$ and $Q_{0,c}$ in different colour bins. Instead, we use the iterative approach developed by \cite{nisbet2018role} and applied to the LoTSS DR1 by \citetalias{2019A&A...622A...2W} for estimating $q(m,c)$ in two stages. Briefly, the first stage of this approach involves identifying an initial estimate of the host galaxies using well established magnitude-only LR techniques (e.g. \citealt{2012MNRAS.423.2407F}). In the second stage, this initial set of host galaxies is split into various colour bins, to allow a starting estimate of $q(m,c)$ to be obtained, which is then used to recompute the LRs, incorporating colour information. This then provides a new set of host galaxy matches, and hence an improved estimate of $q(m,c)$, with this process iterated until the $q(m,c)$ distribution converges. 

\begin{table}
\centering
	\caption{Table of iterated $Q_{0}(c)$ values for ELAIS-N1, Lockman Hole, and Bo\"{o}tes. The colour $\mathrm{c}$ is derived using optical - 4.5\,$\mathrm{\mu}$m magnitude where we use the i- (or I-) band in ELAIS-N1 and Bo\"{o}tes, and the r-band in Lockman Hole. The LR thresholds (LR\textsubscript{th}) derived from the intersection of the completeness and reliability function (see Appendix~\ref{sec:comp_rel}) and used for selecting genuine cross-matches are also listed. Scaled (by excluding sources in region of around stars, i.e. \textsc{flag\_clean}$\neq$ 3) total $Q_{0}(c)$ values are also listed.}
	\label{tab:Q0_c_all}
\begin{tabular}{llll}
\hline\hline
{} & \multicolumn{3}{c}{$Q_{0}(c)$} \\
Colour Bin & ELAIS-N1 & Lockman Hole & Bo\"{o}tes \\
\hline
$\mathrm{c \leq}~ {-0.5}$ & 0.0031 & 0.0013 & 0.0016 \\
${-0.5} ~ \mathrm{< c \leq}~ {-0.25}$ & 0.0034 & 0.0012 & 0.003 \\
${-0.25} ~ \mathrm{< c \leq}~ {0.0}$ & 0.0081 & 0.0041 & 0.0103 \\
${0.0} ~ \mathrm{< c \leq}~ {0.25}$ & 0.0177 & 0.0086 & 0.0204 \\
${0.25} ~ \mathrm{< c \leq}~ {0.5}$ & 0.0301 & 0.0154 & 0.0316 \\
${0.5} ~ \mathrm{< c \leq}~ {0.75}$ & 0.0468 & 0.022 & 0.0465 \\
${0.75} ~ \mathrm{< c \leq}~ {1.0}$ & 0.0562 & 0.0302 & 0.059 \\
${1.0} ~ \mathrm{< c \leq}~ {1.25}$ & 0.0606 & 0.0393 & 0.0608 \\
${1.25} ~ \mathrm{< c \leq}~ {1.5}$ & 0.0566 & 0.044 & 0.0589 \\
${1.5} ~ \mathrm{< c \leq}~ {1.75}$ & 0.0523 & 0.0457 & 0.0555 \\
${1.75} ~ \mathrm{< c \leq}~ {2.0}$ & 0.0557 & 0.045 & 0.0519 \\
${2.0} ~ \mathrm{< c \leq}~ {2.25}$ & 0.0486 & 0.0491 & 0.0504 \\
${2.25} ~ \mathrm{< c \leq}~ {2.5}$ & 0.0489 & 0.0486 & 0.0477 \\
${2.5} ~ \mathrm{< c \leq}~ {2.75}$ & 0.0467 & 0.0481 & 0.0448 \\
${2.75} ~ \mathrm{< c \leq}~ {3.0}$ & 0.0478 & 0.0498 & 0.0472 \\
${3.0} ~ \mathrm{< c \leq}~ {3.25}$ & 0.0481 & 0.0484 & 0.0473 \\
${3.25} ~ \mathrm{< c \leq}~ {3.5}$ & 0.0456 & 0.0493 & 0.0433 \\
${3.5} ~ \mathrm{< c \leq}~ {3.75}$ & 0.0422 & 0.0496 & 0.0357 \\
${3.75} ~ \mathrm{< c \leq}~ {4.0}$ & 0.0388 & 0.0463 & 0.0325 \\
$\mathrm{c >}~ {4.0}$ & 0.076 & 0.1359 & 0.0546 \\
optical-only & 0.0044 & 0.0068 & 0.001 \\
4.5-only & 0.1129 & 0.1771 & 0.1107 \\
no-magnitude & 0.0019 & 0.0061 & 0.001 \\
\hline
Total $Q_{0}(c)$ & 95.3\% & 97.2\% & 91.6\% \\
\makecell[l]{Total $Q_{0}(c)$ with \\ \textsc{flag\_clean}$\neq$3} & 96.2\% & 97.4\% & 94.2\% \\
\hline
LR threshold & 0.056 & 0.055 & 0.22 \\
\hline
\end{tabular}
\end{table}

In practice, in the first stage, we generated a set of initial counterparts to the radio sources using only the magnitude information by cross-matching the radio sources to both the 4.5\,$\mathrm{\mu}$m detected and optical detected \footnote{We define a source as being detected in a given filter if the S/N $>$ 3 inside the 2$\arcsec$ aperture in that filter.} sources (separately). For the optical dataset, we use the PS1 i-band in ELAIS-N1, the NDWFS I-band in Bo\"{o}tes, and the SpARCS r-band in Lockman Hole. While there exist i-band data from RCSLenS in Lockman Hole, the survey coverage had gaps in the field between different pointings due to the survey strategy employed. Therefore, we compromise slightly on the choice of optical filter for LR analysis in favour of area coverage. The method of \cite{2012MNRAS.423.2407F} was then used to compute $Q_{0}$ in each filter. We list the $Q_{0}$ values in the optical and 4.5\,$\mathrm{\mu}$m bands for each field from this first stage in Table~\ref{tab:Q0_run1}. The differences in these $Q_{0}$ values are largely driven by the relative depths of the optical and Spitzer surveys between the three fields.

The $Q_{0}$ values were then used to derive the corresponding $q(m)$ distributions following the method of \cite{2012MNRAS.423.2407F}. The final part of the first stage then involves computing the LRs for all optical and 4.5\,$\mathrm{\mu}$m detected sources (separately) within 10$\arcsec$ of a radio source. An optical or 4.5\,$\mathrm{\mu}$m detected source was accepted as a cross-match if the LR was above the threshold in that particular filter; in this first stage, the LR threshold is simply estimated as the value for which a fraction Q$_{0}$ of cross-matches were accepted in that band. If multiple sources within 10$\arcsec$ were above the LR threshold, the source with the highest LR (in either the optical or the 4.5\,$\mathrm{\mu}$m band) was retained as the most-probable cross-match. The main output of the first stage generates a first-pass set of multi-wavelength counterparts.

In the second stage, the counterparts generated from the first stage were divided into colour categories to provide an initial estimate of $Q_{0}(c)$ ($= N_{c}/N_{LOFAR}$) and $q(m,c)$. Colour bins were derived from the (optical - 4.5\,$\mu$m) colour, provided the source is detected with S/N $>$ 3 in both bands. These sources were then split into 20 colour bins, as listed in Table~\ref{tab:Q0_c_all}. In addition to these, some sources are only detected (S/N $>$ 3) in either the optical or the 4.5\,$\mathrm{\mu}$m band. For these sources, we define two additional colour categories: optical-only and 4.5-only sources. Finally, as mentioned earlier, due to the nature of the detection method using $\chi^{2}$ images, there are sources that have a low S/N in both the i (or r) and 4.5\,$\mathrm{\mu}$m filters but appear in the catalogue due to detections in other bands. These sources were placed in a final colour category, the `no-magnitude' category, for which we manually set a first-pass value for the cross-match fraction of $Q_{0} = $ 0.001 and use the corresponding sky density of all sources in this bin to compute the LRs.

\begin{figure*}
    \centering
    \includegraphics[width=0.90\textwidth]{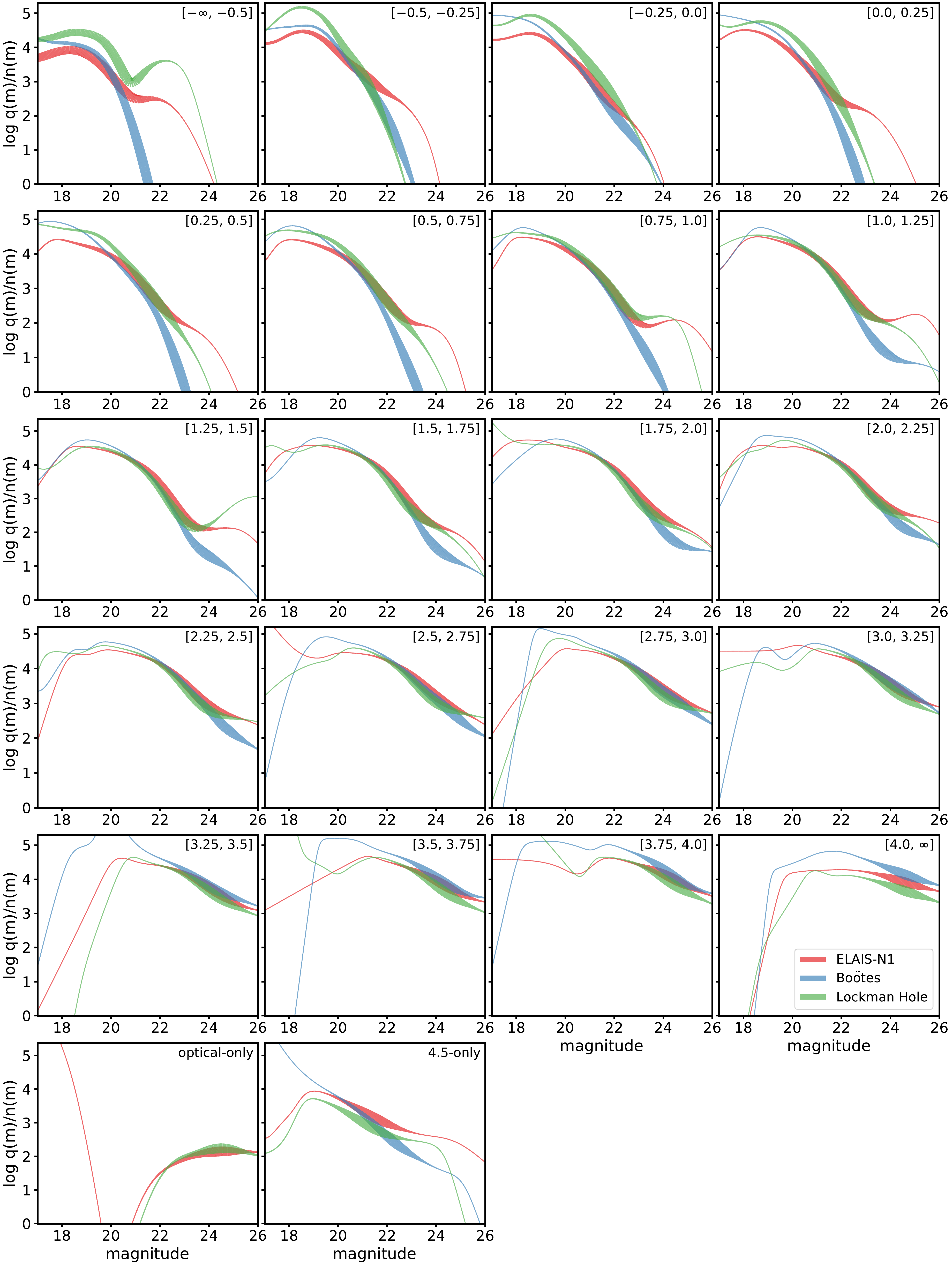}
    \caption{$\mathrm{q(m,c)/n(m,c)}$ ratio distributions versus magnitude, smoothed using a KDE. The x-axis displays the optical magnitude in each colour bin, except for the $\mathrm{4.5}$-only category where the 4.5\,$\mathrm{\mu}$m magnitude is used. The width of the lines corresponds to the number of radio sources within that magnitude bin (hence, the thicker lines indicate well-constrained regions of parameter space). The optical magnitudes plotted are the same as those chosen for the LR analysis (i-band for ELAIS-N1, I-band for Bo\"{o}tes and r-band for Lockman Hole). Although these filters are different, no attempt at filter or colour transformation is made (see Sect.~\ref{sec:lr_results}). Even without these corrections, the distributions agree well between the three fields, especially considering the log scaling of the y-axis.} 
    \label{fig:qm_nm_c} 
\end{figure*}

\begin{figure*}
    \centering
    \includegraphics[width=\textwidth]{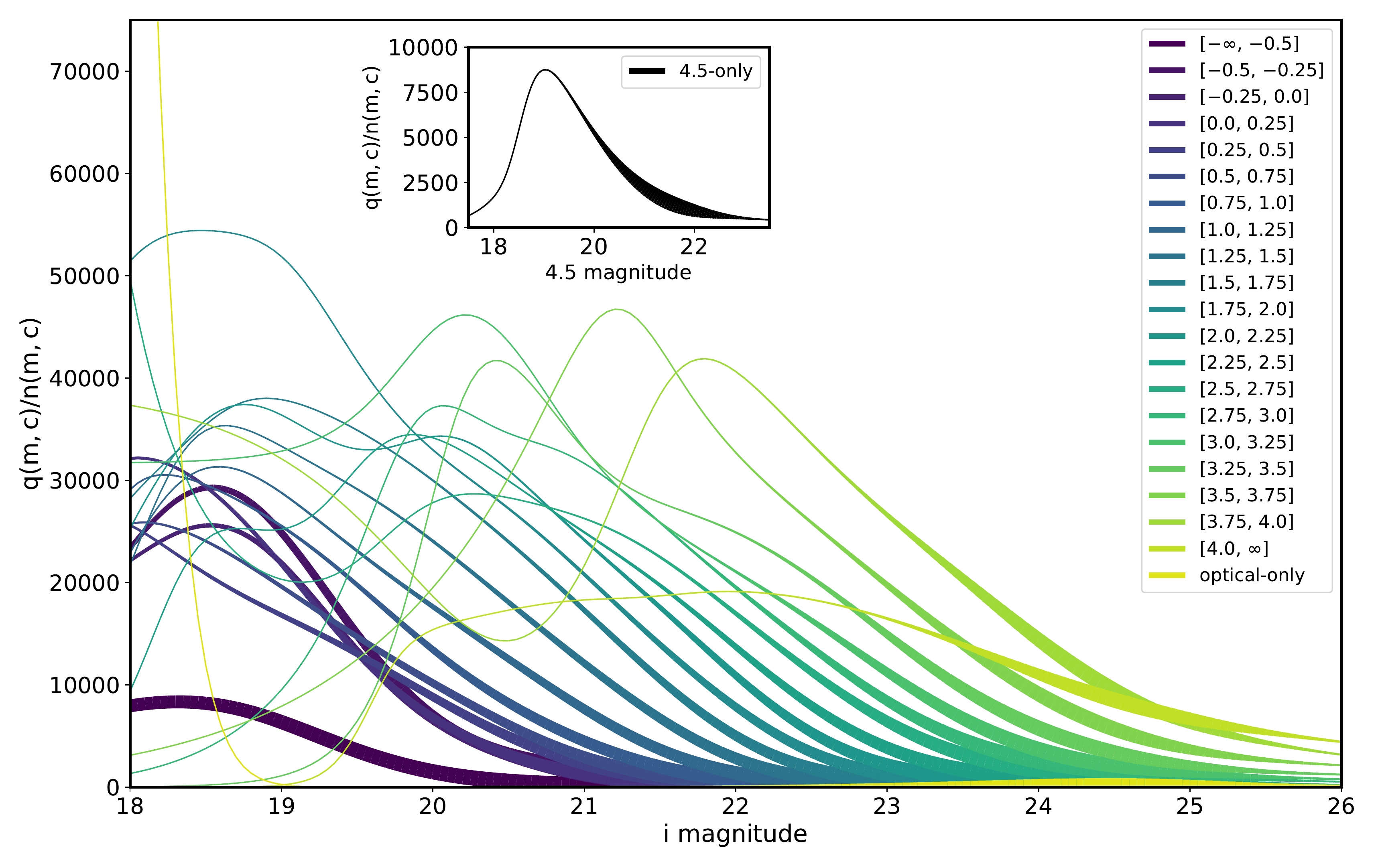}
    \caption{Plot of the $\mathrm{q(m,c)/n(m,c)}$ ratio distributions versus magnitude across the (i~-~4.5) colour bins in ELAIS-N1. The inset shows the same for the 4.5-only bin. The ratios are computed in bins of 0.5 mag (and smoothed using a KDE), with the thickness of the lines corresponding to the number of sources within a given magnitude bin (i.e. thicker lines represent better constrained regions of parameter space). The numbers in the legend correspond to the bin edges in (i~-~4.5) colour space. The evolution of the peak and thickness of the curve across the colour bins indicate that radio galaxies are more likely to be hosted by redder galaxies, especially at faint magnitudes.} 
    \label{fig:qm_nm_c_EN1}
\end{figure*}

After the division into the colour categories, $n(m,c)$ and $q(m,c)$ can be determined trivially. We again smooth these distributions using a Gaussian KDE of width 0.5mag. The LR analysis was then repeated in the same manner as stage one where for each source in the multi-wavelength catalogue that is within 10$\arcsec$ of a radio source, the $n(m)$ and $q(m)$ distribution corresponding to the colour bin of that source is used to compute new LRs. A new LR threshold was determined using the completeness and reliability of the cross-matches (see Appendix~\ref{sec:comp_rel} for a detailed description), improving upon the estimate from the first stage, with the highest LR match above the threshold retained in each case to produce a new set of cross-matches. The process in the second stage was iterated until the cross-matches converged (i.e. no changes in the sources cross-matched between two consecutive iterations), which was typically within 5 iterations. The total $Q_{0}$ is then simply given by summing over the contributions from each colour category, that is, $Q_{0} = \sum_{c} Q_{0}(c)$. Iteration of the LRs can progressively drive down the $Q_{0}(c)$ values to zero in the rarest bins. To avoid this, we set a minimum $Q_{0}(c) = $ 0.001 for any colour bin.

\subsubsection{LR method results}\label{sec:lr_results}
The colour bins and the corresponding final iterated $Q_{0}(c)$ values are provided in Table~\ref{tab:Q0_c_all}. The colour $\mathrm{c}$ is the same optical - mid-IR colour that was used for the LR analysis. Table~\ref{tab:Q0_c_all} also lists the iterated LR threshold values derived from the intersection of the completeness and reliability plots (see Appendix~\ref{sec:comp_rel}). The full sample in all fields achieves a completeness and reliability $>$ 99.7\% (see Fig.~\ref{fig:comp_rel_prob} for ELAIS-N1). Visual inspection of low LR matches, and an analysis of the completeness and reliability of sources with LRs close to the LR threshold, gives confidence that the LR thresholds chosen result in genuine cross-matches (see Appendix~\ref{sec:comp_rel} for full details).

The total $Q_{0}(c)$, given by summing the contribution from each colour category, gives an identification fraction of $\sim$ 95\%, 97\% and 92\% for ELAIS-N1, Lockman Hole, and Bo\"{o}tes, respectively.  Interestingly, Lockman Hole has a higher total $Q_{0}(c)$ than ELAIS-N1, which contains IR data to a similar depth, but much deeper optical data. This difference can be understood by considering the $Q_{0}$ of the 4.5\,$\mu$m band and its coverage, in particular, that of the deeper SERVS data between the two fields. Although the optical data in ELAIS-N1 is much deeper than in Lockman Hole, the 4.5\,$\mu$m data dominates the identification fraction (see Table~\ref{tab:Q0_run1}). The SERVS 4.5\,$\mu$m data in both ELAIS-N1 and Lockman Hole reach a similar depth (as listed in Table~\ref{tab:en1_lh_description}) and achieve the same $Q_{0, 4.5}$. However, Lockman Hole benefits from having SERVS coverage (and therefore this high identification rate) over $\sim$5.6 \mbox{deg$^{2}$}, compared to only $\sim$2.4 \mbox{deg$^{2}$} in ELAIS-N1, resulting in the difference in the total $Q_{0}$. The overall $Q_{0}$ values shown in Table~\ref{tab:Q0_c_all} are significantly higher than the total cross-identification fraction of 71\% achieved in the shallower LoTSS DR1 \citepalias{2019A&A...622A...2W}. 

\begin{figure}
    \centering
    \includegraphics[width=\columnwidth]{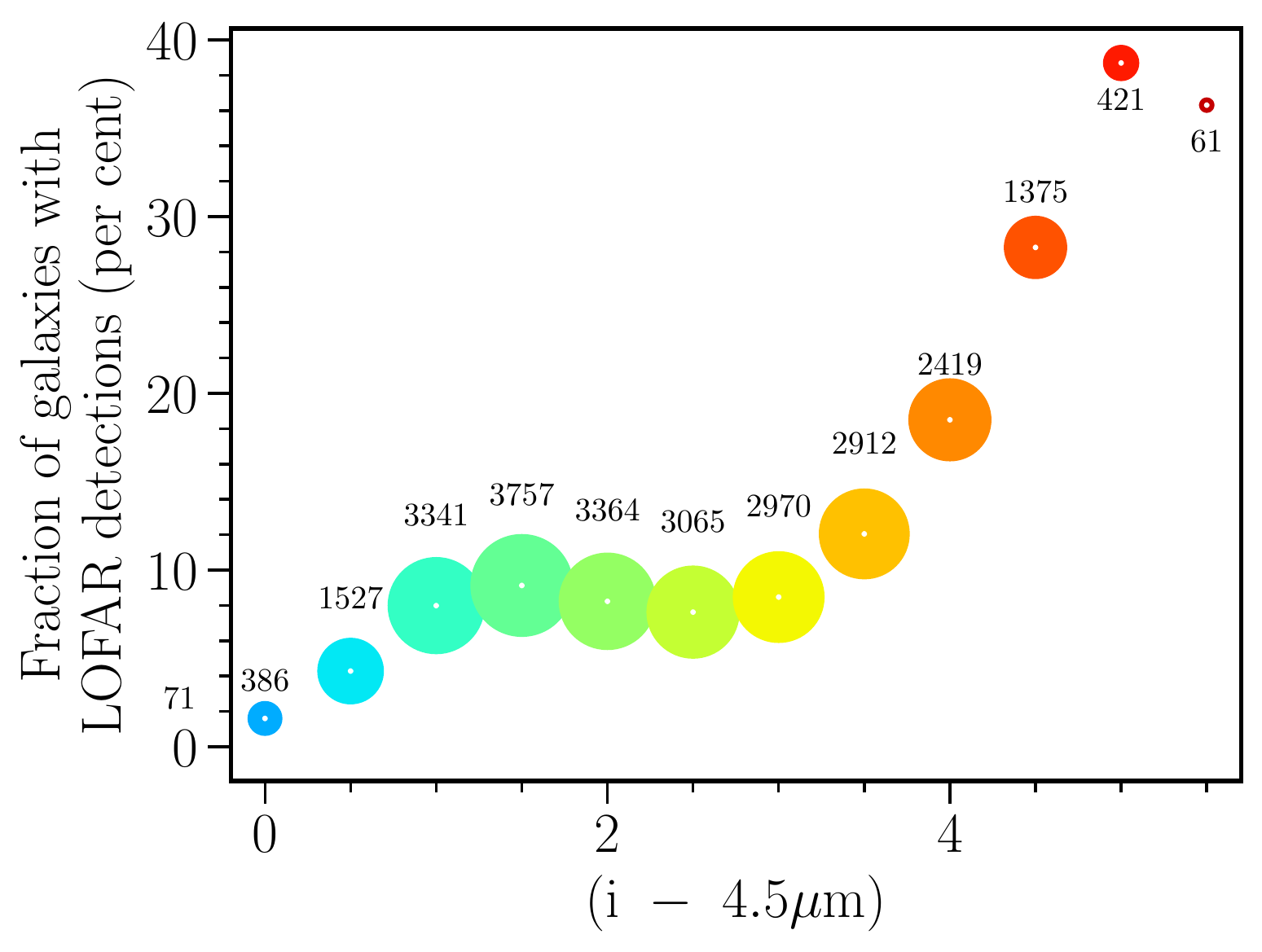}
    \caption{Fraction of all multi-wavelength sources that host a LOFAR source in ELAIS-N1 as a function of (i~-~4.5) colour. The size of the data points corresponds to number of LOFAR sources within that colour bin (indicated by the adjacent number), and the colour of the points is a proxy for counterpart colour. The reddest galaxies are more than an order of magnitude more likely to host a radio source than the bluest of galaxies.} 
    \label{fig:frac_macthes_col_bin}
\end{figure}

\begin{figure}
    \centering
    \includegraphics[width=\columnwidth]{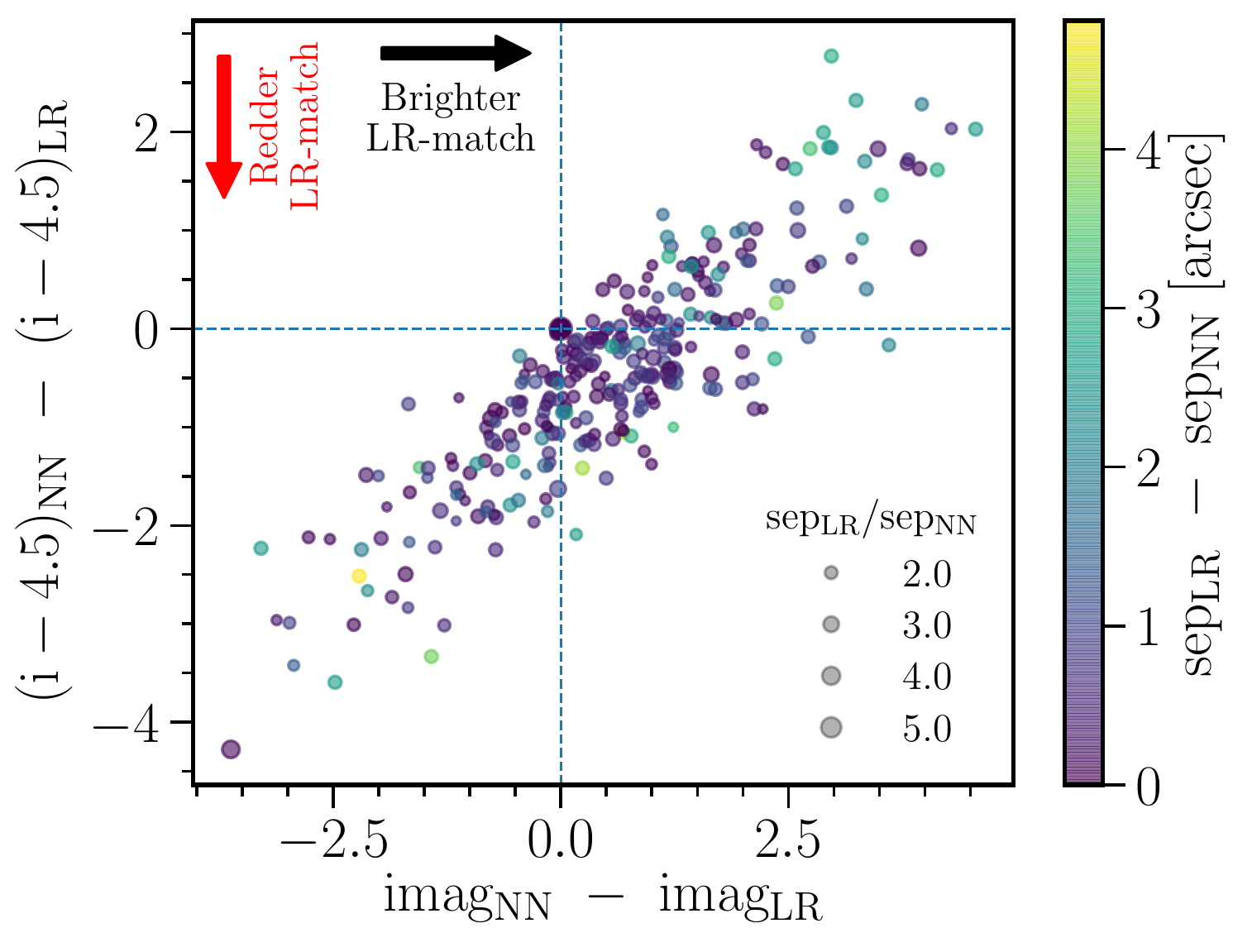}
    \caption{Comparison between the LR method and a simple NN cross-match for radio sources selected by the decision tree to accept the LR-ID in ELAIS-N1. The plot shows colour (i - 4.5$\mu$m) difference versus i-band magnitude difference between a NN search and the LR method. Negative y-axis values correspond to a redder LR match compared to the NN match, and positive x-axis values correspond to a brighter LR match, as indicated by the arrows. For the vast majority ($\approx$98\%) of the radio sources used for this comparison (see Sect.~\ref{sec:lr_results}), the LR match and the NN match are the same (indicated by the large point at (0,0)). The offset from the origin shows that, where these differ, the LR method preferentially selects sources which are either redder or brighter (or both) than the NN match. The colour of the points corresponds to the difference in separation between the LR match and the NN match ($\rm{sep_{LR}-sep_{NN}}$). The size of the points corresponds to the ratio of the separations between the LR method and the NN match (i.e. larger points indicate larger ratios; see plot legend). For radio sources where the LR match is different to the NN match, the separation to the two sources are similar.\label{fig:nn_lr_comp}}
\end{figure}

The iterated $Q_{0}(c)$ values show remarkable agreement across the three fields, especially between ELAIS-N1 and Bo\"{o}tes which both use the similar optical filters. This agreement can be visualised using the iterated (and KDE smoothed) $\log q(m,c)/n(m,c)$ ratio distributions, which are shown in Fig.~\ref{fig:qm_nm_c} as a function of magnitude. We show the distributions for ELAIS-N1 (red), Bo\"{o}tes (blue), and Lockman Hole (green) across all the colour bins. The x-axis for all colour bins is the optical magnitude (i.e. i (and I) band for ELAIS-N1 and Bo\"{o}tes, r-band for Lockman Hole), except for the 4.5-only bin where the 4.5\,$\mathrm{\mu}$m magnitude is used. The bin edges for the (optical - 4.5) colour bins (as in Table~\ref{tab:Q0_c_all}) are shown at the top right corner in each panel. The thickness of the curves corresponds to the number of sources within that magnitude bin, such that the distributions and statistics are reliable where lines are thick, and with thin lines corresponding to poorly constrained regions of parameter space, often influenced by the tails of the KDE smoothing. We note that for Lockman Hole, the comparison to the other two fields is not exactly like-for-like due to the different filters. This relates not only to an x-axis shift in colours, but also the selection of sources in each colour bin; for example, sources in Lockman Hole with 3.5 < (i~-~4.5) $<$ 3.75 have a typical colour of (r~-~i)~$\sim$~1, and hence (r~-~4.5) $>$ 4, so would appear in the c $>$ 4.0 colour category instead. The key note of importance here is that even without the filter transformation for Lockman Hole, the distributions agree well between the three fields, across the colour bins. This agreement is expected as the $q(m,c)$ distribution represents the genuine host galaxy population of radio sources in magnitude and colour space, which should be consistent between the three fields with similar radio survey properties.

In Fig.~\ref{fig:qm_nm_c_EN1}, we again show the iterated (and KDE smoothed) $q(m,c)/n(m,c)$ ratio distribution for all colour bins in ELAIS-N1, all on one plot. The numbers in the legend show bin edges in the (i~-~4.5) colour space, same as in Fig.~\ref{fig:qm_nm_c}. The evolution of the curves going from blue to redder bins indicates that redder galaxies are more likely to host radio sources, especially at faint magnitudes.

This colour dependence on the identification rate can also be visualised by considering the fraction (percentage) of all multi-wavelength sources that host a LOFAR source as a function of the (i~-~4.5) colour, as shown in Fig.~\ref{fig:frac_macthes_col_bin} for ELAIS-N1. The size of the markers indicates the number of LOFAR sources within that colour bin. The sharp rise in the fraction of matches with colour again shows that redder galaxies are more likely to host a LOFAR source as compared to the general galaxy population.

Compared to the shallower radio data available in LoTSS-DR1 (e.g. see Fig.~3 of \citetalias{2019A&A...622A...2W}), we note a rise in the fraction of LOFAR sources at blue (1 < (i~-~4.5) < 2) colours in these deep fields (as seen in Fig.~\ref{fig:frac_macthes_col_bin}), and an increase in the $q(m,c)/n(m,c)$ ratios for blue bins (as shown in Fig.~\ref{fig:qm_nm_c_EN1}). These trends compared to \citetalias{2019A&A...622A...2W} are probably due to the significant increase in depth of the radio data, where the faint radio source population ($\lesssim$~1mJy at 150MHz) starts to be dominated by radio quiet quasars and star-forming galaxies (see Fig.~4 of \citealt{2008MNRAS.388.1335W}), which are typically found in bluer galaxies.

In Fig.~\ref{fig:nn_lr_comp}, we compare the counterparts identified by the LR method with those that would be selected by a simple NN match for those radio sources that are selected by the decision tree as being suitable for the LR method in ELAIS-N1 (27056 sources). The plot shows the difference in the (i - 4.5$\mu$m) colour versus i-band magnitude difference between the NN match and the LR match. For $>$ 98\% of the sources chosen for this analysis, the LR match is also the NN match, as indicated by the cluster of points at x,y $=$ (0,0). For 500 sources, the selected LR-ID is not the same as the NN match: the deviation of these sources from the origin illustrates the role of the LR method. In all of these cases, the LR is either redder or brighter (or both) than the NN match. In some cases where the LR match is bluer, it is always brighter with typically larger counterpart separations than the NN matches (see Fig.~\ref{fig:nn_lr_comp}).

\begin{figure*}
    \centering
    \includegraphics[width=\textwidth]{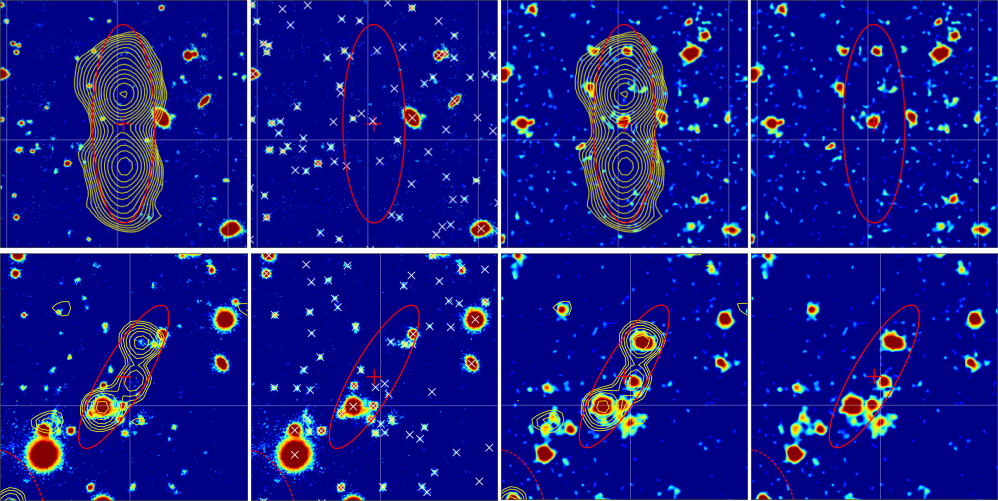}
    \caption{Example set of images used for visual classification of two sources (in rows) using LOFAR Galaxy Zoo (LGZ). The radio source to be classified is in the red ellipse with its \textsc{PyBDSF} radio position marked by a red cross. The first frame shows the optical image with radio contours overlaid. The second frame shows the same optical image now without the radio contours, but with white crosses to mark multi-wavelength catalogue detections. The third frame shows the 4.5\,$\mathrm{\mu}$m image with radio contours overlaid. The fourth frame shows the same 4.5\,$\mathrm{\mu}$m image but without radio contours to aid in host galaxy identification. \textbf{Top:} An example of a large radio source. \textbf{Bottom:} An example of a blended radio source initially sent to LGZ, where the radio emission (contours) arises from three distinct physical sources that have been incorrectly grouped together into one \textsc{PyBDSF} source (red ellipse). This source was flagged as a blend during the LGZ process by the majority of volunteers and was appropriately sent to the expert-user workflow for de-blending.}
    \label{fig:lgz_example1}
\end{figure*}

\subsection{Visual classification and source association}\label{sec:general_vis}
The LR technique is not suitable for cross-identification of sources with significantly extended (large) or complex radio emission. For such sources, visual classification must be used to identify the multi-wavelength counterparts. In addition, it is more likely that for large and complex sources, the individual radio components of a given physical source may not be grouped together correctly by \textsc{PyBDSF} - whether that be extended emission (e.g. from radio lobes) not being grouped as a single source or multiple physical sources being grouped (blended) into a single radio source. To perform correct associations and then identifications for these sources, we use a combination of LOFAR Galaxy Zoo (described in Sect.~\ref{sec:general_lgz}) and an expert-user workflow (described in Sect.~\ref{sec:deblend}) based on a source's end-point from the decision tree (Sect.~\ref{sec:workflow}).

\subsubsection{LOFAR Galaxy Zoo}\label{sec:general_lgz}
For this task of visual classification, we use the Zooniverse framework that was adapted for the LoTSS DR1: LOFAR Galaxy Zoo (LGZ; \citetalias{2019A&A...622A...2W}). LGZ is a web based interface for performing source association and host galaxy identification by visually inspecting a given radio source using the radio data and corresponding multi-wavelength images. The user can then perform identification and association by selecting appropriate radio or optical sources on the images and answering questions about the source. The details of the LGZ interface and the choice of images and options provided to the user are almost identical to the LoTSS DR1, and are described by \citetalias{2019A&A...622A...2W}, and hence not reproduced here. We briefly summarise the interface and the capabilities of LGZ, highlighting differences from the LoTSS DR1 approach. As with LoTSS DR1, the LGZ sample was only made available to members of the LOFAR consortium.

The user was presented with four sets of images when classifying a source. An example of the images presented for two radio sources are shown in Fig.~\ref{fig:lgz_example1}. The first frame shows an optical image with contours of radio emission. The second frame shows the same optical image without the radio contours but with white crosses to indicate a detection in the multi-wavelength catalogue. The third frame shows the radio contours overlaid on the Spitzer 4.5\,$\mathrm{\mu}$m image. For the deep fields, we introduce an additional fourth frame, which is the same as the third frame but without the radio contours to aid in visual inspection. On all four frames, the \textsc{PyBDSF} source in question is marked with a solid red ellipse and a red cross, while other \textsc{PyBDSF} sources are marked with a dashed red ellipse. The instructions given to the user for the task remain the same as LoTSS DR1. Using these four images, the user must first select any additional radio source components (i.e. dashed red ellipses) that are associated with the radio source in question. Then, the user must select all plausible multi-wavelength identifications (if any). Finally, the user must answer the following questions: Is this an artefact? Is this a radio source blend? Is the image too zoomed in? Are any of the images missing?

Each radio source sent to LGZ was classified by at least five astronomers and the output from LGZ was converted into a set of quality flags for the association and identification steps; the consensus from these classifications and flags was used to form the source associations and identifications. The details of the flags used to decide the associations are as described in \citetalias{2019A&A...622A...2W}. The questions in the final step of LGZ were asked to enable the selection of sources for which source association and/or identification could not be fully carried out and therefore may require further inspection. Sources flagged as artefacts by a majority (more than 50\%) of the users were removed from the \textsc{PyBDSF} catalogue. Sources flagged as `image too zoomed in' or, as `blends' by more than 40\% of users were associated separately by a single expert in the expert-user workflow (see Section~\ref{sec:deblend}).

If the LGZ consensus was for source association, a new source was generated by combining its constituent \textsc{PyBDSF} sources and the constituent \textsc{PyBDSF} sources were then removed from the final catalogue. We generate other radio source properties, similar to the ones in the \textsc{PyBDSF} catalogue (e.g. total flux, size, position, etc.) for this new source. We refer the reader to \citetalias{2019A&A...622A...2W} for the details of this process of source association. We also note here that the LGZ association and identification takes precedence over LR identification. For example, consider a radio-AGN split by \textsc{PyBDSF} into three sources, one \textsc{PyBDSF} source consisting of only the compact core and a \textsc{PyBDSF} source for each of the two lobes. In such a case, it is likely that the LR method would have identified the genuine host galaxy belonging to the compact core, but the extended lobes would have been sent to LGZ, where the three components would be associated together and the host galaxy identified for the new source; this over-rides the LR identification.

\subsubsection{Expert-user workflow}\label{sec:deblend}
While testing sources that went to visual classification from initial versions of the decision tree, it was immediately apparent that there was a significant increase in the occurrence of blends of radio sources compared to LoTSS DR1, due to the deeper radio data. It would be very inefficient to simply send such sources to LGZ. We therefore first attempt to select sources (see Fig.~\ref{fig:en1_flowchart}) that could potentially be `blends' and send them directly to the expert-user workflow, which has de-blending functionality. Other potential blends were sent to this workflow as an output from either the pre-filter workflow (see Sect.~\ref{sec:prefilt}) or from LGZ.

In the expert-user workflow, non-static LGZ style images were provided to a single expert, but also with information from the \textsc{PyBDSF} Gaussian component catalogue displayed. The expert user has the ability to split each \textsc{PyBDSF} source into its constituent Gaussians, which can then be associated (if needed) to generate multiple new sources. Then, multi-wavelength identification (or lack thereof) can be performed for the newly generated sources. For these de-blended sources, the final catalogue contains other radio source properties as in the \textsc{PyBDSF} catalogue, in this case generated from the \textsc{PyBDSF} Gaussian catalogue (see \citetalias{2019A&A...622A...2W}). We note that not all sources sent as potential blends to the expert-user workflow were genuine blends; for such sources, no de-blending was performed but the host galaxy identification was still carried out as part of this workflow. For a small number of cases, there are more potential distinct physical sources of emission than fitted \textsc{PyBDSF} Gaussians. In such cases, we only de-blend the \textsc{PyBDSF} source to the number of Gaussians available, selecting the most appropriate host galaxies that contributed the majority of the flux to the available Gaussians.

In addition, the expert-user workflow also has a zoom in or out functionality, and so the sources flagged as `image too zoomed in' in LGZ (160, 96, and 60 sources in ELAIS-N1, Lockman Hole, and Bo\"{o}tes, respectively) were re-classified by a single expert with re-generated images using the initial LGZ classification as a starting point. The expert-user workflow was also used to identify radio source host galaxies that were missing from the multi-wavelength catalogues, and in addition, used to perform a final inspection of some large-offset LR-IDs, and all radio sources without an identification (hereafter; no-IDs), as detailed in Sect.~\ref{sec:uncat_hosts}~--~\ref{sec:noid}. The expert-user workflow is adapted from the `too zoomed in' and `deblend' workflows developed for LoTSS DR1 and we refer the reader to \citetalias{2019A&A...622A...2W} for a full description of this workflow.

\begin{table*}
\centering
	\caption{Output of pre-filter workflow. Percentages are calculated based on the number of \textsc{PyBDSF} catalogue sources in the multi-wavelength overlap area (listed in Table~\ref{tab:prop_final-opt_pybdsf}). Sources flagged as `Blend', `Too zoomed in' or `Uncatalogued host' are sent to the expert-user workflow for classification.}
	\label{tab:pref_out}
\begin{tabular}{lllllll}
\hline\hline
\multicolumn{1}{l}{Outcomes} & \multicolumn{2}{c}{ELAIS-N1} & \multicolumn{2}{c}{Lockman Hole} & \multicolumn{2}{c}{Bo\"{o}tes} \Tstrut \\
{} & Number & Fraction & Number & Fraction & Number & Fraction \Bstrut \\
\hline
LGZ & 346 & 1.11\% & 121 & 0.41\% & 94 & 0.5\% \\
Accept LR match & 410 & 1.32\% & 110 & 0.37\% & 43 & 0.23\% \\
No plausible match & 739 & 2.38\% & 555 & 1.86\% & 320 & 1.71\% \\
Too zoomed in & 23 & 0.07\% & 6 & 0.02\% & 8 & 0.04\% \\
Artefact & 72 & 0.23\% & 15 & 0.05\% & 7 & 0.04\% \\
Uncatalogued host\tablefootmark{a} & 97 & 0.31\% & 77 & 0.26\% & 102 & 0.54\% \\
Blend\tablefootmark{b} & 77 & 0.25\% & 18 & 0.06\% & 4 & 0.02\% \\
\hline
Total & 1764 & 5.68\% & 902 & 3.03\% & 578 & 3.08\% \\
\hline
\end{tabular}
\tablefoot{
\tablefoottext{a}{Uncatalogued Host: sources where the host galaxy was not detected in the multi-wavelength catalogue. These were later manually added using the expert-user workflow and forced photometry (see Sect.~\ref{sec:uncat_hosts}).}\\
\tablefoottext{b}{Slightly different \textsc{PyBDSF} parameters adopted (accidentally) for ELAIS-N1 compared to Lockman Hole and Bo\"{o}tes, result in more sources being initially separated into different \textsc{PyBDSF} components in Lockman Hole and Bo\"{o}tes, and hence fewer pre-filter `Blends' (but a higher proportion of sources needing the expert-user workflow; see Table~\ref{tab:final_radio_stats}).}
}
\end{table*}

\subsubsection{Pre-filter workflow}\label{sec:prefilt}
In some cases, the radio source or the LR identification properties alone were not sufficient to decide if a source should be sent to LR, LGZ, or the expert user workflow for identification. Rather than send all such sources to LGZ, which is by far the most time consuming process as it requires classification of each source by five volunteers, we instead perform quick visual sorting (pre-filtering) of some stages of the decision tree prior to deciding the most appropriate workflow for counterpart identification. The aim of this pre-filtering step was to quickly assess whether: (i) the best candidate ID selected by LR is unambiguously correct (regardless of whether the LR is above or below the LR\textsubscript{th}); (ii) the source needs to be sent to LGZ (this was the option used in case of any doubt, to enable a consensus decision to then be taken); (iii) the source is correctly associated but has no plausible multi-wavelength counterpart; (iv) the source is a blend, to be sent to the expert-user workflow; (v) the source is an artefact; (vi) the host galaxy detection is missing in the multi-wavelength catalogue (these sources are also sent to the expert-user workflow); or (vii) the image is too zoomed in (also sent to the expert-user workflow). In practice, for sources sent to the pre-filter workflow, static optical and 4.5\,$\mathrm{\mu}$m images showing the radio contours and the current best LR match and LR value (if any) were generated and categorised by a single expert for all three fields using a \textsc{Python} based interface. The categorised sources were then sent to the appropriate workflows, as shown in Table~\ref{tab:pref_out}. In cases where the host galaxy is missing from the multi-wavelength catalogue, we manually added these to the multi-wavelength catalogue using the process described in Sect.~\ref{sec:uncat_hosts}.

\subsubsection{Missing host galaxies in multi-wavelength catalogue}\label{sec:uncat_hosts}
During the visual classification steps, we noticed that the host galaxies of a small but non-negligible fraction of radio sources were present in our optical or IR mosaics but missing from our multi-wavelength catalogues (hereafter, `uncatalogued hosts'; see Table~\ref{tab:pref_out}). There were a few key reasons for this lack of detections; for example, the host galaxy being too close to bright stars where detections were typically missing (especially sources within the optical mask region), and missed `Spitzer-only' sources that were blended in the lower resolution Spitzer data. We therefore attempt to select the missing host galaxies (uncatalogued hosts) and manually add them to our multi-wavelength catalogues in each field as follows.

These sources with uncatalogued hosts were selected from each of pre-filter, expert-user, and LGZ workflows. In the pre-filter workflow, this was one of the options available (see Sect.~\ref{sec:prefilt}). For LGZ, one of the outputs is the \textsc{Badclick} flag which indicates the number of volunteers who have clicked on an host galaxy position that is not in the multi-wavelength catalogue. Through visual inspection, we found that radio sources with \textsc{Badclick} $>$ 2 typically correspond to a host galaxy which was missing in the multi-wavelength catalogue but which was sufficiently visible in the LGZ images to be identified by the volunteers. These radio sources with uncatalogued hosts were then sent to the expert-user workflow, where a single expert performed the identification and generated the coordinates of the uncatalogued hosts. Similarly, for sources that were directly sent to the expert-user workflow (e.g. as potential blends), the clicked position of the host galaxy (if uncatalogued) was also generated at the same time.

In processing host galaxy click positions from the expert-user workflow, we define uncatalogued hosts as those where the separation between the host galaxy click position and the multi-wavelength catalogue is more than 1$\arcsec$. These uncatalogued hosts are then added by either searching in the full Spitzer-detected catalogue (which picks up Spitzer-only sources that were not added to the merged catalogue) or, if they are not found there, by performing forced photometry (in all filters) at the positions of the uncatalogued hosts.

\begin{table*}
    \centering
    \caption{Description of the ``NoID'' flag values. Flag $= 0$ indicates that an identification is present, and the higher flag values indicate the reason for the lack of an identification.\label{tab:noid_def}}
    \begin{tabular}{lllll}
    \hline\hline
    \multicolumn{1}{l}{Flag} & \multicolumn{1}{l}{Description} & \multicolumn{3}{c}{Field}\\
    {} & {} & ELAIS-N1 & Lockman Hole & Bo\"{o}tes \\
    \hline
0 & Source has an ID & 30839 & 30402 & 18579 \\
1 & Radio source position accurate & 407 & 392 & 217 \\
2 & Radio position accurate; possible faint ID but below catalogue limit & 164 & 158 & 213 \\
3 & Extended (radio position may be inaccurate); no plausible ID & 100 & 103 & 54 \\
4 & Radio position lies under another un-associated object & 34 & 12 & 33 \\
5 & Extended source, one or more potential IDs, but none unambiguous & 66 & 95 & 83 \\
	\hline
    \end{tabular}
\end{table*}

\begin{table*}
\centering
	\caption{The number of radio sources in the source-associated radio-optical cross-matched catalogue and the number and fraction of sources that have an identification (or lack thereof), split by the identification method. ID fractions are calculated based on the total number of radio sources (listed at the bottom of the table) in the source-associated and cross-matched radio catalogue.\label{tab:final_radio_stats}}
\begin{tabular}{lllllll}
\hline\hline
\multicolumn{1}{l}{} & \multicolumn{2}{c}{ELAIS-N1} & \multicolumn{2}{c}{Lockman Hole} & \multicolumn{2}{c}{Bo\"{o}tes} \\
{} & Number & Fraction & Number & Fraction & Number & Fraction \\ 
\hline
LR & 26701 & 84.5\% & 24851 & 79.7\% & 16151 & 84.2\% \\
LGZ & 1966 & 6.2\% & 2395 & 7.7\% & 1058 & 5.5\% \\
Expert-user & 2172 & 6.9\% & 3156 & 10.1\% & 1370 & 7.1\% \\
Total-ID & 30839 & 97.6\% & 30402 & 97.6\% & 18579 & 96.9\% \\
No-ID & 771 & 2.4\% & 760 & 2.4\% & 600 & 3.1\% \\
\hline
\multicolumn{1}{l}{Total} & \multicolumn{2}{c}{31610} & \multicolumn{2}{c}{31162} & \multicolumn{2}{c}{19179} \\
\hline
\end{tabular}
\end{table*}

\subsubsection{Cleaning and inspection of large-offset LR matches}\label{sec:final_cleaning}
A small number of sources (140, 101, and 27 in ELAIS-N1, Lockman Hole, and Bo\"{o}tes, respectively) had a counterpart identified by the LR method that was significantly offset ($>$ 3$\arcsec$) from its radio source. Such a large offset is surprising, casting doubt on whether the LR-ID is accurate; we therefore visually inspected all of these sources via the expert-user workflow to either confirm that the multi-wavelength ID found by LR method is the genuine host, or to assign the correct host galaxy (where possible). Roughly half of these sources were confirmed to be reliable; the other half were typically associated with extended sources which should not have been selected for statistical cross-matching. For these, the correct counterpart (or lack thereof) was assigned by the expert user.

\subsubsection{Investigation of sources without an identification}\label{sec:noid}
The outputs from all of the various identification methods were joined to generate a cross-matched radio catalogue, with the correct source associations. Sources without an identification were then visually inspected in an expert-user workflow to confirm that the lack of an ID was correct and to indicate the reason for the lack of identification. For a small fraction of sources, this was found to be in error, typically due to the source being either an artefact, a blend, or a potential host galaxy missing from the catalogue, but which had not satisfied the criteria in LGZ output for selection. Such sources were then sent to the expert-user workflow (except artefacts, which were removed) to resolve the association and identification. 

For sources genuinely without an ID, a flag (\textsc{``NoID''}) was assigned to indicate the reason for the lack of an identification. The flag values and their definitions are listed in Table~\ref{tab:noid_def} along with the numbers in each category per field. In addition to studying the nature of these sources, an advantage of assigning the NoID flag is that with upcoming spectroscopic surveys (e.g. WEAVE-LOFAR; \citealt{2016sf2a.conf..271S}), a fibre could well be positioned at the position of those radio sources with secure positions to obtain spectra of (and of any emission lines from) the host galaxies where existing optical to MIR imaging data is too faint. 

A large fraction (typically $>$~70\%) of the radio sources without an identification were un-resolved (or barely resolved) sources with a secure radio source position. The host galaxy, however, was below the survey depths of our multi-wavelength dataset (albeit in some cases, low significance emission may be present). The second biggest fraction consisted of extended radio sources, with large positional uncertainties and poorly defined positions; some of these had no plausible ID whereas others had one or more plausible IDs, but none reliable enough to be chosen.

\section{Final cross-matched catalogues}\label{sec:final_radio_cata}
The final cross-matched and associated catalogue in ELAIS-N1 contains 31610 radio sources, with host galaxies identified for 97.6\% of these. Similarly, there are 31162 sources in Lockman Hole with host galaxies found for 97.6\%, and 19179 sources in Bo\"{o}tes with host galaxies identified for 96.9\%. These properties, along with the number of sources (and the fraction) identified by each method, are listed in Table~\ref{tab:final_radio_stats}.

Compared to similar cross-matching efforts in the literature, for example in the ELAIS-N1 field by \cite{ocran2019gmrten1_ids} using 610~MHz GMRT observations, we find a higher cross-identification rate by $>$~5\%. We also note the larger (by $\sim$3\%) fraction of sources requiring visual classification (expert-user and LGZ) in Lockman Hole, and a similar decrease in the fraction of sources where the LR identification was accepted compared to the other two fields. This is likely due to the slight difference in the \textsc{PyBDSF} source extraction parameters used for Lockman Hole (and for Bo\"{o}tes), where a significantly larger fraction of the sources were fitted with multiple Gaussian components, resulting in ambiguity in the decision tree and requiring source association or de-blending. The effect of the difference in the \textsc{PyBDSF} source extraction parameters is less prominent in Bo\"{o}tes, likely due to the shallower radio data depth. It is important to note that these differences should not affect the final source-associated, cross-matched catalogues, but simply result in a difference in the method of the identification: visual classifications were more often used to form the correct source associations for sources split into multiple Gaussians.

The LOFAR Deep Fields value-added catalogue released contains properties of the correctly associated radio sources, and the multi-wavelength counterpart identifications and properties (where available). 

The radio source properties are as follows:
\begin{itemize}
    \item The IAU source identification (``Source\_Name'') based on source position.
    \item Radio source position and uncertainties (``RA'', ``E\_RA'', ``Dec'', and ``E\_Dec'').
    \item Radio source peak and total flux densities and corresponding uncertainties (``Peak\_flux'', ``E\_Peak\_flux'', ``Total\_flux'', ``E\_Total\_flux'').
    \item Ellipse shape parameters and corresponding uncertainties (``Maj'', ``Min'', ``PA'', ``E\_Maj'', ``E\_Min'', ``E\_PA''). These are blank for associated sources; see below for properties of associated sources. For de-blended sources, these are taken from the \textsc{PyBDSF} Gaussian catalogue.
    \item A code to define the source structure (``S\_Code''; `S' = single-Gaussian, `M' = multi-Gaussian, `Z' = associated/compound source)
    \item Overlap bit flag indicating the coverage of the multi-wavelength surveys at the radio source position (``\textsc{flag\_overlap\_radio}''). See Table~\ref{tab:prop_final-opt_pybdsf} for the recommended flag values.
    \item Bright star masking flag indicating masked and un-masked regions in the Spitzer- and optical-based bright star mask (``\textsc{flag\_clean\_radio}''), based on radio position.
\end{itemize}

Associated sources have additional radio source properties given by:
\begin{itemize}
    \item Ellipse shape parameters for associated sources (``LGZ\_Size'', ``LGZ\_Width'', ``LGZ\_PA'')
    \item Gaussian de-convolved shape parameters (``DC\_Maj'', ``DC\_Min'', ``DC\_PA'')
    \item Number of \textsc{PyBDSF} source components associated (``Assoc'')
    \item Quality flag of the association (``Assoc\_Qual'')
\end{itemize}

The multi-wavelength identification (if any) and host galaxy properties are as follows:
\begin{itemize}
    \item Unique identifier of the ID to the multi-wavelength catalogue (``ID'')
    \item multi-wavelength ID source position (``ALPHA\_J2000'', ``DELTA\_J2000'')
    \item Aperture and extinction corrected fluxes (and flux errors) from our recommended aperture size <band>\_flux\_corr and <band>\_fluxerr\_corr in $\mu$Jy.
    \item Aperture and extinction corrected magnitude (and magnitude errors) from our recommended aperture size <band>\_mag\_corr and <band>\_magerr\_corr in the AB system.
    \item Overlap bit flag indicating the coverage of the multi-wavelength surveys at the counterpart source position (``\textsc{flag\_overlap}''). See Table~\ref{tab:prop_final-opt_pybdsf} for the recommended flag values.
    \item Bright star masking flag indicating masked and un-masked regions in the Spitzer- ($= 3$) and optical- ($= 1$) based bright star mask (``\textsc{flag\_clean}'')
    \item The maximum LR match (if an ID is present, and if the ID is obtained from the LR method; ``lr\_fin'')
    \item multi-wavelength ID position based $\mathrm{E(B-V)}$ reddening values from~\cite{1998Schlegeldustmap} dust map (``EBV'').
    \item Flag indicating reason for lack of identification (``NoID''; `0' = an identification exists). Flag definitions are listed in Table~\ref{tab:noid_def}.
\end{itemize}

Additional columns pertaining to the photometric redshifts, rest-frame colours, absolute magnitudes, and stellar masses are described in \citetalias{duncan2020_inpress}, and columns relating to the far-infrared data are described in \cite{mccheyne2020_inpress}. For full details of all columns presented, please see the accompanying data release documentation.

\begin{figure}
    \centering
    \includegraphics[width=\columnwidth]{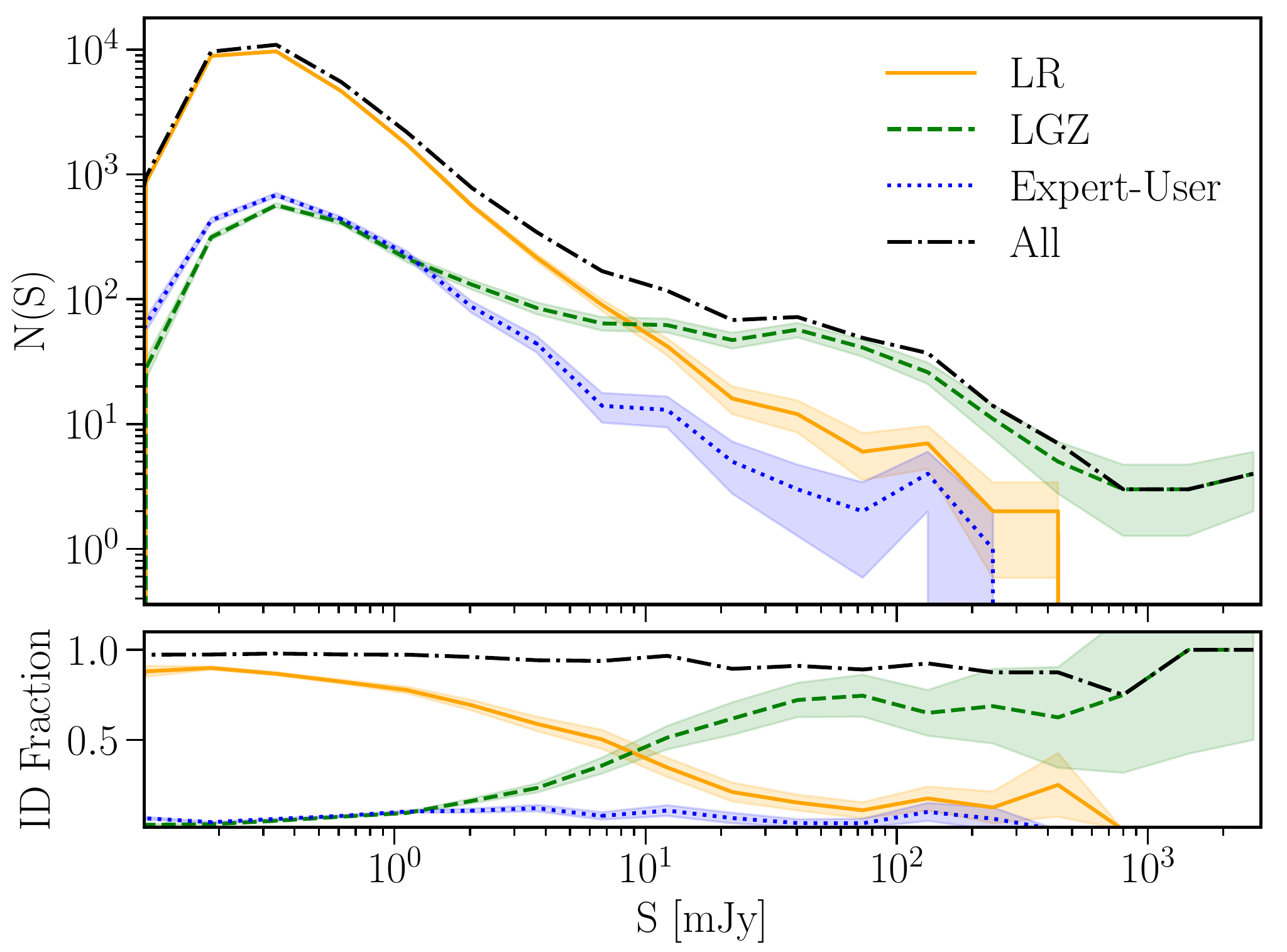}
    \caption{\textit{Top:} The number of radio sources with identifications as a function of radio flux density and the identification method used (LR, LGZ or `expert user workflow'). The flux density distribution of all sources with identification is shown by the dot-dashed black line. \textit{Bottom:} The identification fraction as a function of the flux density, also split by the identification method. The identification fraction is computed based on the total number of radio sources (with or without an identification). The LGZ method dominates the identification rate above $\sim$10mJy, with the LR method dominating below this. The filled regions show Poisson error estimates.}
    \label{fig:NS_id_fraction}
\end{figure}

Fig.~\ref{fig:NS_id_fraction} shows the number of sources (top panel) and the identification fraction (bottom panel) as a function of the radio flux density and the identification method. The fraction of sources requiring LGZ for identification decreases from 100\% at the brightest fluxes down to well below 5\% at the faintest fluxes. There is a transition of the dominant method of identification at $\sim$10mJy from LGZ to the LR method, which accounts for $\gtrsim$90\% of the sources at the faintest fluxes. This higher identification rate by the statistical method showcases the power of the ancillary data available in these deep fields compared to the shallower LoTSS DR1. The expert-user method plays a sub-dominant role across most of the flux density range but, as a result of the depth of the radio data and consequently increasing number of blends, begins to dominate the identification rates achieved from the visual methods at the faintest fluxes, and therefore corresponds to a significant number of sources within our sample (as shown in Table~\ref{tab:final_radio_stats}). The trend of decreasing overall identification rate (see Fig.~\ref{fig:NS_id_fraction}) with decreasing radio flux densities noted by \citetalias{2019A&A...622A...2W} (down to $\sim$ 1mJy in \citetalias{2019A&A...622A...2W}) in LoTSS-DR1 (see Fig.~8 of \citetalias{2019A&A...622A...2W}) is not observed in these deep fields. This is expected as this decrease in identification rate was attributed to the shallow PanSTARSS and WISE data available for cross-matching by \citetalias{2019A&A...622A...2W}. In the LoTSS Deep Fields, although the typical redshift of sources probed increases with decreasing radio flux density, the significantly deeper multi-wavelength data available allows us to effectively identify counterparts down to lower radio flux densities than LoTSS-DR1, where the LR method starts to dominate the identification rates.

\begin{figure*}[!htbp]
    \centering
    \includegraphics[width=\textwidth]{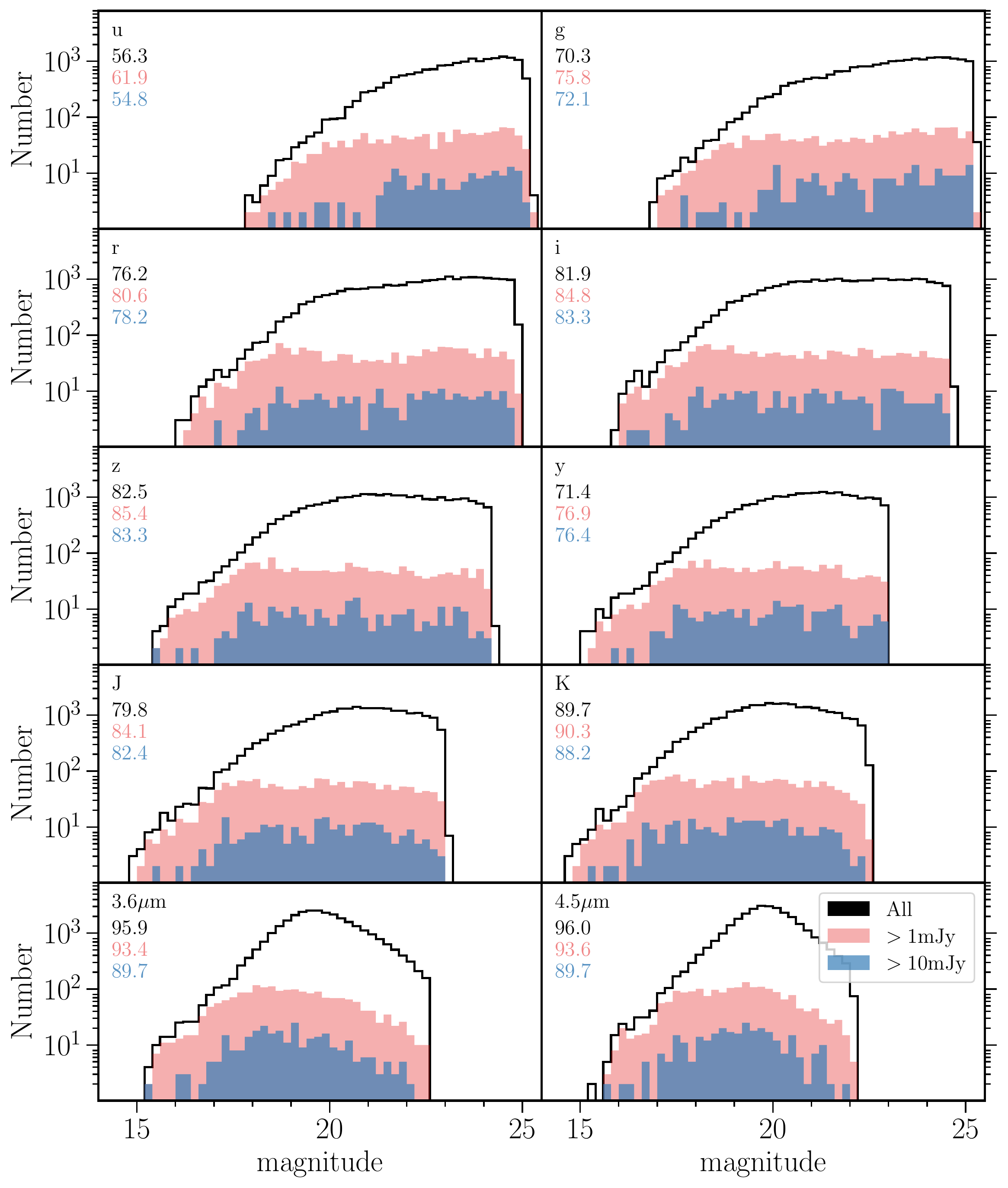}
    \caption{\label{fig:mag_dist_matches}Magnitude distributions of the host galaxies of radio sources in optical (SpARCS and PanSTARRS) to IR bands in ELAIS-N1. Each panel also shows the same for the subset of radio sources with radio flux densities $>$1mJy (pink) and $>$10mJy (blue). The numbers in each panel corresponds to the fraction of all radio sources with a counterpart detected in that band, for all sources, and for $>$1mJy and $>$10mJy sources. The identification fraction increases with wavelength from $\sim$56\% in u-band up to 96\% in 4.5\,$\mu$m.} 
\end{figure*}

\section{Properties of host galaxies}\label{sec:science}
In Fig.~\ref{fig:mag_dist_matches}, we show the magnitude distribution of all counterparts identified in ELAIS-N1 for a range of optical (SpARCS and PS1) to IR bands (UKIDSS and SWIRE). Also shown in shaded regions are the distributions for the subset of radio sources with 150~MHz radio flux densities $>$\,1mJy (pink) and $>10$\,mJy (blue). The top value listed in each panel is the percentage of all radio sources in ELAIS-N1 that have a counterpart detected within that given band. The second and third values provide the corresponding percentages for the number of radio sources with radio flux densities $>$\,1mJy and $>$\,10mJy, respectively, that have a counterpart detected in that band.

For the 3.6 and 4.5\,$\mu$m channels, the magnitude distribution of the counterparts is clearly peaked at a magnitude of 19--20, and declines towards fainter magnitudes. This is well within the detection limit of the Spitzer survey, illustrating that the vast majority of radio counterparts are detected by SWIRE and SERVS, as indicated by the high (96\%) identification rates in these two channels. In the NIR filters, the distributions show a broad peak around 20th--21st magnitude, turning over close to the magnitude limit of the UKIDSS survey. In contrast the distributions in the bluer (optical) filters show no signs of turning down at the faintest magnitudes probed, consistent with the lower identification rates achieved in these filters being limited by the depth probed by the available optical surveys. This trend in the shape of the magnitude distributions can also be seen by the increase in identification rate achieved with wavelength, increasing from $\sim$~56\% to 96\% from the u-band to the 4.5\,$\mu$m band. Even in the bluer filters, however, the faint end of the distributions flatten, as compared to the well-known monotonic increase of the number counts of all galaxies towards fainter magnitudes, indicating that radio galaxies are preferentially hosted in brighter galaxies (a result which motivates the LR approach). 

Interestingly, the distributions for the $>$1mJy sources peak at brighter magnitudes than those of `All' sources, suggesting that higher flux density radio sources even more strongly favour brighter host galaxies. Comparing the difference between the distributions of `All' sources and $>$\,1mJy sources, it is clear that at faint optical and IR magnitudes, the radio sources with flux densities below 1\,mJy dominate the population. In contrast, at bright optical and IR magnitudes, the majority of the radio population has flux densities above 1\,mJy. This trend may be driven by a shift in the dominant radio source population below $\sim$1\,mJy, where we expect a significant fraction of both nearby star-forming galaxies and high redshift (obscured) radio quiet quasars \citep{2008MNRAS.388.1335W}, which are likely hosted by fainter optical galaxies.

\begin{figure}
    \centering
    \includegraphics[width=\columnwidth]{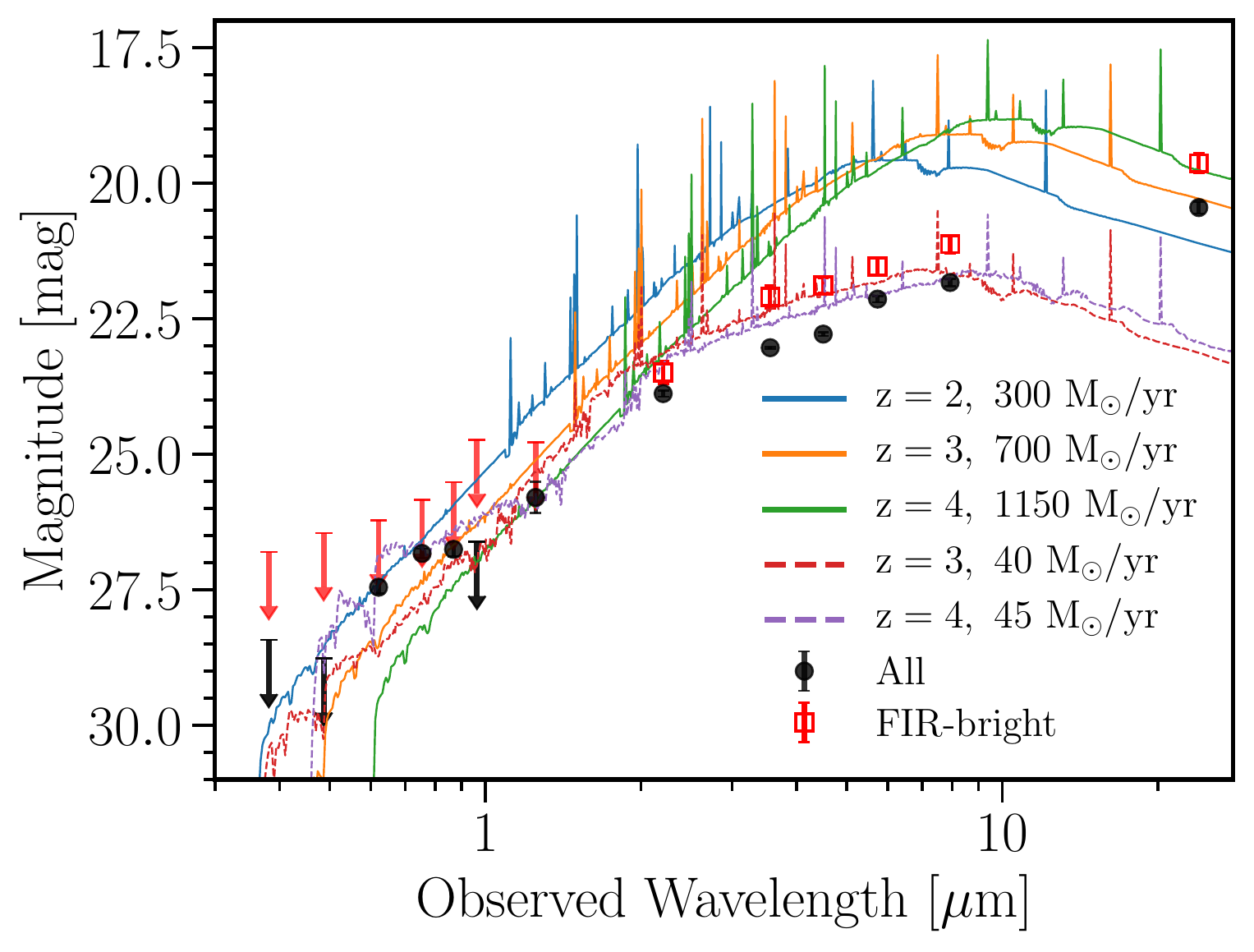}
    \caption{Optical to mid-IR stacked SED of the unidentified radio sources with secure radio positions in ELAIS-N1 (571 sources; black points). The stack for the subset of the `FIR-bright' sources (50 sources) is shown as red squares. Overlaid on top are typical star-forming galaxy templates, computed using models of \citet{bc2003ssp} at $z = 2, 3$ and 4 (blue, orange and green solid lines, respectively), scaled to the median SFR detectable for these sources at the depth of the LOFAR data in ELAIS-N1. The dashed lines show two-component templates of galaxies with an old stellar population undergoing a recent burst of star-formation (see text) at $z = 3$ and 4 (red and purple, respectively).} 
    \label{fig:noid_sed}
\end{figure}

\subsection{Radio sources without an identification}\label{sec:noid_sed}
We now investigate the potential nature of host galaxies of radio sources without an identification, focusing on ELAIS-N1, by stacking the optical to mid-IR images. To do this, we first select LOFAR sources without an identification that have a secure radio position (``No\_ID'' = 1 or 2; 571 sources). Most ($>$~90\%) of these radio sources have radio flux densities of $\mathrm{S_{150MHz} < 1mJy}$; at these flux densities, as discussed above, the radio population is a mix of different source types. A median stack based on the radio positions is then performed, with photometry extracted in the optical-NIR (3$\arcsec$ aperture) and Spitzer (4$\arcsec$ aperture) filters, corrected to total magnitudes to match the catalogues. For the 24~$\mu$m MIPS band, we extract photometry from the 10$\arcsec$ aperture and perform aperture corrections based values computed by \cite{2007PASP..119..994E}.

The resulting stacked SED for all the unidentified sources (filled black circles) is shown in Fig.~\ref{fig:noid_sed}. Also shown in blue, orange and green solid lines are templates of galaxies dominated by a recent burst of star-formation occurring 100\,$\rm{Myrs}$ ago, at $z = 2, 3$ and $4$, respectively, computed using the stellar population models of \citet{bc2003ssp}, the \citet{imf2003chabrier} initial mass function (IMF) and the \citet{calzetti2000attentuation} attenuation curve. The star-formation rate (SFR) normalisation of these templates is fixed to the SFR required to produce the median radio flux density observed of these sources, assuming all of the radio emission is associated with star-formation. This is calculated using the relation between the 150~MHz luminosity ($L_{\mathrm{150}}$) and SFR derived by \cite{gurkan2018lofar_sfr}, assuming a spectral index $\alpha = 0.7$, and requires a SFR of 300\,$\rm{M_{\odot}yr^{-1}}$ at $z = 2$, 700\,$\rm{M_{\odot}yr^{-1}}$ at $z = 3$ and 1150\,$\rm{M_{\odot}yr^{-1}}$ at $z = 4$. As evident from the stacked SED, we do not observe enough emission if the radio emission is entirely due to star formation, particularly at near- and mid-infrared wavelengths, to match the star-forming galaxy templates and the difference in flux by over an order of magnitude cannot simply be explained by extinction (especially in the IRAC bands).

To further investigate the potential role of the star-forming galaxies within this radio source population, we examine the FIR emission from these sources. 50 of these 571 No-ID sources ($\sim$7\%) have significant FIR emission (hereafter `FIR-bright' sources); these are defined as sources with a 250$\mu$m flux density (F$_{\rm{250{\mu}m}}$) $> 15$~mJy. When stacked separately (red squares), these FIR-bright sources are found to be around 1-1.5 mag brighter in the mid-IR bands, however still nearly an order of magnitude fainter in flux than the star-forming templates at $z = 2-4$. Thus, even though the FIR measurements suggest that some star-formation may be on-going in these sources, we conclude that the radio emission in the majority of these sources is not dominated by star-formation.

Instead, to provide an illustration of the SED that a typical radiative-mode (or high-excitation radio galaxy; HERG) AGN might have, we consider a two component model, including both an old stellar population and a period of recent star-formation (since high-redshift radiative-mode AGN are found to lie close to the star-forming main sequence; e.g. \citealt{mainieri2011agnsfr,bonzini2015sfragnms,suh2019agnsfr}). Specifically, the red and purple dashed lines on Fig.~\ref{fig:noid_sed} show a $10^{10}~\rm{M_{\odot}}$ old stellar population (formation at $z = 12$ with an exponentially declining star-formation rate with a characteristic time of 150 Myr) with a burst of star-formation within the past 50 Myr with a SFR of 40 and 45 $\rm{M_{\odot}\,yr^{-1}}$ (consistent with being on the star-forming main sequence; \citealt{speagle2014sfrms}) at $z = 3$ and 4, respectively. These illustrative SEDs broadly trace the stacked data points, suggesting that a significant fraction of the unidentified radio source population is likely dominated by high-redshift obscured/radiative-mode AGN, which are expected to contribute significantly to source counts at S$_{\rm{150MHz}} < $ 1\,mJy \citep{2008MNRAS.388.1335W}. We also observe a significant detection at 24~$\mu$m for the `All' stack, indicative of hot dust emission; this emission could potentially arise from the hot torus surrounding an AGN (i.e. a HERG-like AGN) that is expected to peak at rest-frame wavelength of $\sim$ 10~$\mu$m \citep[e.g.][]{2004MNRAS.355..973S}. The median 150~MHz luminosity for the `All' sources, assuming a spectra index $\alpha = 0.7$, is $\rm{log(L_{150})} = 25.0~\rm{W~H^{-1}}$ if at $z = 3$, and $\rm{log(L_{150})} = 25.3~\rm{W~H^{-1}}$ if at $z = 4$. These are significantly fainter than the radio galaxies known at these redshifts (e.g. \citealt{jarvis2009hzrg,saxena2018hzrg}).

Considering the stacked magnitudes, it is interesting to note that the inclusion of the HSC-SSP DR2 optical data, with target depths of 27.5, 27.1, and 26.8 mag in the g, r, and i-bands, respectively, should allow us to identify a large fraction of the currently optically faint counterparts in ELAIS-N1. The nature of this optically faint radio source population will be investigated quantitatively in future work.

\section{Conclusions}\label{sec:conclusions}
In this paper, we have presented the value-added catalogue of multi-wavelength counterparts to the radio sources detected in the first LoTSS Deep Fields Data Release by \citet{tasse2020_inpress} and \citet{sabater2020_inpress}, covering the ELAIS-N1, Lockman Hole, and Bo\"{o}tes fields. The value-added radio-optical cross-matched catalogues presented contain 81\,951 radio sources, with counterparts identified and matched aperture optical to infrared properties presented for 79\,820 sources ($>$~97\%), covering $\sim$~26 \mbox{deg$^{2}$} in total, across multiple sight-lines.

To achieve this, we first built new multi-wavelength catalogues in both ELAIS-N1 and Lockman Hole, consisting of forced, matched aperture photometry across 20 and 16 bands, respectively, from UV to mid-IR. These catalogues were built using deep $\chi^{2}$ detection images, using information from the optical and IR bands, to maximise the catalogue completeness and generate clean, robust photometry and colours; this provides a significant improvement to catalogues existing in literature for photometric redshifts and SED fitting. In this paper, we also present and release these multi-wavelength catalogues and accompanying optical to IR mosaics.

The counterparts to the radio sources are identified using a combination of the statistical Likelihood Ratio method and visual classification schemes. We use the LR method that incorporates both magnitude and colour information, as described in~\cite{nisbet2018role} and~\cite{2019A&A...622A...2W}, to maximise the identification rate and increase the robustness of the cross-matching. The deep ancillary data available in these fields allows us to achieve an identification rate of up to 97\% using the LR method alone. The LR method however is not suitable for large or complex radio sources; such sources require visual classification instead, which is mainly performed using the LOFAR Galaxy Zoo framework developed for LoTSS-DR1 \citep{2019A&A...622A...2W}. To determine sources that can be identified using the LR method and those that require visual classification, we adapted and further developed the decision tree used in LoTSS-DR1. The high LR identification rates allowed us to require any source without a LR identification to undergo visual inspection.

The cross-matching effort leads to multi-wavelength identifications for 97.6\%, 97.6\%, and 96.9\% of sources in ELAIS-N1, Lockman Hole, and Bo\"{o}tes, respectively. The colour properties of host galaxies show that the reddest of galaxies are more than an order of magnitude more likely to host a LOFAR source than the bluest of galaxies. This is also visualised by the magnitude distributions of the host galaxies in different filters, which show that we are able to identify most of the LOFAR sources in the mid-IR. In contrast, deeper optical data are required to achieve higher identification rates and probe the optically faint counterparts to beyond the peak of host-galaxy magnitude distributions in these optical filters.

The scientific potential of the catalogues presented in this paper is further increased by the availability of photometric redshifts. Rest-frame colours and photometric redshifts for both the multi-wavelength catalogues and the radio-optical cross-matched catalogues in the three fields are presented in \citetalias{duncan2020_inpress}. This enables one to determine physical properties of host galaxies (such as luminosities, stellar masses, star-formation rates, etc.), which are used to perform source classification as presented in \citetalias{best2020_inpress}.

We performed a stacking analysis of the radio sources without an identification (but with secure radio positions) and compared the resultant average SED to typical star-forming and passive AGN templates. This revealed that the unidentified radio source population is likely dominated by a significant fraction of obscured AGN at moderate to high redshift ($z > 3$). For future LoTSS Deep Fields data release, the inclusion of deeper optical data from the HSC-SSP DR2 in ELAIS-N1, reaching depths of 27.5, 27.1, and 26.8 mag in the g-, r-, and i-bands, respectively, should allow us to identify counterparts to a majority of the currently unidentified radio sources.

We are continuing to acquire and calibrate more LOFAR observations of the first three deep fields, ultimately aiming to achieve a target sensitivity of 10$~\mathrm{\mu Jy~beam^{-1}}$. The increase in source density offered by the deeper radio data will require more efficient and automated methods for de-blending, such as the XID+ software \citep{hurley2017xid} used for FIR Herschel data, with the capability of modelling the extended nature of radio sources. Furthermore, future data releases will also include deep radio imaging of the North Ecliptic Pole (NEP) field, reaching comparable sensitivity to the first three deep fields. This will be complemented by the ongoing and planned multi-wavelength observations from the next generation of telescopes such as Euclid and eROSITA and will achieve optical and IR depths capable of identifying the host galaxies currently undetected in the first three deep fields.

\section*{Acknowledgements}
This paper is based (in part) on data obtained with the International LOFAR Telescope (ILT) under project codes LC0\_015, LC2\_024, LC2\_038, LC3\_008, LC4\_008, LC4\_034 and LT10\_01. LOFAR \citep{2013A&A...556A...2V} is the Low Frequency Array designed and constructed by ASTRON. It has observing, data processing, and data storage facilities in several countries, which are owned by various parties (each with their own funding sources), and which are collectively operated by the ILT foundation under a joint scientific policy. The ILT resources have benefitted from the following recent major funding sources: CNRS-INSU, Observatoire de Paris and Université d'Orléans, France; BMBF, MIWF-NRW, MPG, Germany; Science Foundation Ireland (SFI), Department of Business, Enterprise and Innovation (DBEI), Ireland; NWO, The Netherlands; The Science and Technology Facilities Council, UK; Ministry of Science and Higher Education, Poland. We thank the anonymous referee for their useful comments and suggestions which have improved the content and presentation of the paper.

RK acknowledges support from the Science and Technology Facilities Council (STFC) through an STFC studentship via grant ST/R504737/1. PNB and JS are grateful for support from the UK STFC via grant ST/R000972/1. We thank Boris H\"{a}u{\ss}ler for providing a wrapper script for resampling and co-adding images (using \textsc{SWarp}) that was adapted and developed further.
MB acknowledges support from INAF under PRIN SKA/CTA FORECaST and from the Ministero degli Affari Esteri della Cooperazione Internazionale - Direzione Generale per la Promozione del Sistema Paese Progetto di Grande Rilevanza ZA18GR02. RB acknowledges support from the Glasstone Foundation. MB acknowledges support from the ERC-Stg DRANOEL, no 714245 and from INAF under PRIN SKA/CTA ‘FORECaST’. RKC acknowledges funding from the John Harvard Distinguished Science Fellowship. JHC and BM acknowledge support from the UK STFC under grants ST/R00109X/1, ST/R000794/1 and ST/T000295/1. KJD, WLW and HJAR  acknowledge support from the ERC Advanced Investigator programme NewClusters 321271. AG acknowledges support from the Polish National Science Centre (NCN) through the grant 2018/29/B/ST9/02298. MJH acknowledges support from STFC via grant ST/R000905/1. MJ acknowledges support from the National Science Centre, Poland under grant 2018/29/B/ST9/01793. MJJ acknowledges support from the UK Science and Technology Facilities Council [ST/N000919/1] and the Oxford Hintze Centre for Astrophysical Surveys which is funded through generous support from the Hintze Family Charitable Foundation. MKB and AW acknowledge support from the National Science Centre, Poland under grant no. 2017/26/E/ST9/00216. VHM thanks the University of Hertfordshire for a research studentship [ST/N504105/1]. IP acknowledges support from INAF under the SKA/CTA PRIN “FORECaST” and under the PRIN MAIN STREAM “SAuROS” projects. WLW also acknowledges support from the CAS-NWO programme for radio astronomy with project number 629.001.024, which is financed by the Netherlands Organisation for Scientific Research (NWO).

The Pan-STARRS1 Surveys (PS1) and the PS1 public science archive have been made possible through contributions by the Institute for Astronomy, the University of Hawaii, the Pan-STARRS Project Office, the Max-Planck Society and its participating institutes, the Max Planck Institute for Astronomy, Heidelberg and the Max Planck Institute for Extraterrestrial Physics, Garching, The Johns Hopkins University, Durham University, the University of Edinburgh, the Queen's University Belfast, the Harvard-Smithsonian Center for Astrophysics, the Las Cumbres Observatory Global Telescope Network Incorporated, the National Central University of Taiwan, the Space Telescope Science Institute, the National Aeronautics and Space Administration under Grant No. NNX08AR22G issued through the Planetary Science Division of the NASA Science Mission Directorate, the National Science Foundation Grant No. AST-1238877, the University of Maryland, Eotvos Lorand University (ELTE), the Los Alamos National Laboratory, and the Gordon and Betty Moore Foundation.

Part of this work was carried out on the Dutch national e-infrastructure with the support of the SURF Cooperative through grant e-infra 160022 \& 160152. The LOFAR software and dedicated reduction packages on https://github.com/apmechev/GRID\_LRT were deployed on the e-infrastructure by the LOFAR e-infragroup, consisting of J.\ B.\ R.\ Oonk (ASTRON \& Leiden Observatory), A.\ P.\ Mechev (Leiden Observatory) and T. Shimwell (ASTRON) with support from N.\ Danezi (SURFsara) and C.\ Schrijvers (SURFsara). This research has made use of the University of Hertfordshire high-performance computing facility (\url{http://uhhpc.herts.ac.uk/}) and the LOFAR-UK computing facility located at the University of Hertfordshire and supported by STFC [ST/P000096/1]. This research made use of {\sc Astropy}, a community-developed core Python package for astronomy \citep{astropy:2013, astropy:2018} hosted at \url{http://www.astropy.org/}, of {\sc Matplotlib} \citep{hunter2007matplotlib}, of {\sc APLpy}, an open-source astronomical plotting package for Python hosted at \url{http://aplpy.github.com/}, and of {\sc topcat} and {\sc stilts} \citep{taylor2005topcat,taylor2006stilts}.

This work has made use of data from the European Space Agency (ESA) mission {\it Gaia} (\url{https://www.cosmos.esa.int/gaia}), processed by the {\it Gaia} Data Processing and Analysis Consortium (DPAC, \url{https://www.cosmos.esa.int/web/gaia/dpac/consortium}). Funding for the DPAC has been provided by national institutions, in particular the institutions participating in the {\it Gaia} Multilateral Agreement.
This work is based on observations obtained with MegaPrime/MegaCam, a joint project of CFHT and CEA/DAPNIA, at the Canada-France-Hawaii Telescope (CFHT) which is operated by the National Research Council (NRC) of Canada, the Institut National des Sciences de l'Univers of the Centre National de la Recherche Scientifique (CNRS) of France, and the University of Hawaii. This research used the facilities of the Canadian Astronomy Data Centre operated by the National Research Council of Canada with the support of the Canadian Space Agency. RCSLenS data processing was made possible thanks to significant computing support from the NSERC Research Tools and Instruments grant program. This work is based in part on observations made with the Spitzer Space Telescope, which was operated by the Jet Propulsion Laboratory, California Institute of Technology under a contract with NASA.

Herschel is an ESA space observatory with science instruments provided by European-led Principal Investigator consortia and with important participation from NASA. This research has made use of data from HerMES project (\url{http://hermes.sussex.ac.uk/}). HerMES is a Herschel Key Programme utilising Guaranteed Time from the SPIRE instrument team, ESAC scientists and a mission scientist. The HerMES data was accessed through the Herschel Database in Marseille (HeDaM - \url{http://hedam.lam.fr}) operated by CeSAM and hosted by the Laboratoire d'Astrophysique de Marseille. This work is based in part on observations made with the Galaxy Evolution Explorer (GALEX). GALEX is a NASA Small Explorer, launched in 2003 April. We gratefully acknowledge NASA's support for construction, operation, and science analysis for the GALEX mission, developed in cooperation with the Centre National d'Etudes Spatiales of France and the Korean Ministry of Science and Technology.
\bibliographystyle{aa}
\bibliography{lofar_deepfields_ids}

\begin{appendix}
\section{Aperture to total magnitudes}\label{sec:appendix1:apcorr}
We describe here the corrections that need to be applied to go from raw aperture magnitudes (provided for 8 apertures) to total magnitudes, corrected for aperture and Galactic extinction effects. For all fields, we recommend using the $\mathrm{band\_mag\_corr}$ which are corrected for aperture and extinction using the 3$\arcsec$ for optical-NIR bands and the 4$\arcsec$ for the Spitzer IRAC bands. These corrections are performed by

\begin{equation}\label{eq:mag_corr}
\begin{split}
    \mathrm{band\_mag\_corr} =  \mathrm{MAG\_APER\_band\_ap} - 2.5\log F_{\mathrm{band,ap}} ~ \\ - ~\mathrm{EBV} \times A_{\mathrm{band}}/E(B-V)
\end{split}
\end{equation}
where $\mathrm{band}$ is the filter, ap is the aperture size, $F_{\mathrm{band,ap}}$ is the aperture correction factor for a given band and aperture. We apply the values listed in Table~\ref{tab:apcorr} derived using the method described in Sect.~\ref{sec:aper_corr} (the stellar based aperture corrections are also provided in Table~\ref{tab:apcorr}). EBV is the reddening value computed based on source position and the \cite{1998Schlegeldustmap} dust map, and, $A_{\mathrm{band}}/\mathrm{E(B-V)}$ are filter dependent extinction factors listed in Table~\ref{tab:en1_lh_description} and Table~\ref{tab:bootes_extcorr}. Eq.~\ref{eq:mag_corr} can also be used for any other band or aperture size combination to derive an aperture and Galactic extinction corrected magnitude.

\begin{table}[!htbp]
    \centering
    \caption{Filter dependent extinction correction factors per unit reddening, $\mathrm{A_{band}}/E(B-V)$ and 3$\sigma$ depths from 3\arcsec apertures for Bo\"{o}tes.\label{tab:bootes_extcorr}}
    \begin{tabular}{ccc}
    \hline
    Band & $\mathrm{A_{band}}/E(B-V)$ & $3 \sigma$ depth \\ {} & {} & [mag] \\
    \hline
    FUV & 28.637 & 26.3 \\
    NUV & 8.675 & 26.7 \\
    u & 4.828 & 25.9 \\
    Bw & 4.216 & 26.2 \\
    R & 2.376 & 25.2 \\
    I & 1.697 & 24.6 \\
    z & 1.423 & 23.4 \\
    z\_Subaru & 1.379 & 24.3 \\
    y & 1.185 & 23.4 \\
    J & 0.811 & 23.1 \\
    H & 0.515 & 22.5 \\
    K & 0.338 & 20.2 \\
    Ks & 0.348 & 21.8 \\
    3.6\,$\mathrm{\mu}$m & 0.184 & 23.3 \\
    4.5\,$\mathrm{\mu}$m & 0.139 & 23.1 \\
    5.8\,$\mathrm{\mu}$m & 0.105 & 21.6 \\
    8\,$\mathrm{\mu}$m & 0.074 & 21.6 \\
    \hline
    \end{tabular}
\end{table}

\begin{table*}
\centering
\caption{List of aperture corrections derived for each filter in ELAIS-N1 and Lockman Hole, using the method described in Sect.~\ref{sec:aper_corr}. The raw aperture fluxes released in the catalogues should be divided by the values listed to correct the fluxes for aperture effects. The galaxies-based aperture corrections are appropriate for moderately distant galaxies and are applied to both the value-added catalogues and multi-wavelength catalogues in our recommended aperture sizes. Stellar-based corrections (not applied to catalogues) are derived using stars in GAIA-DR2 with 18 $<$ Gmag $<$ 20.\label{tab:apcorr}}
\begin{tabular}{cccccccc|ccccccc}  
\hline\hline
\multicolumn{1}{c}{Survey-Filter} & \multicolumn{7}{c|}{Galaxies-based corrections} & \multicolumn{7}{c}{Stellar-based corrections}\\
\multicolumn{1}{c}{} & \multicolumn{7}{c|}{Fraction of flux in aperture (arcsec)} & \multicolumn{7}{c}{Fraction of flux in aperture (arcsec)} \\
\multicolumn{1}{c}{ } & 1 & 2 & 3 & 4 & 5 & 6 & 7 & 1 & 2 & 3 & 4 & 5 & 6 & 7 \Bstrut\\
\hline
\textbf{ELAIS-N1} & {} & {} & {} & {} & {} & {} & {} & {} & {} & {} & {} & {} & {} & {} \\
SpARCS-u & 0.39 & 0.76 & 0.9 & 0.95 & 0.97 & 0.98 & 0.99 & 0.41 & 0.79 & 0.91 & 0.96 & 0.98 & 0.99 & 1.0 \\
PS1-g & 0.25 & 0.6 & 0.78 & 0.87 & 0.92 & 0.95 & 0.97 & 0.27 & 0.62 & 0.79 & 0.88 & 0.93 & 0.96 & 0.98 \\
PS1-r & 0.25 & 0.6 & 0.77 & 0.85 & 0.91 & 0.94 & 0.97 & 0.3 & 0.64 & 0.8 & 0.88 & 0.92 & 0.95 & 0.97 \\
PS1-i & 0.23 & 0.55 & 0.74 & 0.83 & 0.89 & 0.93 & 0.96 & 0.31 & 0.65 & 0.8 & 0.87 & 0.92 & 0.95 & 0.97 \\
PS1-z & 0.23 & 0.54 & 0.73 & 0.84 & 0.9 & 0.94 & 0.96 & 0.33 & 0.66 & 0.81 & 0.88 & 0.92 & 0.95 & 0.97 \\
PS1-y & 0.2 & 0.49 & 0.68 & 0.79 & 0.86 & 0.91 & 0.95 & 0.29 & 0.59 & 0.75 & 0.84 & 0.89 & 0.93 & 0.95 \\
HSC-g & 0.55 & 0.83 & 0.91 & 0.95 & 0.97 & 0.98 & 0.99 & 0.56 & 0.84 & 0.92 & 0.95 & 0.97 & 0.98 & 0.99 \\
HSC-r & 0.46 & 0.8 & 0.9 & 0.94 & 0.96 & 0.97 & 0.98 & 0.49 & 0.82 & 0.91 & 0.95 & 0.97 & 0.98 & 0.99 \\
HSC-i & 0.61 & 0.87 & 0.92 & 0.95 & 0.97 & 0.98 & 0.99 & 0.64 & 0.9 & 0.94 & 0.96 & 0.97 & 0.98 & 0.99 \\
HSC-z & 0.37 & 0.71 & 0.85 & 0.91 & 0.95 & 0.97 & 0.98 & 0.47 & 0.8 & 0.91 & 0.95 & 0.97 & 0.98 & 0.99 \\
HSC-y & 0.49 & 0.78 & 0.87 & 0.92 & 0.95 & 0.97 & 0.98 & 0.63 & 0.86 & 0.92 & 0.94 & 0.96 & 0.97 & 0.98 \\
HSC-NB921 & 0.35 & 0.69 & 0.84 & 0.91 & 0.94 & 0.97 & 0.98 & 0.63 & 0.88 & 0.94 & 0.97 & 0.98 & 0.99 & 0.99 \\
J & 0.39 & 0.78 & 0.92 & 0.96 & 0.98 & 0.99 & 1.0 & 0.45 & 0.82 & 0.93 & 0.97 & 0.99 & 0.99 & 1.0 \\
K & 0.32 & 0.7 & 0.88 & 0.95 & 0.98 & 0.99 & 1.0 & 0.42 & 0.79 & 0.92 & 0.96 & 0.98 & 1.0 & 1.0 \\
IRAC-3.6\,$\mathrm{\mu}$m & 0.1 & 0.34 & 0.56 & 0.73 & 0.83 & 0.89 & 0.93 & 0.13 & 0.4 & 0.64 & 0.79 & 0.88 & 0.93 & 0.95 \\
IRAC-4.5\,$\mathrm{\mu}$m & 0.11 & 0.34 & 0.56 & 0.72 & 0.84 & 0.9 & 0.94 & 0.13 & 0.4 & 0.63 & 0.78 & 0.88 & 0.94 & 0.96 \\
IRAC-5.8\,$\mathrm{\mu}$m & 0.08 & 0.26 & 0.45 & 0.62 & 0.74 & 0.84 & 0.91 & 0.11 & 0.33 & 0.53 & 0.68 & 0.8 & 0.89 & 0.95 \\
IRAC-8.0\,$\mathrm{\mu}$m & 0.07 & 0.25 & 0.43 & 0.58 & 0.68 & 0.77 & 0.85 & 0.09 & 0.3 & 0.5 & 0.64 & 0.73 & 0.82 & 0.9 \\
\hline
\textbf{Lockman Hole} & {} & {} & {} & {} & {} & {} & {} & {} & {} & {} & {} & {} & {} & {} \\
SpARCS-u & 0.31 & 0.69 & 0.87 & 0.94 & 0.97 & 0.98 & 0.99 & 0.35 & 0.73 & 0.88 & 0.95 & 0.97 & 0.99 & 0.99 \\
SpARCS-g & 0.21 & 0.54 & 0.76 & 0.87 & 0.93 & 0.96 & 0.98 & 0.33 & 0.72 & 0.88 & 0.94 & 0.97 & 0.98 & 0.99 \\
SpARCS-r & 0.24 & 0.57 & 0.78 & 0.89 & 0.94 & 0.97 & 0.98 & 0.49 & 0.82 & 0.92 & 0.96 & 0.97 & 0.98 & 0.99 \\
SpARCS-z & 0.29 & 0.64 & 0.83 & 0.91 & 0.96 & 0.98 & 0.99 & 0.52 & 0.82 & 0.91 & 0.95 & 0.97 & 0.98 & 0.99 \\
RCSLenS-g & 0.23 & 0.57 & 0.78 & 0.88 & 0.93 & 0.96 & 0.98 & 0.48 & 0.83 & 0.93 & 0.96 & 0.98 & 0.99 & 0.99 \\
RCSLenS-r & 0.25 & 0.58 & 0.78 & 0.89 & 0.94 & 0.97 & 0.98 & 0.55 & 0.86 & 0.94 & 0.97 & 0.98 & 0.99 & 0.99 \\
RCSLenS-i & 0.33 & 0.68 & 0.85 & 0.93 & 0.97 & 0.99 & 0.99 & 0.61 & 0.89 & 0.95 & 0.97 & 0.99 & 0.99 & 1.0 \\
RCSLenS-z & 0.28 & 0.62 & 0.8 & 0.89 & 0.94 & 0.97 & 0.99 & 0.56 & 0.84 & 0.92 & 0.96 & 0.97 & 0.98 & 0.99 \\
J & 0.29 & 0.67 & 0.86 & 0.95 & 0.98 & 1.0 & 1.0 & 0.46 & 0.82 & 0.93 & 0.97 & 0.98 & 0.99 & 1.0 \\
K & 0.32 & 0.7 & 0.88 & 0.96 & 0.99 & 1.01 & 1.01 & 0.47 & 0.82 & 0.92 & 0.96 & 0.98 & 0.99 & 1.0 \\
IRAC-3.6\,$\mathrm{\mu}$m & 0.1 & 0.33 & 0.56 & 0.73 & 0.83 & 0.89 & 0.93 & 0.13 & 0.4 & 0.63 & 0.79 & 0.88 & 0.92 & 0.95 \\
IRAC-4.5\,$\mathrm{\mu}$m & 0.11 & 0.34 & 0.56 & 0.72 & 0.83 & 0.9 & 0.94 & 0.13 & 0.4 & 0.63 & 0.77 & 0.87 & 0.93 & 0.96 \\
IRAC-5.8\,$\mathrm{\mu}$m & 0.08 & 0.27 & 0.46 & 0.61 & 0.74 & 0.84 & 0.91 & 0.1 & 0.32 & 0.52 & 0.67 & 0.78 & 0.88 & 0.94 \\
IRAC-8.0\,$\mathrm{\mu}$m & 0.07 & 0.25 & 0.44 & 0.58 & 0.68 & 0.77 & 0.85 & 0.09 & 0.29 & 0.49 & 0.62 & 0.7 & 0.79 & 0.88 \\
\hline
\end{tabular}
\end{table*}

\section{Bright star masking}\label{sec:appendix2:starmask}
The radii masked around stars as a function of the GAIA DR2 G-band magnitude, based on both the optical-NIR and the Spitzer detections, for the three fields, are listed in Table~\ref{tab:star_mask_radii}. The bright star masking, based on both the optical-NIR and the Spitzer detected catalogues is applied to the final, merged multi-wavelength catalogue in each field, with the process described in Sect.~\ref{sec:star_masking}.

\begin{table*}
\centering
\caption{Radii masked around bright stars (in arcsec) as a function of their GAIA DR2 G-band magnitude using both the optical-NIR and Spitzer detected catalogues.\label{tab:star_mask_radii}}
\begin{tabular}{lllllll}
\hline
\multicolumn{1}{c}{G-magnitude} & \multicolumn{2}{c}{ELAIS-N1} & \multicolumn{2}{c}{Lockman Hole} & \multicolumn{2}{c}{Bo\"{o}tes} \Tstrut \\
\multicolumn{1}{c}{[mag]} & \multicolumn{2}{c}{[arcsec]} & \multicolumn{2}{c}{[arcsec]} & \multicolumn{2}{c}{[arcsec]}\\
{} & Optical-NIR & Spitzer & Optical-NIR & Spitzer & Optical-NIR & Spitzer \Tstrut \Bstrut \\
\hline
${16.0} < \mathrm{Gmag} \leq {16.5}$ & 15 & 12 & 13 & 10 & 25 & 21\Tstrut \\
${15.0} < \mathrm{Gmag} \leq {16.0}$ & 18 & 14 & 20 & 15 & 31 & 26 \\
${14.0} < \mathrm{Gmag} \leq {15.0}$ & 23 & 16 & 23 & 18 & 39 & 33 \\
${13.0} < \mathrm{Gmag} \leq {14.0}$ & 27 & 22 & 35 & 20 & 52 & 44 \\
${12.0} < \mathrm{Gmag} \leq {13.0}$ & 33 & 27 & 40 & 25 & 62 & 52 \\
${11.0} < \mathrm{Gmag} \leq {12.0}$ & 40 & 31 & 55 & 40 & 83 & 70 \\
${10.0} < \mathrm{Gmag} \leq {11.0}$ & 55 & 50 & 80 & 45 & 95 & 80 \\
$\mathrm{Gmag} \leq {10.0}$ & 65 & 60 & 130 & 55 & 101 & 85 \\
\hline
\end{tabular}
\end{table*}

\section{Likelihood Ratio thresholds}\label{sec:comp_rel}
\begin{figure}
    \centering
    \resizebox{\hsize}{!}{
    \includegraphics[width=\columnwidth,keepaspectratio]{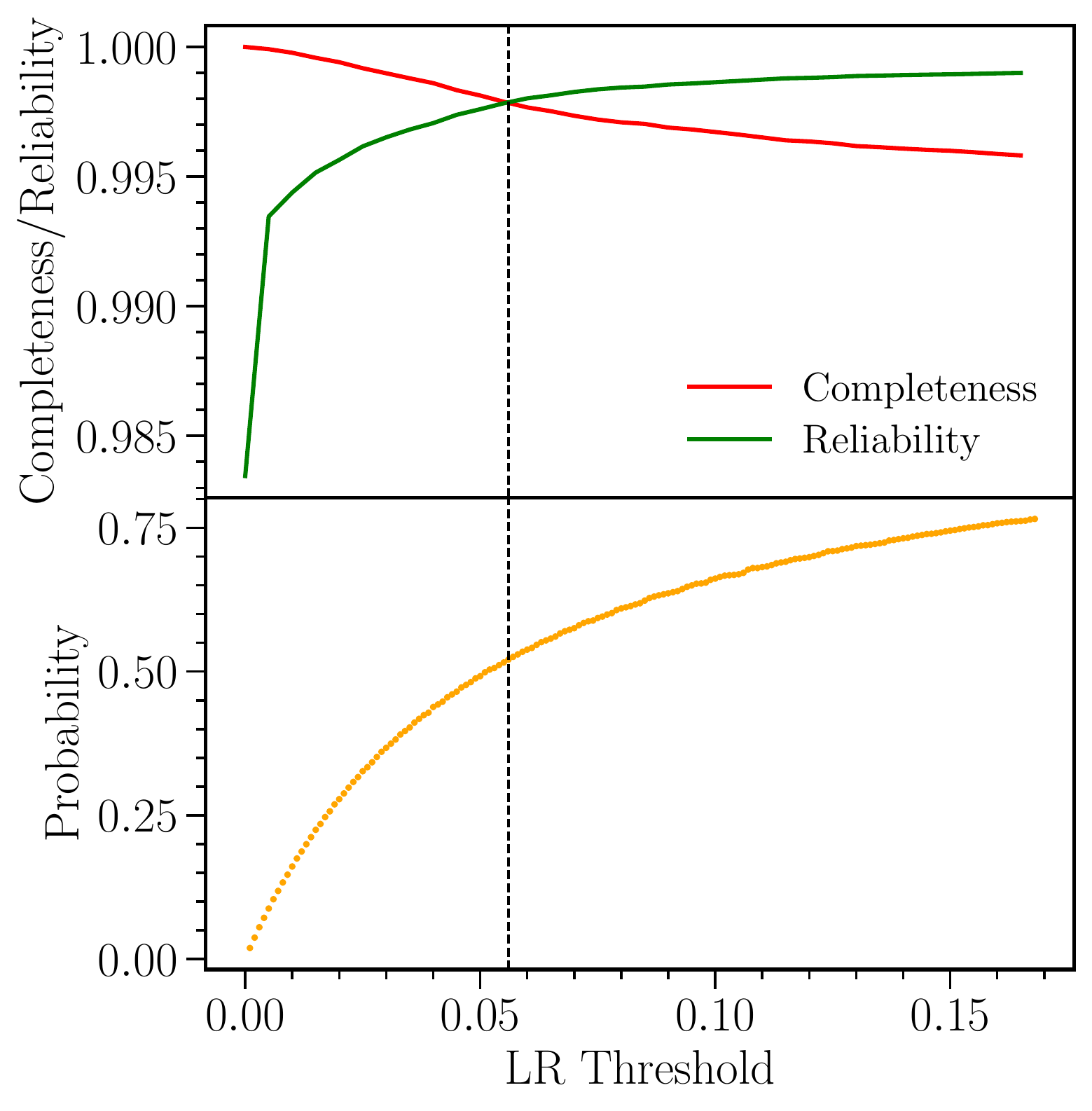}}
    \caption{\label{fig:comp_rel_prob}\textit{Top:} Completeness and Reliability curves as a function of the LR threshold (LR\textsubscript{th}) in ELAIS-N1. The LR\textsubscript{th} ($= 0.056$; vertical line) is chosen as the cross-over point of the two curves, achieving both completeness and reliability $>$ 99.7\%. \textit{Bottom:} The probability of a LR-match being a genuine cross-match as a function of the LR threshold, as it is lowered (see text and \citealt{nisbet2018role} for more detail). At the adopted LR\textsubscript{th}, the LR matches have a probability of just over 50\% of being genuine counterparts, confirming that this threshold maximised completeness with limited loss of reliability.}
\end{figure}

The LR value for each potential counterpart to be the genuine counterpart of a radio source is computed using Equation~\ref{eq:lr} as described in Sect.~\ref{sec:general_lr}. We determine the LR threshold ($LR_{\rm th}$) above which to accept a match as being the genuine counterpart as follows. For a given LR threshold $LR_{\rm th}$, one can compute the completeness $C(LR_{\rm th})$ and reliability $R(LR_{\rm th})$ as
\begin{equation}\label{eq:comp}
  C(LR_{\rm th}) = 1-\frac{1}{Q_0 N_{\rm radio}} \sum_{LR_i < LR_{\rm th}}
  \frac{Q_0 \, LR_i}{Q_0 \, LR_i + (1 - Q_0)}
, \end{equation}

\begin{equation}\label{eq:rel}
  R(LR_{\rm th}) = 1-\frac{1}{Q_0 N_{\rm radio}} \sum_{LR_i \ge LR_{\rm th}}
  \frac{1 - Q_0}{Q_0\, LR_i + (1 - Q_0)}
,\end{equation}
where $N_{\rm radio}$ is the number of radio sources in the catalogue and $LR_{i}$ is the LR of the ith radio source \citep{1977A&AS...28..211D,2003MNRAS.346..627B}. The completeness sums over the lower LR values and is defined as the fraction of real identifications that are accepted. The reliability sums over the LR values above the threshold and is defined as the fraction of accepted identifications that are correct. 

\begin{figure}
    \centering
    \includegraphics[width=\columnwidth,keepaspectratio]{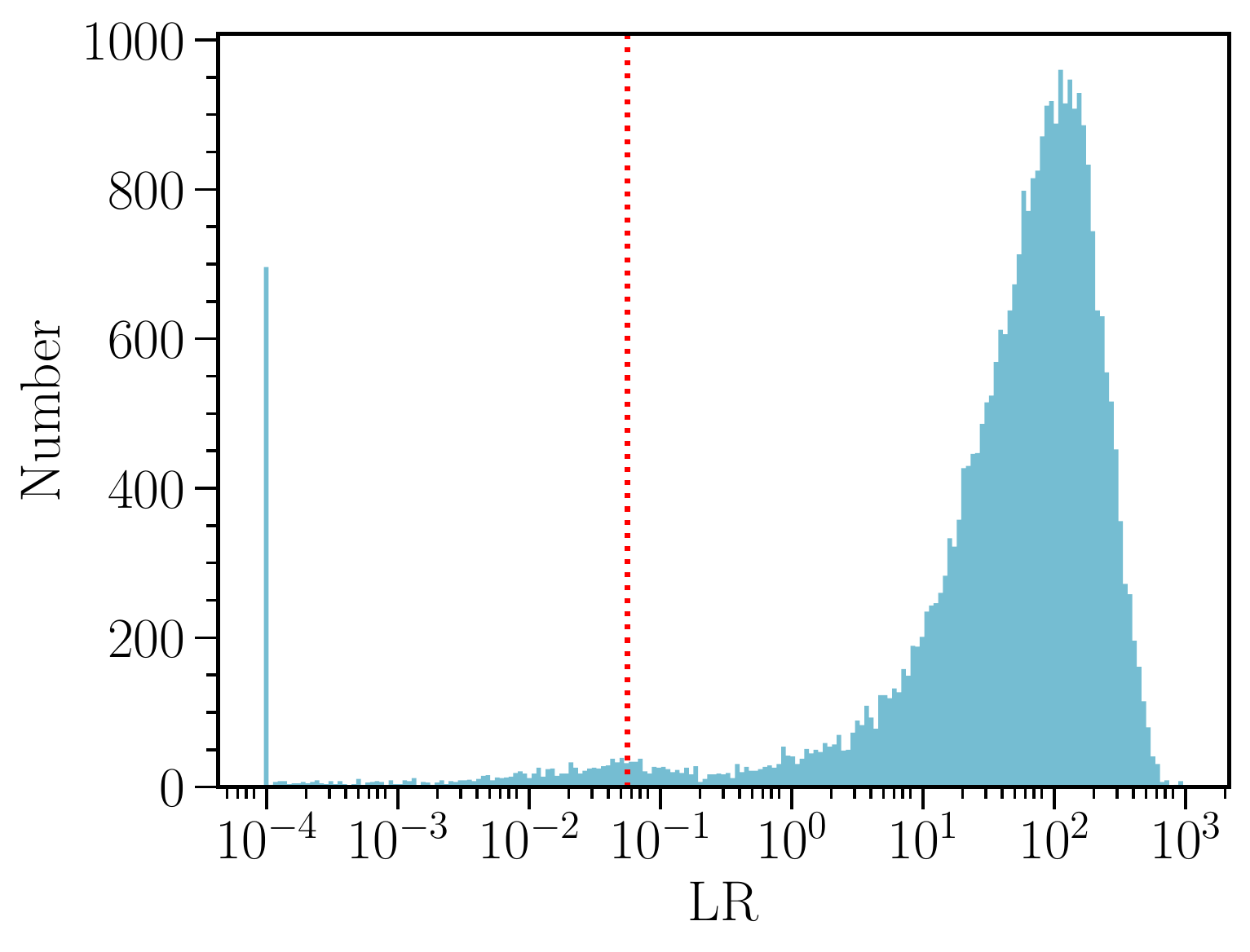}
    \caption{\label{fig:lr_hist}Histogram of the LR values in ELAIS-N1 with equal spaced bins in log-space. The vertical red line shows the LR threshold chosen (0.056). Some sources can have exceptionally low LR values, hence, sources with LR $\lesssim 10^{-4}$ are placed in the first bin to aid visual inspection.}
\end{figure}

We determine an appropriate threshold by using the cross-over point between the completeness and reliability curves as shown in Fig.~\ref{fig:comp_rel_prob} (\textit{top}) for ELAIS-N1 (see \citetalias{2019A&A...622A...2W} for further discussion). In ELAIS-N1, the LR\textsubscript{th} $= 0.056$ chosen, returns a cross-matching completeness and reliability in excess of 99.7\%. The division of the radio sources into colour bins drives down the LR values and hence the LR thresholds compared to the magnitude-only run, resulting in LR thresholds below unity. We inspect the sources with LR values near the LR\textsubscript{th} by visual examination and find that the LR\textsubscript{th} chosen results in genuine counterparts. The choice of the LR threshold can also be validated by considering the additional LR-matches, along with the change in completeness and reliability as the LR threshold is lowered, following the method described in detail by \citet{nisbet2018role}. In summary, using Equations~\ref{eq:comp} and \ref{eq:rel}, the number of genuine matches above a threshold $T$, is given by $N_{radio}~Q_{0}~C(T)$, and the total number of matches above the threshold $T$ is given by $N_{radio}~Q_{0}~C(T)/R(T)$. Then, if the threshold is lowered from $T$ to $T-\Delta T$, this will result in a set of additional matches, some of which will be genuine counterparts. The probability of the additional matches added being genuine when the threshold is changed from $T$ to $T-\Delta T$ is given as
\begin{equation}\label{eq:prob_th}
P_{genuine}(T \to T-\Delta T) = \dfrac{C(T) - C(T-\Delta T)}{\dfrac{C(T)}{R(T)} - \dfrac{C(T-\Delta T)}{R(T-\Delta T)}}.
\end{equation}
This probability $P_{genuine}(T\to T-\Delta T)$, as a function of the LR threshold is also shown in Fig.~\ref{fig:comp_rel_prob} (\textit{bottom}) for ELAIS-N1. The plot shows that at the LR threshold value chosen, the probability of the LR match being a genuine counterpart is $>$ 52\%, suggesting that the counterparts with LR values below unity (but above the threshold) are more likely to be genuine matches than false identifications. A similar analysis in Lockman Hole and Bo\"{o}tes yields probabilities of $\sim$65\% and $\sim$70\%, respectively, at the respective LR thresholds.

An appropriate choice for the LR threshold can also be visualised using a histogram of the LR values, as shown in Fig.~\ref{fig:lr_hist} for ELAIS-N1 considering the \textsc{PyBDSF} catalogue sources. There are $\approx$ 1700 sources with LRs $<$ LR\textsubscript{th}, which were sent to visual inspection. This corresponds to a fraction of $\sim$ 0.05, consistent with the final iterated $Q_{0} \sim$ 0.95 obtained using the cross-over point of the completeness and reliability curves (Fig.~\ref{fig:comp_rel_prob}). The plot also shows that most of the sources have LR values significantly higher than LR\textsubscript{th}. Moreover, $>$ 99\% of the \textsc{PyBDSF} sources with a LR > LR\textsubscript{th} have LR values above 2 $\times$ LR\textsubscript{th} (= 0.11); the probability of the cross-match identified being genuine at this point is $\sim$ 70\% (using Fig.~\ref{fig:comp_rel_prob}).

\end{appendix}
\end{document}